\documentclass[aps, pra,reprint,onecolumn,groupedaddress]{revtex4-2}
\usepackage[hidelinks]{hyperref}

\usepackage{graphicx}
\usepackage{xcolor}
\usepackage{amsmath}
\usepackage{amsfonts}
\usepackage{amssymb}
\usepackage{float}
\usepackage{multirow}
\usepackage{booktabs}
\setcitestyle{round, numbers,sort&compress}

\begin{document}

\makeatletter
\renewcommand\@biblabel[1]{#1.}
\makeatother

\makeatletter
\def\maketitle{
\@author@finish
\title@column\titleblock@produce
\suppressfloats[t]}
\makeatother

\title{Fast and converged classical simulations of evidence for the utility of quantum computing before fault tolerance}
\author{Tomislav Begu\v{s}i\'{c}}\thanks{These authors contributed equally to this work.}
\author{Johnnie Gray}\thanks{These authors contributed equally to this work.}
\author{Garnet Kin-Lic Chan}
\email{gkc1000@gmail.com}
\affiliation{Division of Chemistry and Chemical Engineering, California Institute of Technology, Pasadena, California 91125, USA}

\begin{abstract}
A recent quantum simulation of observables of the kicked Ising model on 127 qubits implemented circuits that exceed the capabilities of exact classical simulation. We show that several approximate classical methods, based on sparse Pauli dynamics and tensor network algorithms, can simulate these observables orders of magnitude faster than the quantum experiment, and can also be systematically converged beyond the experimental accuracy. Our most accurate technique combines a mixed Schr\"odinger and Heisenberg tensor network representation with the Bethe free entropy relation of belief propagation to compute expectation values with an effective wavefunction-operator sandwich bond dimension ${>}16,000,000$, achieving an absolute accuracy, without extrapolation, in the observables of ${<}0.01$, which is converged for many practical purposes. We thereby identify inaccuracies in the experimental extrapolations and suggest how future experiments can be implemented to increase the classical hardness.
\end{abstract}

\maketitle

\section*{Introduction}
As quantum computers mature,
it is critical to benchmark their performance against classical simulations. 
One approach is to compare against exact classical simulations, such as state-vector methods which store the wavefunction amplitudes of the $n$ qubits as a $2^n$-dimensional vector, or exact tensor network (TN) contractions~\cite{
boixoSimulationLowdepthQuantum2017,
chenClassicalSimulationIntermediateSize2018,
huangClassicalSimulationQuantum2020,
wuStrongQuantumComputational2021,
panSolvingSamplingProblem2022,
morvanPhaseTransitionRandom2023}, which have an exponential cost in circuit tree-width~\cite{Markov_Shi:2008}.
However, as qubit counts and circuit depths increase, quantum experiments are now beginning to exceed the capabilities of exact classical simulations. 

\begin{figure}
    \centering
    \includegraphics[width=0.8\textwidth]{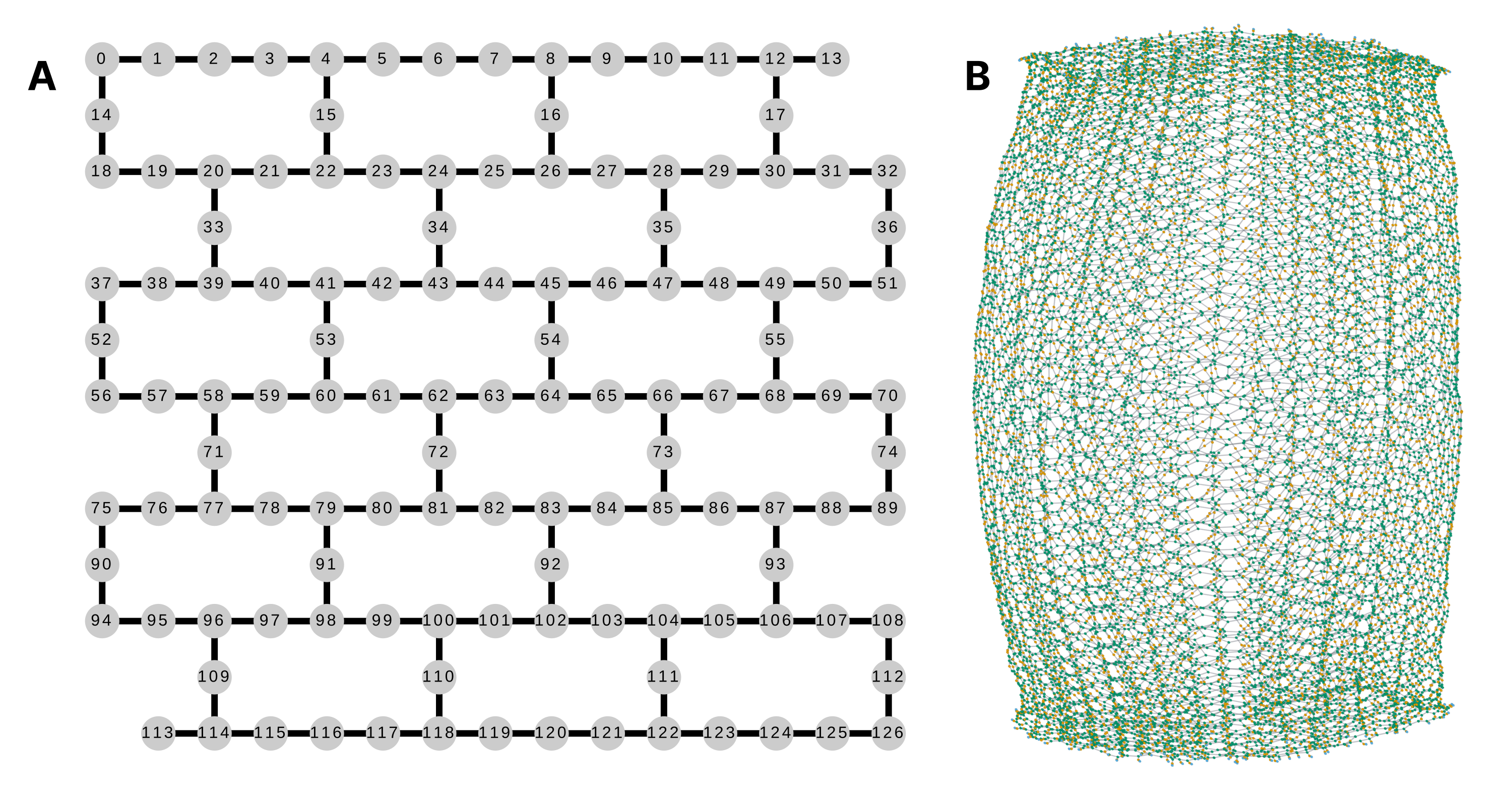}
    \caption{\textbf{Schematic overview of the simulation.}
    \textbf{A}:
    Qubit layout of the quantum device used in Ref.~\cite{kim2023evidence}.
    \textbf{B}:
    Visualization of the full tensor network representation of the quantum circuit expectation value for 20 steps. Green tensors are entangling RZZ gates, orange tensors are RX rotations.
    }
    \label{fig:big-overview}
\end{figure}

A recent quantum experiment of this kind simulated the kicked Ising model on IBM's 127-qubit processor~\cite{kim2023evidence}.
Using zero-noise extrapolation~\cite{temme2017error} of the experimental data, Ref.~\cite{kim2023evidence}
presented evidence for the utility of quantum computing before fault tolerance. 
Starting from the ferromagnetic state, the experiment implemented the circuit 
\begin{equation}
    U = \large( \prod_{\langle j, k\rangle} e^{i \pi Z_j Z_k /4}  \prod_{j} e^{-i \theta_h X_j /2} \large)^T
    \label{eq:circuit}
\end{equation}
consisting of $T$ steps of Clifford $ZZ$ and non-Clifford $X$ rotations. In Eq.~\ref{eq:circuit}, $\theta_h$ is a rotation angle that is varied, $j$ and $k$ run over all 127 qubits, and $\langle j, k\rangle$ denotes neighbors on the heavy hexagon lattice of the \textit{Eagle} quantum processor \texttt{ibm\_kyiv} (Fig.~\ref{fig:big-overview}A). It was argued in Ref.~\cite{kim2023evidence} that the expectation values $\langle O \rangle = \langle 0 | U^{\dagger} O U |0 \rangle$ of different Pauli observables $O$ were not only beyond the reach of exact classical simulations, but furthermore could not be computed to the experimental accuracy with certain approximate classical techniques, such as matrix product state (MPS) or isometric 2D tensor network (isoTNS) simulations \cite{zaletel2020isometric}.

However, in a recent unpublished, preliminary note by some of us~\cite{Begusic2023} using an approximate classical method termed sparse Pauli dynamics, we showed that we could compute the experimental expectation values with an accuracy comparable to the experimental extrapolation, in a time faster than the experiment. The same conclusion was reached in multiple preprints from other groups~\cite{tindall2023efficient, Kechedzhi2023, Shao2023, liao2023simulation, rudolph2023classical}, including some of the authors of the original experiment~\cite{Anand2023}. Nonetheless, as pointed out in Ref.~\cite{Anand2023}, the remaining errors in the classical simulations, as well as in the extrapolated experimental data, still left considerable uncertainty in the precise values of the observables targeted in the original experiment.

Here we show that multiple approximate classical methods can be converged for all examples studied in \cite{kim2023evidence}, including the 20-step dynamics involving 127 qubits, well beyond the results in the initial preprints~\cite{tindall2023efficient, Kechedzhi2023, Anand2023, Shao2023} and with a remaining uncertainty substantially less than the experimental extrapolation error bars. To this end, we provide a full account, and further extend the application, of the Clifford-based sparse Pauli dynamics (SPD) in~\cite{begusic2023simulating} to larger scale simulations.
We further introduce new tensor network techniques that use lazy belief propagation, in the Schr\"odinger picture (for projected entangled pair states (PEPS)), in the Heisenberg picture (for projected entangled pair operators (PEPO)), and in a mixed Schr\"odinger/Heisenberg picture (PEPS/PEPO expectation value). An important component in these improved tensor network simulations is the application of the Bethe free entropy belief propagation formula, extending the ideas in~\cite{pancottiOnestepReplicaSymmetry2023}, that allows us to completely avoid the expensive tensor network contraction and use large bond dimensions in the mixed Schr\"odinger/Heisenberg picture.
We carefully benchmark the errors associated with these techniques, and introduce a uniform norm metric to compare the quality of the different approximations. In our most accurate simulations of the largest circuits in the mixed picture (Fig.~\ref{fig:big-overview}B), and using the free entropy formula, we effectively evaluate a tensor network representation of the expectation value with bond dimension 16,777,216. Without resorting to any extrapolations, the associated absolute error we achieve in the targeted observable is conservatively estimated as $< 0.01$. For many practical applications, this may be considered fully converged.

\section*{Results}

\subsection{Approximate classical methods for quantum circuit expectation values}

Various classical strategies have been developed to approximate quantum circuit expectation values with a cost that depends on characteristics of the quantum circuit. For example, Clifford-based approaches \cite{Aaronson_Gottesman:2004,Bravyi_Gosset:2016,Bennink_Pooser:2017,Bravyi_Howard:2019, rall2019simulation}, can treat circuits with large numbers of qubits and highly entangled states as long as there is only a small number of non-Clifford gates. Similarly, approximate TN methods provide accurate results so long as the circuit generates a limited amount of entanglement. 
In practice, the goal with such approximate techniques is to numerically converge the results with respect to a parameter controlling the accuracy. Below we describe approximate simulation methods in both the near-Clifford and tensor network categories, and the innovations we use to converge the expectation values of the kicked Ising experiment.

\subsubsection{Sparse Pauli dynamics}

The SPD approach \cite{Begusic2023} was recently introduced as a variant of Clifford perturbation theory \cite{begusic2023simulating}. It is based on representing a target observable in the Heisenberg picture as a sum of Pauli operators $O = \sum_{P \in \mathcal{P}} a_P P$, where $\mathcal{P}$ is some subset of $n$-qubit Paulis and $a_P$ are the corresponding complex coefficients. A Pauli rotation gate $U_{\sigma}(\theta) = e^{-i\theta \sigma / 2}$, defined by a real rotation angle $\theta$ and a Hermitian $n$-qubit Pauli operator $\sigma$, transforms any $n$-qubit Pauli operator $P$ according to
\begin{equation}
    U_{\sigma}(\theta)^{\dagger} P U_{\sigma}(\theta) = 
    \begin{cases}
    \cos(\theta) P + i \sin(\theta) \sigma P, \, &\{\sigma, P\} = 0, \\
    P, &[\sigma, P] = 0.
    \end{cases}
    \label{eq:Pauli_rotation}
\end{equation}
Consequently, for each Pauli operator $P \in \mathcal{P}$ that anticommutes with $\sigma$ of the rotation gate, the representation of the observable $O$ must be expanded to $\mathcal{P}^{\prime} = \mathcal{P} \cup \{\sigma P\}$. This will, in general, lead to an exponential growth of the number of terms in the sum. However, if the rotation gate is Clifford, meaning that $\theta = k \pi /2$ for integer $k$, only one of the terms in the right-hand side of Eq.~$\ref{eq:Pauli_rotation}$ will remain and the number of Paulis will not increase. In SPD, we make use of this by initially transforming the non-Clifford Pauli rotation gates and the observable by the Clifford gates of the circuit. Similarly, we rewrite the Pauli rotation angles as $\theta = \theta^{\prime} + k \pi /2$ with $\theta^{\prime} \in (-\pi/4, \pi/4]$, where the term proportional to $\pi/2$ is treated as a separate Clifford gate. To truncate the exponentially growing Pauli series representing the Heisenberg-evolved observable, in Ref.~\cite{Begusic2023} we employed a perturbative criterion for adding or removing Pauli operators. Here, we formulate SPD using a threshold-based criterion, i.e., by truncating the Pauli representation of the observable to those Paulis whose coefficients are greater than the prescribed threshold (see Methods for full details). Then the accuracy of the simulation is systematically improved by reducing this threshold until convergence. (We note that recently, other related Pauli-based approaches have been proposed, with truncation based on Fourier expansion or Hamming weights of the Pauli operators \cite{Nemkov_Fedorov:2023, Shao2023, fontana2023classical, rudolph2023classical}).

\subsubsection{Tensor network simulations}

\begin{figure*}[t!]
    \centering
    \includegraphics[width=\linewidth]{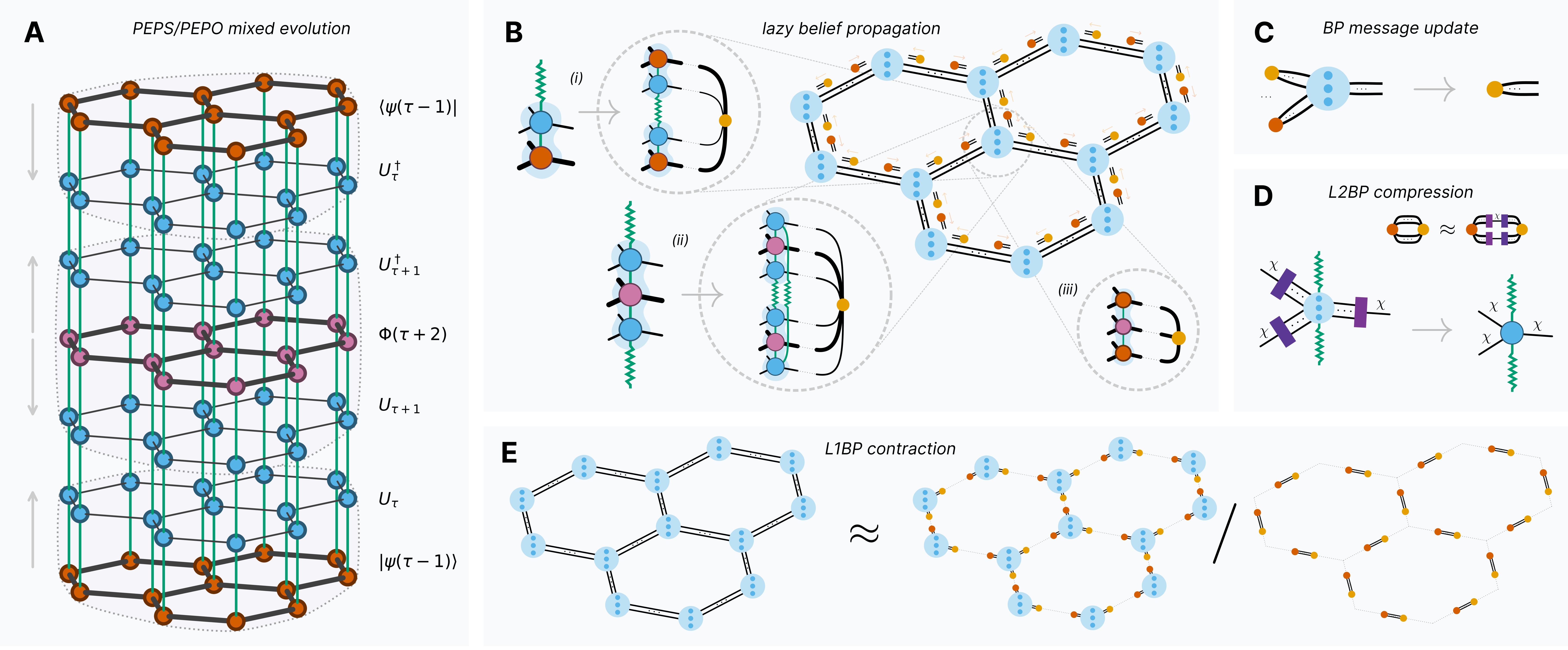}
    \caption{
    \textbf{
    Tensor network simulation overview.
    } Full discussion in main text and Methods.
    \textbf{A}: schematic of mixed PEPS/PEPO evolution on toy hexagonal lattice. The PEPS (red) is evolved using the next unitary layer, then compressed using L2BP. The PEPO (pink) is evolved and compressed similarly. 
    Green bonds: `physical' indices connecting a single site.
    Once all unitary layers have been incorporated,
    the remaining PEPS/PEPO/PEPS sandwich is contracted using L1BP.
    \textbf{B}: belief propagation (BP) on the lattice. 
    Blue circles: tensors on sites (which represent lazy contraction of tensors from multiple layers, represented by   internal dots). Each bond has left and right tensor messages (yellow/red pairs of dots).    
    We run BP algorithms on three types of tensor networks which differ in the definition of the site-tensors
    (shown in insets). (i) Site-tensor when compressing the PEPS and its unitary update, minimizing the 2-norm error (L2BP method).
    (ii) Site-tensor when  compressing the PEPO and its unitary  update, minimizing the 2-norm error (L2BP method).
    (iii) Site-tensor when using the Bethe free-entropy formula to evaluate the PEPS-PEPO-PEPS expectation value (L1BP method).
    \textbf{C}: 
    Lazy BP contraction for updating a single message. 
    \textbf{D}: 
    Lazy 2-norm belief propagation (L2BP) \emph{compression} of a local site.
    From converged messages, pairs of projectors (purple rectangles), from the upper right approximation, are found for each bond.
    Projectors adjacent to each site are then contracted in, yielding single tensors with maximum dimension $\chi$.
    \textbf{E}: 
    Lazy 1-norm belief propagation (L1BP) tensor network \emph{contraction}.
    Given converged messages, the global contracted value is estimated from contractions of tensors and messages.
    }
    \label{fig:tnlbp}
\end{figure*}

In our TN simulations we allow ourselves to partition the unitary between the evolution of the states, and that of the observable $O$. Thus, in general, 
we evolve both a PEPS representation forwards to step $\tau$, $|\psi(0)\rangle\rightarrow |\psi(\tau)\rangle$, and a PEPO backwards $\Phi(T + 1)\rightarrow\Phi(\tau + 1)$ (note $\Phi(T+1) = O$). 
In the limit that $\tau = T$, only the PEPS is evolved (Schr\"odinger picture) and the entanglement is entirely stored in the state. Conversely, if $\tau = 0$, only the PEPO is evolved (Heisenberg picture), and the entanglement is stored in the operator.
By allowing other values of $\tau$ (mixed picture)
we can limit ourselves to moderate entanglement growth in both the state and operator.
A schematic overview of the TN methods is shown in Fig.~\ref{fig:tnlbp} (further details in Methods and supplementary materials). Below, we refer to the case of purely Schr\"odinger evolution ($\tau=T$) as the `PEPS method', the case of purely Heisenberg evolution ($\tau=0$) as the `PEPO method', and the case of half and half ($\tau = T / 2$) as the `MIX method'. It is this MIX method that we find converges fastest and we show is the most accurate.

We use two key \emph{belief propagation}~(BP) tools to perform these evolutions efficiently, and crucially, evaluate the final overlap $\langle \psi (\tau) | \Phi(\tau + 1) | \psi(\tau) \rangle$ at bond dimensions, $\chi$, that are far beyond exact contraction. 

The first is 2-norm BP compression. 2-norm BP compression is a version of simple update compression~\cite{jiangAccurateDeterminationTensor2008,lubasch2014algorithms,lubasch2014unifying}, where tensor network bonds are compressed by the singular value decomposition of the contraction of pairs of tensors, and the rest of the tensor network (the environment) is represented as a product of gauges. To obtain these gauges, the BP iteration is applied to the doubled tensor network used to compute the 2-norm (i.e. Fig.~\ref{fig:tnlbp}B(i) for PEPS compression and Fig.~\ref{fig:tnlbp}B(ii) for PEPO compression).
As well summarized in 
Refs.~\cite{alkabetzTensorNetworksContraction2021,tindallGaugingTensorNetworks2023}, at the convergence of the BP procedure, the BP messages correspond to working in the `super orthogonal' gauge defined in Ref.~\cite{ranOptimizedDecimationTensor2012}. The key difference with the usual simple update formulation is that the BP message iteration allows for this gauge to be computed without modifying the original tensors. This then allows for the lazy computation of gauges without forming intermediate tensors of large bond dimension~\cite{wangTensorNetworkMessage2023,guoBlockBeliefPropagation2023a}. We refer to this lazy implementation of 2-norm BP compression as L2BP. 
On a lattice such as heavy-hex with coordination number 3, it scales as $\mathcal{O}(\chi^4)$.

The second tool, which we refer to as lazy 1-norm belief propagation (L1BP) \emph{contraction}, is an extension to quantum expectation values of recent work connecting BP techniques and TNs~\cite{pancottiOnestepReplicaSymmetry2023}.
The essential idea is to approximate the final overlap contraction between states and operators as the exponential of the Bethe free entropy~\cite{mezardInformationPhysicsComputation2009}. To obtain this, we first compute converged BP messages of the 
sandwich tensor network of the operators and states (Fig.~\ref{fig:tnlbp}B(iii)), and the value of the contraction is approximated by a simple function of the contractions between tensors and converged BP messages (Fig.~\ref{fig:tnlbp}E).
In the quantum setting, we apply this relation despite the tensor network being non-positive.
L1BP contraction can be utilized to compute local and non-local operators (such as high weight Paulis) as well as the norms of $|\psi\rangle$ and $\Phi$.
On the heavy-hex lattice, assuming both $\psi$ and $\Phi$ have bonds of size $\chi$ (this refers to the MIX method), it scales as $\mathcal{O}(\chi^6)$.
Using it we here evaluate the expectation value of a PEPS/PEPO expectation value where the PEPS and PEPO each have bond dimension $\chi=256$, for a combined effective TN expectation value bond dimension of $\chi=256^3=16,777,216$.
L1BP contraction is exact for tree geometries, but is only a heuristic approximation away from such geometries. Consequently, we verify the accuracy of the L1BP contraction for our expectation values in multiple ways in the simulations below.

For clarity, we briefly note some differences to other approaches. In Ref.~\cite{tindall2023efficient} the authors perform PEPS evolution using 2-norm belief propagation to compress after each gate application. This is subtly different from our PEPS evolution, where
due to our lazy approach it is natural to evolve and compress one or more full layers of gates simultaneously, which we find improves accuracy.
A second difference is that Ref.~\cite{tindall2023efficient} evaluates the observables using the approximate local density matrices computed in the BP approach, rather than our L1BP expression. We discuss the implications of this for the accuracy of 
 $\langle Z_{62} \rangle$ at $\theta_h > \pi / 4$ later.
{Finally, in another paper that appeared at the same time as our preprint, Ref.~\cite{liao2023simulation}, the authors perform PEPO Heisenberg evolution using simple update~\cite{jiangAccurateDeterminationTensor2008}, before contracting the final PEPO exactly at $\tau=0$. In our PEPO method we apply entire layers of gates simultaneously, and the use of L1BP contraction enables us to directly contract $\Phi$ with $\log_2(\chi) / 2$ layers remaining, avoiding the extra compression steps in Ref.~\cite{liao2023simulation}.
}

\subsection{Simulation of expectation values of the kicked Ising circuit}

\begin{figure}
    \centering
    \includegraphics[width=0.85\textwidth]{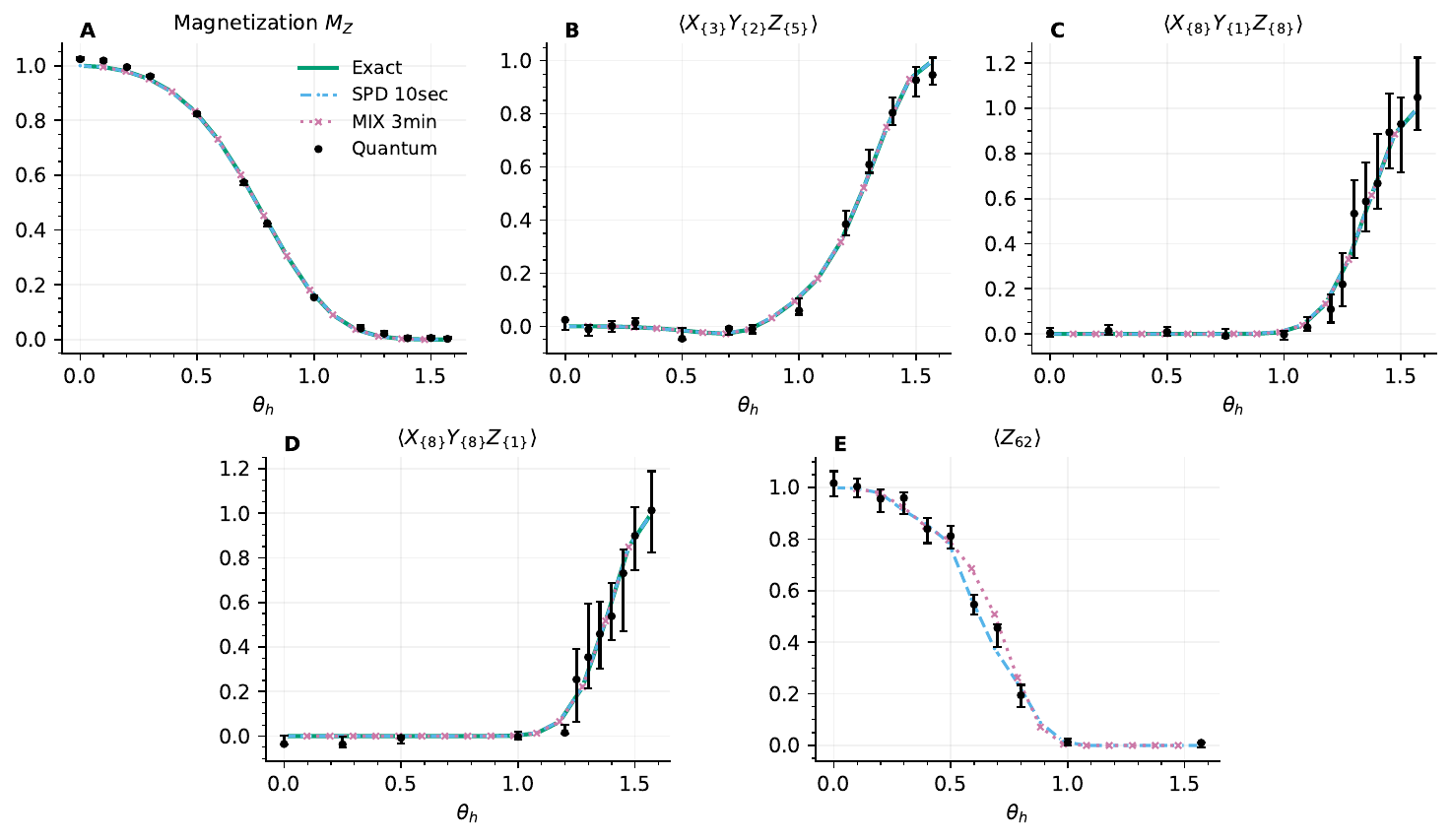}
    \caption{\textbf{Expectation values of Ref.~\cite{kim2023evidence}}.
    \textbf{A}:
    magnetization; 
    \textbf{B}:
    $X_{\{3\}}Y_{\{2\}}Z_{\{5\}} \equiv X_{13,29,31}Y_{9,30}Z_{8,12,17,28,32}$; 
    \textbf{C}:
    $X_{\{8\}}Y_{\{1\}}Z_{\{8\}} \equiv X_{37,41,52,56,57,58,62,79}Y_{75}Z_{38,40,42,63,72,80,90,91}$, all after 5 steps; \textbf{D}: 
    $X_{\{8\}}Y_{\{8\}}Z_{\{1\}} \equiv X_{37,41,52,56,57,58,62,79}Y_{38,40,42,63,72,80,90,91}Z_{75}$ for a 5-step circuit with an additional $X$-rotation layer, and 
    \textbf{E}:
    $\langle Z_{62} \rangle$ after 20 steps. The results presented include the exact benchmarks calculated using TN contraction (``Exact'', \textbf{A}--\textbf{D}), SPD results obtained within on average 10 seconds per point on a single core of a laptop computer (``SPD 10sec''), MIX TN results on a single GPU under 3 minutes (``MIX 3min''), and the zero-noise extrapolated quantum experiment (``Quantum''). All data apart from the quantum simulations were computed at $\theta_h$ between zero and $\pi/2$ in increments of $\pi/32$. Thresholds used for the SPD simulations are reported in Table~\ref{tab:Thresholds1}.
    }\label{fig:SPD_Quick}
\end{figure}

We start with a brief recapitulation of the results in the unpublished note~\cite{Begusic2023} recomputed using the slightly modified SPD algorithm in the current work, as well as from fast but approximate TN simulations. In Fig.~\ref{fig:SPD_Quick} we show these classical simulations of the experimental data of Ref.~\cite{kim2023evidence}.
The observables include the magnetization $M_Z = \sum_j\langle Z_j \rangle/n$ and two high-weight Pauli observables ($X_{13,29,31}Y_{9,30}Z_{8,12,17,28,32}$ and $X_{37,41,52,56,57,58,62,79}Y_{75}Z_{38,40,42,63,72,80,90,91}$) after 5 steps ($T=5$ in Eq.~\ref{eq:circuit}), another weight-17 Pauli observable $X_{37,41,52,56,57,58,62,79}Y_{38,40,42,63,72,80,90,91}Z_{75}$ for a circuit consisting of 5 steps and an additional layer of $X$ rotations, and $\langle Z_{62} \rangle$ after 20 steps of the circuit in Eq.~\ref{eq:circuit}. 
Exact benchmarks, computed with an exact TN simplification and contraction method \cite{gray2022hyper}, are shown for the first four examples. (We note that the exact result shown in Fig.~\ref{fig:SPD_Quick}D was not computed in the original Ref.~\cite{kim2023evidence}).

The data generated using SPD are simulated within on average 10 seconds per point on a single core of a laptop computer, and are compatible with the experimental zero-noise extrapolated data. As observed in~\cite{Begusic2023}, this is three orders of magnitude faster than the reported quantum wall-clock run time for Figs.~\ref{fig:SPD_Quick}D,~E (4 h and 9.5 h, respectively) and faster than the hypothesized run time of IBM's Eagle processor without any classical processing steps (estimated at about 5~min for Fig.~\ref{fig:SPD_Quick}E). 
{
We also show data from the MIX TN method with $\chi=64$. The bond dimension here is chosen such that the most expensive computations (Figs.~\ref{fig:SPD_Quick}A, E) complete in less than 3 minutes of computation time on a single consumer GPU (Figs.~\ref{fig:SPD_Quick}B-D only require about 10 seconds per point). We see again that the TN simulations generate data compatible with the experimental zero-noise extrapolated data faster than the run time of the quantum processor.
}

\begin{figure}
    \centering
    \includegraphics[width=\textwidth]{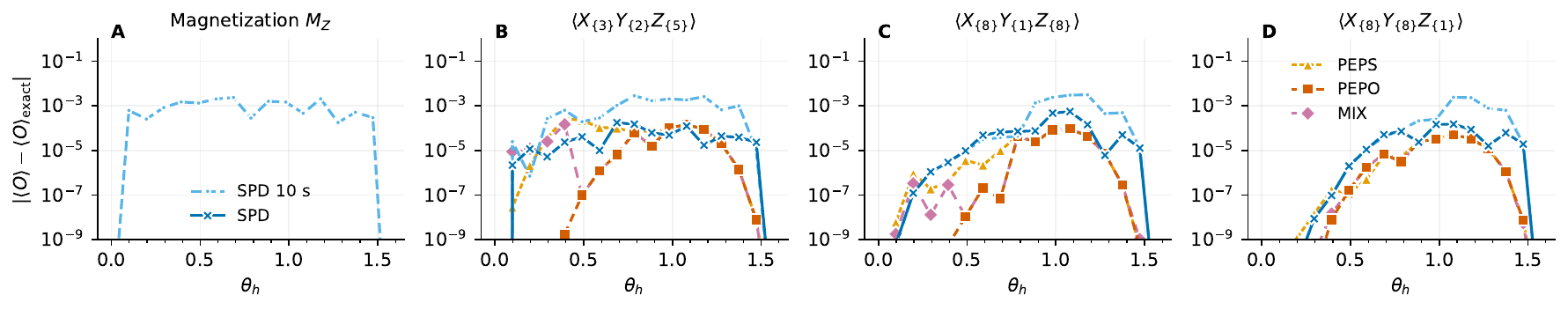}
    \caption{\textbf{Absolute errors of approximate methods for the observables presented in Fig.~\ref{fig:SPD_Quick}.} We show the SPD results obtained within on average 10 seconds per point (``SPD 10 s''), converged SPD results (``SPD''), and the three TN methods (\textbf{B}--\textbf{D}): PEPS, PEPO, and MIX (see text). Thresholds used for the converged SPD simulations are reported in Table~\ref{tab:Thresholds1}.}
    \label{fig:Errors}
\end{figure}

We next turn to computing more accurate data using the approximate SPD and TN approaches. To do so, in Fig.~\ref{fig:Errors} we first show the absolute errors of these approximate simulation methods for the examples in Figs.~\ref{fig:SPD_Quick}A--D corresponding to shallow circuits, where we have access to an exact result. Our fast SPD results already match the exact values very well, with a maximum error over all points on the order of $10^{-3}$. The 
magnetization can be computed numerically exactly with SPD (zero threshold) within $\sim 15$ minutes per point on a single core, while the error of the 
other three observables can be evaluated with a maximum error around $10^{-4}$ (Fig.~\ref{fig:Errors}B--D, dark blue line), where the most expensive points take up to 45 min, 5 h, and 6 h (see Table~\ref{tab:timings-spd}), respectively, on 4 cores in our Python implementation. For these shallow depth circuits, the TN calculations with $\chi=64$ (as discussed, taking a maximum of 3 minutes on the consumer GPU) already result in no truncation, thus the error is solely due to the 1-norm BP contraction approximation. These errors are shown for the PEPS method, the PEPO method, and the MIX method for the observables computed in Figs.~\ref{fig:Errors}B-D; in all cases, they are less than $10^{-4}$.

\begin{figure}
    \centering
    \includegraphics[width=0.85\textwidth]{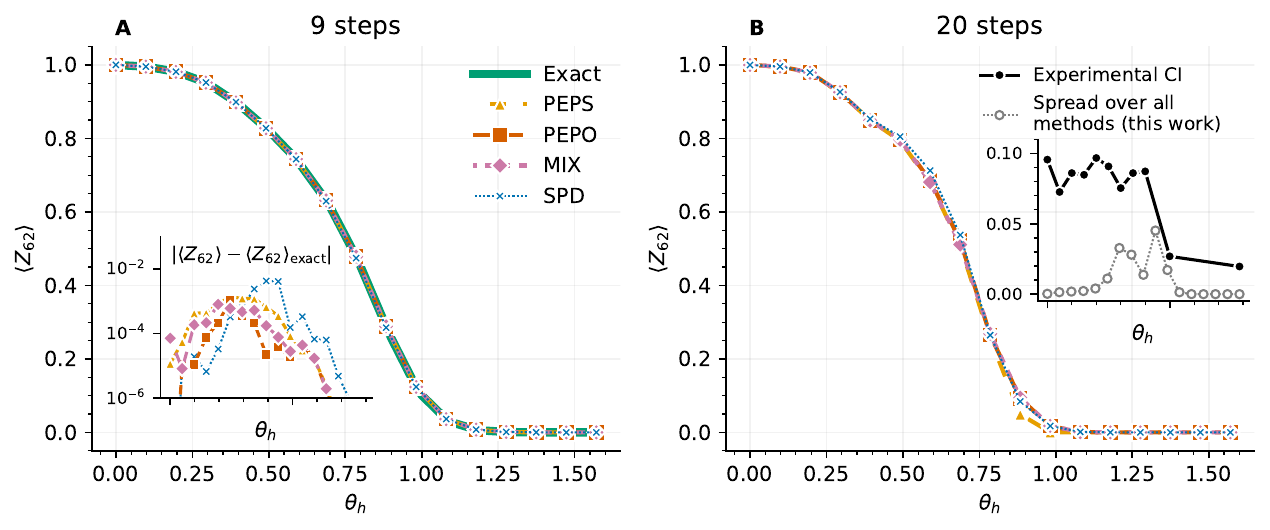}
    \caption{\textbf{Expectation value of $\mathbf{\langle Z_{62} \rangle}$.} \textbf{A}: after 9 steps; \textbf{B}: after 20 steps of the circuit in Eq.~\ref{eq:circuit}, as calculated with SPD and PEPS, PEPO, and MIX TN methods. For the 9-step result, we compare these to the exact benchmark; for the 20-step result, we compare the spread of these classically simulated expectation values with the confidence intervals (CI) of the extrapolation procedure used in the quantum simulation experiment \cite{kim2023evidence}. 
    All data were computed at $\theta_h$ between zero and $\pi/2$ in increments of $\pi/32$. Thresholds used for SPD are reported in Table~\ref{tab:Thresholds2} and bond dimensions used for TN simulations are in Table~\ref{tab:tn-chis}.}
    \label{fig:Z_62}
\end{figure}

We now discuss converging the most difficult observable in Ref.~\cite{kim2023evidence}, namely $\langle Z_{62} \rangle$ with a 20-step circuit. 
For orientation, we first give results for a simpler 9-step circuit for this observable where we can still compute an exact benchmark. The data are presented in Fig.~\ref{fig:Z_62}.
For the 9-step expectation value (Fig.~\ref{fig:Z_62}A), all results agree to within $0.01$ with the exact TN benchmark. At our employed parameters (see Table~\ref{tab:Thresholds2} for SPD thresholds, Table~\ref{tab:tn-chis} for TN $\chi$ values) the largest errors appear at the highly non-Clifford points in the middle of the studied $\theta_h$ range. 
For the 20-step circuit (Fig.~\ref{fig:Z_62}B, SPD thresholds in Table~\ref{tab:Thresholds2}, up to $\chi=320$ for TN methods, see Table~\ref{tab:tn-chis}) we no longer have exact data, but we can still examine the maximum spread across all the methods. We find this to be less than $0.045$. Notably, at each computed $\theta_h$, the spread in the classical results is substantially smaller than the error bar in the experimental extrapolation at a similar $\theta_h$.

\begin{figure}
    \centering
    \includegraphics[width=\textwidth]{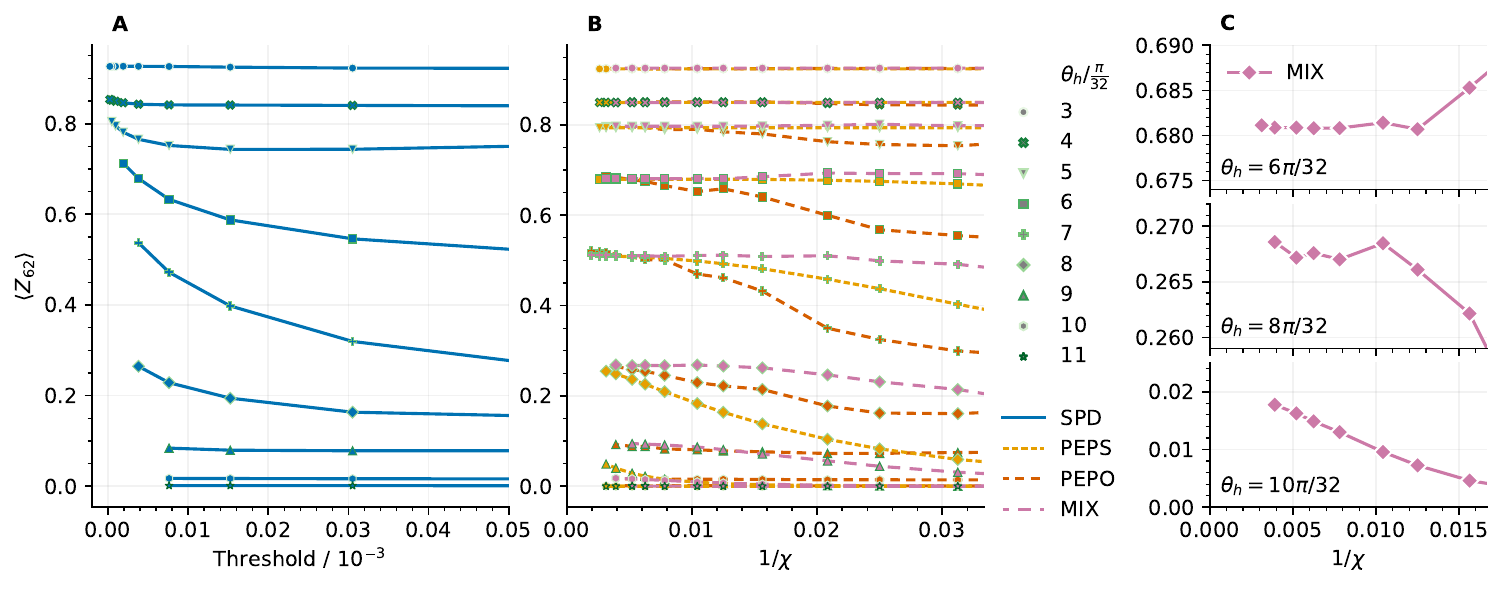}
    \caption{\textbf{Convergence of $\mathbf{\langle Z_{62} \rangle}$ after 20 steps.} \textbf{A}: SPD with respect to the threshold; \textbf{B}: the PEPS, PEPO, and MIX methods with respect to bond dimension $\chi$. The results are presented for a range of $\theta_h$ values indicated by different markers.
    \textbf{C}: Zoom in on the convergence of the MIX method at $\theta_h = 6\pi/32$,  $8\pi/32$ and $10\pi/32$ (see Fig.~\ref{fig:convergence-extrapolation-error-mix} for other values of $\theta_h$).}
    \label{fig:convergence}
\end{figure}

In Fig.~\ref{fig:convergence} we show the convergence with respect to threshold (for SPD) and bond dimension (for the TN methods). 
At intermediate $\theta_h$, the SPD data appears to be less converged than the TN data at the smallest accessible values of the threshold.
Interestingly, both Heisenberg-picture methods, namely PEPO and SPD, exhibit a non-monotonic convergence behavior (see Fig.~\ref{fig:convergence-wide_xrange}), similar to the Heisenberg-picture matrix-product operator (MPO) simulations of Ref.~\cite{Anand2023}. In turn, PEPS and (to some extent) MIX simulations exhibit a monotonic convergence with respect to $\chi$. Fig.~\ref{fig:convergence}C shows that the MIX results are the most converged among the three TN methods. In the supplementary material, we provide three estimates of the residual error of the MIX method with respect to $\chi$: these estimates range from well less than $10^{-2}$, to perhaps as small as $10^{-3}$.

\begin{figure}
    \centering
    \includegraphics[width=\textwidth]{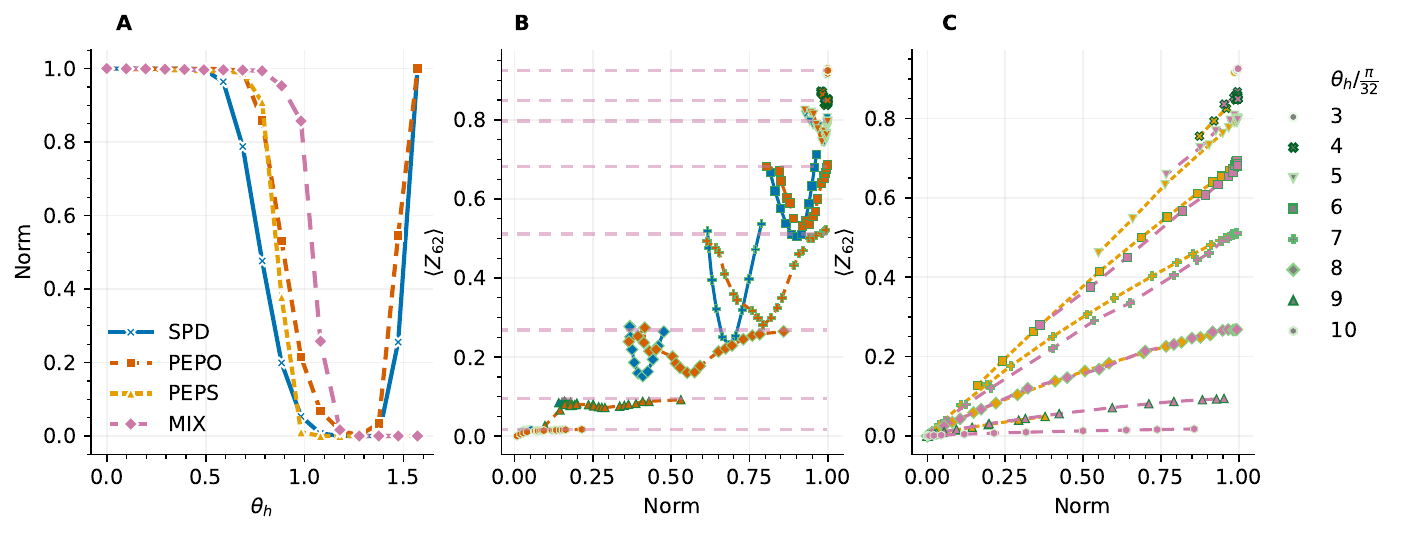}
    \caption{\textbf{Convergence analysis of norm and $\mathbf{\langle Z_{62} \rangle}$ after 20 steps.} Norm as a function of angle $\theta_h$ (\textbf{A}) and the convergence of $\langle Z_{62} \rangle$ after 20 steps for SPD and PEPO (\textbf{B}), and PEPS and MIX (\textbf{C}) methods with respect to the norm. $\theta_h$ values are indicated by different markers. Horizontal lines in panel \textbf{B} correspond to the highest-$\chi$ MIX results at respective $\theta_h$.}
    \label{fig:convergence-norm}
\end{figure}

It is useful to define a metric that allows us to compare the results of different approximate methods on a more equal footing. For this, we use the norm as a common convergence parameter and an approximate measure of fidelity. For PEPS, we use the wavefunction norm $N_{\psi} = \lVert \psi \rVert$, for operator-based methods (PEPO and SPD) we use the Frobenius norm $N_O = \sqrt{\text{Tr}(O^{\dagger} O)/2^n}$, and for the mixed-picture approach we use $N_{\rm MIX} = N_{\psi} N_O$. These are shown in Fig.~\ref{fig:convergence-norm}A as functions of the rotation angle $\theta_h$ in the 20 step simulation. Note that achieving an accurate fidelity is expected to be much harder than computing an expectation value, thus this represents a stringent test of the accuracy of the classical methods.
We find that the operator-based approaches (SPD, PEPO) follow a similar trend, where the norm is close to $1$ for Clifford and near-Clifford angles, and low or even near-zero for intermediate values of $\theta_h$. Conversely, the PEPS approach yields near-zero norms at large $\theta_h$, even near the Clifford angle $\pi/2$. The MIX approach achieves norms $\sim 1$ for $\theta_h < 10\pi/32$, and outperforms all other approaches for $\theta_h < 13\pi/32$, although it
still yields near-zero norm near $\pi/2$. However, both the PEPS and MIX approaches correctly find that the expectation values vanish at the high-$\theta_h$ points, indicating that any complicated entanglement generated by the circuit is not reflected in the expectation value.
 In Figs.~\ref{fig:convergence-norm}B, C, we plot the convergence of $\langle Z_{62}\rangle$ against norm.
Similar to in Fig.~\ref{fig:convergence-wide_xrange}, we observe non-monotonic convergence in the expectation value for the Heisenberg-based methods, while PEPS and MIX show smooth and monotonic convergence with norm. Notably, the MIX method achieves a large norm for all non-zero $\langle Z_{62}\rangle$.

\begin{figure}
    \centering
    \includegraphics[width=0.9\textwidth]{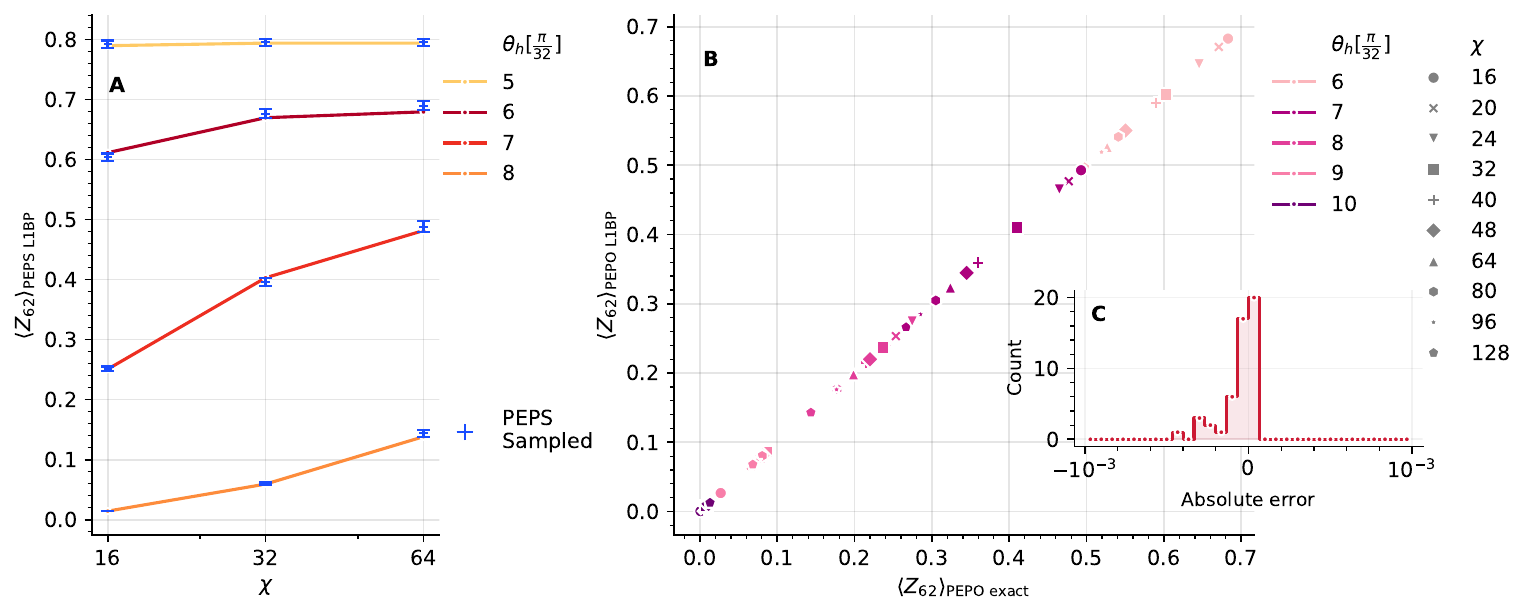}
    \caption{
    \textbf{Verifying L1BP contraction for PEPS and PEPO evolution.}
    \textbf{A}: Comparison of $\langle Z_{62} \rangle$ computed by L1BP contraction (lines) and via Monte Carlo sampling (markers with error-bars) for the PEPS method at 20 steps at a range of $\chi$ and $\theta_h$.
    The sampled points are computed from 10,000 shots each.
    \textbf{B}: Expectation values $\langle Z_{62} \rangle$ for a 20-step circuit computed with L1BP contraction ($y$-axis) and exact ($x$-axis) contraction at different values of $\theta_h$ and for different bond dimensions $\chi$. 
    \textbf{C}: histogram of the absolute errors of all points shown. Note to enable exact contraction we evolve the PEPO all the way to time 0, the final layer is not compressed leading to a contribution of $\times4$ to $\chi$. This is slightly different to the PEPO method elsewhere in the text.
    }
    \label{fig:pepo-verify}
\end{figure}

The above studies support the convergence of our simulation methods with respect to the threshold/bond dimension, and demonstrate in particular that estimates from the MIX TN method are particularly well converged with respect to $\chi$. However, as discussed earlier, the TN methods implemented here include an additional approximation in the form of the 1-norm belief propagation (L1BP) contraction that introduces an uncontrolled error into the expectation value. We now verify the accuracy of this approximation in 
three separate ways. The first verification is that for exactly accessible reference quantities such as the depth 5 and depth 9 observables, or the 31 qubit circuits in Ref.~\cite{Kechedzhi2023}, we can directly see that the \emph{total} error (i.e. including truncation) is $<10^{-3}$ in all cases. 
The second verification is that in the case of the PEPS method, when the bond dimension is too large for exact contraction of the observable, but still small enough that we can contract amplitudes -- i.e. $\langle x | \psi \rangle$ for a given bit-string $x$, we can compute an unbiased Monte Carlo estimation of the observable by sampling from the state.
In all our checks, we find that L1BP contraction is consistent with such estimates, which themselves are accurate to about 0.1-1\% limited by shot noise (see Fig.~\ref{fig:pepo-verify}A). 
The final check we can perform is that for PEPO evolution when the final bond dimension is small enough, $\chi \lesssim 128$, we can exactly contract $\langle 0 | \Phi | 0 \rangle$. In this regime we find a discrepancy with L1BP $\lesssim 5 \times 10^{-4}$, as shown in Fig.~\ref{fig:pepo-verify}B,C. 
From these checks of the L1BP approximation, as well as the convergence studies with respect to $\chi$ and norm, we can conservatively estimate that the MIX TN method, without extrapolation, provides results for the 20-step circuit $\langle Z_{62}\rangle$ to an absolute accuracy of better than $0.01$. The most expensive (i.e. largest $\chi$) MIX calculations required to reach this accuracy took approximately 1 day of computation time on a NVidia A100 GPU. A more complete set of timing data is provided in the supplementary material.

\begin{figure}
    \centering
    \includegraphics[width=\textwidth]{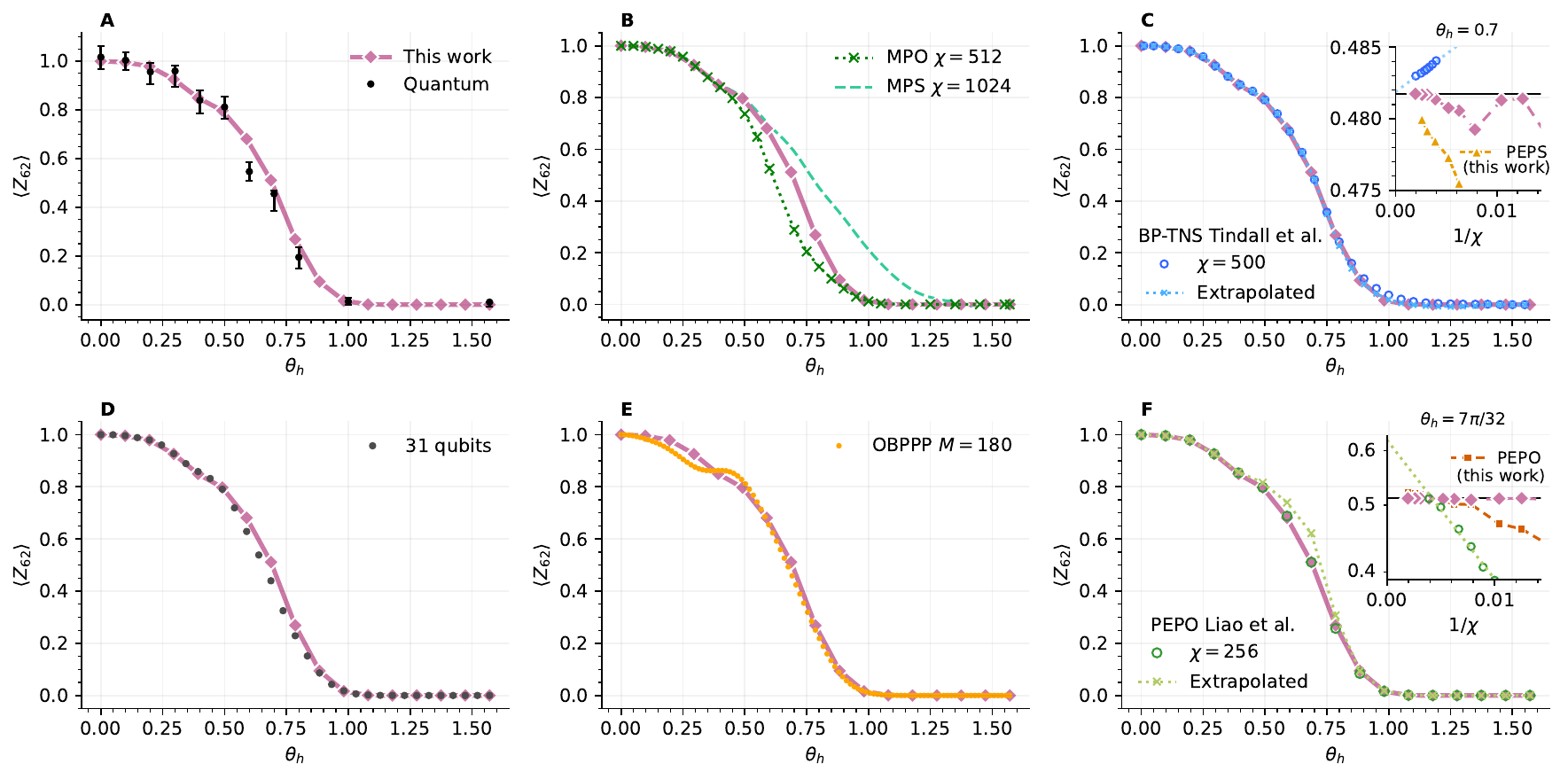}
    \caption{\textbf{Comparison of various simulation results for $\mathbf{\langle Z_{62} \rangle}$.} Expectation values $\langle Z_{62} \rangle$ for a 20-step circuit computed with the reference, MIX method (shown in all panels, labeled ``This work''), 
    \textbf{A}: the zero-noise extrapolated quantum simulation (``Quantum''); \textbf{B}: MPS and MPO approaches \cite{kim2023evidence,Anand2023}; 
    \textbf{C}: BP-TN method of Ref.~\cite{tindall2023efficient} with $\chi = 500$ and the extrapolated $\chi \rightarrow \infty$ result, the inset compares convergence at $\theta_h=0.7$ with our MIX and PEPS methods; 
    \textbf{D}: exact state-vector simulation on a reduced 31-qubit model \cite{Kechedzhi2023}; 
    \textbf{E}: OBPPP approach of Ref.~\cite{Shao2023}; 
    \textbf{F}: PEPO method of Ref.~\cite{liao2023simulation} with $\chi = 256$ and the extrapolated result, the inset compares convergence at $\theta_h=7\pi/32$ with that of our MIX and PEPO methods.
    The estimated error bars of the MIX method (less than 0.01) are smaller than the plot markers.
    Horizontal black lines in the insets of panels C and F correspond to the MIX results at highest available $\chi$, while the dotted cyan (\textbf{C}) and green (\textbf{F}) lines correspond to the extrapolation fits of Refs.~\cite{tindall2023efficient} and \cite{liao2023simulation}, respectively.}
    \label{fig:method-comparison}
\end{figure}

We now use our converged estimates of $\langle Z_{62}\rangle$ from the 20-step circuit using the MIX method to assess the accuracy of the extrapolated experimental results as well as results from other classical simulations. In Fig.~\ref{fig:method-comparison}A we see that most of the experimental data points agree with our results, but
the experimental extrapolation at $6\pi/32$ has a significant error outside of its error bars, and that the data at $7\pi /32$ is also at the limit of its error bars. It is interesting to see that these highly non-Clifford circuits, which are in some ways more challenging classically, are also challenging for the experimental extrapolation. 

In Fig.~\ref{fig:method-comparison}B, we show the MPS and MPO results from~\cite{kim2023evidence} and~\cite{Anand2023}. As already argued in those works, these results are not well converged, and this is confirmed by our accurate data. 

In Fig.~\ref{fig:method-comparison}C, we show results from the BP-TN approach of~\cite{tindall2023efficient}.
The unextrapolated data ($\chi=500$) agrees well with our data up to $8\pi/32$ but deviates slightly at larger angles, where they are slightly above our MIX results.
In Ref.~\cite{tindall2023efficient} $\langle Z_{62} \rangle$ is computed using the approximate local density matrices associated with the 2-norm BP messages, rather than the L1BP contraction. This implicitly renormalizes the PEPS wavefunction after compression, due to how the local density matrices are defined in terms of the BP messages. We find that such a renormalization leads to convergence behavior from above, as observed in their data around $\theta_h \sim 1$. 
On the other hand the MIX method (where we do not renormalize the state) converges from below. As argued from Fig.~\ref{fig:convergence-norm}, the unextrapolated MIX results are likely of higher fidelity than the PEPS results near $\theta_h \sim 1$. 
We note that the extrapolated data from Ref.~\cite{tindall2023efficient} is in good agreement with our unextrapolated MIX data for all $\theta_h$. This is illustrated in the inset which compares convergence at $\theta_h=0.7$, showing extrapolation for the BP-PEPS method in Ref.~\cite{tindall2023efficient}, our own PEPS method (where the state is not renormalized after compression) and the reference MIX result.

In Fig.~\ref{fig:method-comparison}D we see that the reduced 31-qubit model of Ref.~\cite{Kechedzhi2023} is in good agreement with our results at small and large angles, but shows substantial deviation at intermediate $\theta_h$, with a discrepancy of $\sim 0.1$. (Fig.~\ref{fig:31qubit-mix} shows that our MIX approach accurately reproduces the exact 31-qubit result, indicating that the discrepancy in Fig.~\ref{fig:method-comparison}D arises due to the model reduction to 31 qubits).
Next, in Fig.~\ref{fig:method-comparison}E, 
the observable’s back-propagation on Pauli paths (OBPPP) method of Ref.~\cite{Shao2023}, which is itself closely related to SPD, shows good agreement with our data for larger angles of $\theta_h > 5\pi/32$, but interestingly exhibits larger errors at low $\theta_h$, where all other classical simulations, and the quantum simulations agree very well. In the absence of additional convergence data for OBPPP, it is difficult to say whether the agreement at the other points is fortuitous or not.
Finally, in Fig.~\ref{fig:method-comparison}F we compare against the PEPO method described in the recent preprint~\cite{liao2023simulation}, which used simple update style Heisenberg evolution along with exact contraction of the final TN.
We show both the highest $\chi$ result of that paper, as well as their extrapolated result. 
We see that although the largest $\chi$ data is in good agreement with our reference data, the extrapolated data in Ref.~\cite{liao2023simulation} overshoots considerably.
We can trace this to non-monotonic convergence of Heisenberg picture methods (clearly seen in the SPD and PEPO data in Fig.~\ref{fig:convergence-norm}B, see also inset of Fig.~\ref{fig:method-comparison}F), which renders such extrapolations misleading.

\section*{Discussion}

Our work demonstrates that classical algorithms can simulate the expectation values of the quantum circuits corresponding to the kicked Ising dynamics experiment on 127 qubits, not only faster than, but also well-beyond the accuracy of, the current quantum experiments. For this specific experiment, the classical data may be considered for practical purposes to be converged, and this brings to a close the question of quantum utility. However, we also identify that for certain parameters of the quantum circuits, the accuracy in the classical simulation of expectation values is achieved despite a low estimated global fidelity (as measured by the error in the global norm). Identifying physically relevant observables with a stronger sensitivity to the global fidelity, for example, through analyses similar to that in Ref.~\cite{Kechedzhi2023}, will therefore be key to designing future experiments to demonstrate quantum utility. Finally, the advances in classical simulation methods reported here hold promise for many areas of quantum simulation, and highlight the rich landscape of approximate classical algorithms that have yet to be explored for the simulation of quantum circuits and quantum dynamics.

\section*{Methods}

\subsection*{Sparse Pauli dynamics}
Let us consider an observable $O=\sum_{P \in \mathcal{P}} a_P P$ that is a sum of a subset $\mathcal{P}$ of $n$-qubit Pauli operators $P$ with complex coefficients $a_P$. To compute its expectation value
\begin{equation}
    \langle O \rangle = \langle 0 | U^{\dagger} O U | 0 \rangle
\end{equation}
for a quantum circuit $U$, in SPD we apply the circuit gates to the observable in the Heisenberg picture,
\begin{align}
    U^{\dagger} O U = \mathcal{U} O,
\end{align}
where we introduced the Liouville-space unitary operator $\mathcal{U}$. Specifically, we consider circuits composed of Clifford gates and, possibly non-Clifford, Pauli rotation gates $U_j \equiv U_{\sigma_j}(\theta_j) = \exp(-i \theta_j \sigma_j / 2)$. For the rotation gates, we first transform the angles $\theta_j = \theta_j^{\prime} + k \pi/2$, where the integer multiple of $\pi/2$ forms a separate Clifford operator and $|\theta_j^{\prime}| < \pi / 4$. We can now write the Heisenberg-evolved observable as 
\begin{equation}
    \mathcal{U} O= \mathcal{C}_N \mathcal{U}_N \dots \mathcal{C}_1 \mathcal{U}_1 O,
\end{equation}
where $\mathcal{C}_j$ correspond to Clifford operators and $\mathcal{U}_j$ correspond to Pauli rotation gates. Because Clifford operators transform any Pauli operator into another Pauli operator, we can apply them in order to the Pauli rotation gates and observable to arrive at
\begin{equation}
    \mathcal{U} O = \mathcal{\tilde{U}}_N \dots \mathcal{\tilde{U}}_1 \tilde{O},
\end{equation}
where $\mathcal{\tilde{U}}_j$ corresponds to a rotation gate $U_{\tilde{\sigma}_j}(\theta_j)$ with $\tilde{\sigma}_j = \mathcal{\tilde{C}}_N \dots \mathcal{\tilde{C}}_j \sigma_j$ and, similarly, $\tilde{O} = \mathcal{\tilde{C}}_N \dots \mathcal{\tilde{C}}_1 O$. This is also known as Clifford recompilation and has been used in the context of stabilizer-state simulations \cite{Qassim_Emerson:2019}.

In the remainder we assume that the observable and Pauli rotation gates have been transformed as above. Applying a non-Clifford Pauli rotation gate $U_{\sigma}(\theta)$ to the observable $O = \sum_{P \in \mathcal{P}} a_P P$ yields
\begin{equation}
    U_{\sigma}(\theta)^{\dagger} O U_{\sigma}(\theta) = \sum_{P \in \mathcal{P}^{\prime}} a_P^{\prime} P, 
\end{equation}
where the new set of Pauli operators is $\mathcal{P}^{\prime} = \mathcal{P} \cup \{ \sigma P | P \in \mathcal{P}, \{\sigma, P\} = 0\}$ and the updated coefficients are
\begin{equation}
    a_P^{\prime} = \begin{cases}
        a_P \cos(\theta) + i a_{\sigma P} \sin(\theta), \{\sigma, P\} = 0 \\
        a_P, [\sigma, P] = 0.
    \end{cases}
    \label{eq:update_coeffs}
\end{equation}
In general, $2|\mathcal{P}| \ge |\mathcal{P}^{\prime}| \ge |\mathcal{P}|$, i.e., the number of Pauli operators representing the observable grows with each applied gate, leading to the worst-case scaling of $\mathcal{O}(2^N)$.

As an approximation to the exact Heisenberg evolution, we replace the quantum circuit $\mathcal{U}$ by
\begin{equation}
    \mathcal{U}_{\rm SPD} = \Pi_{\delta} \mathcal{U}_N \dots \Pi_{\delta} \mathcal{U}_1,
\end{equation}
where $\Pi_{\delta} (\sum_{P \in \mathcal{P}} a_P^{\prime} P) = \sum_{P \in \mathcal{P}^{\prime}} a_P P$ so that $|a_P| \ge \delta$ for all $P \in \mathcal{P}^{\prime}$. In words, the exact representation of the observable is truncated after each gate to only include the terms whose coefficients are greater than the prescribed threshold $\delta$. This can be considered an alternative to methods that truncate the sum based on the Hamming weight of Pauli operators, i.e., the number of non-identity Pauli matrices appearing in a Pauli operator \cite{kuprov2007polynomially, rakovszky2022dissipation,Shao2023}.

\begin{figure}
    \centering
    \includegraphics[width=0.9\textwidth]{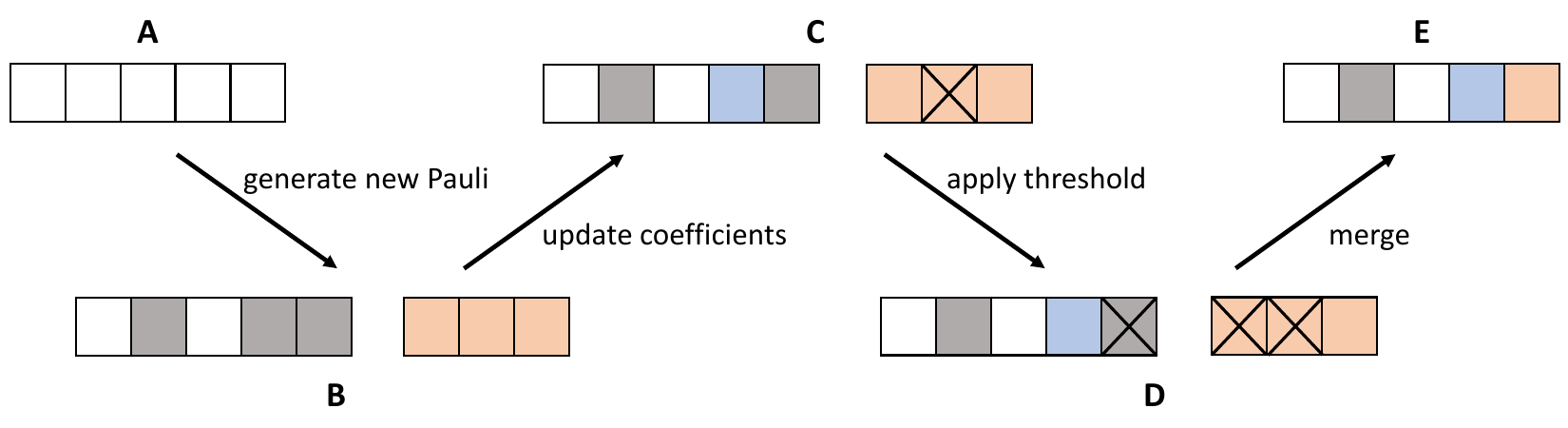}
    \caption{\textbf{Scheme representing the computation of $\mathbf{U_{\sigma}^{\dag} (\sum_{P} a_P P) U_{\sigma}}$ within SPD.} \textbf{A}: The observable is represented by  an array of Pauli operators $P$ with corresponding coefficients $a_P$. \textbf{B}: In SPD, we identify which $P$ commute with $\sigma$ (white) and which do not (gray). For $P$ that anticommute with $\sigma$, we compute $\sigma \cdot P$ (orange). \textbf{C}: Coefficients are updated according to Eq.~\ref{eq:update_coeffs}. This requires that we find Pauli operators $\sigma \cdot P$ that existed in the initial observable, which is accomplished by keeping the Pauli arrays ordered and performing a binary search. New Pauli operators that were not in the representation of the initial observable are shown in orange and not crossed out. \textbf{D}:After all coefficients have been updated, we find Pauli operators whose coefficients are below threshold (additional two crossed out elements in the scheme). \textbf{E}: Finally, the remaining Pauli operators are merged into a new array representing the updated observable.}
    \label{fig:SPDScheme}
\end{figure}

We store Pauli operators as $2n$ bit strings $(z_1, \dots, z_n, x_1, \dots, x_n) \in \{0,1\}^{2n}$ and integer phases $q$ so that $P = (-i)^{q} \prod_{i=1}^{n} Z_i^{z_i} X_i^{x_i}$. In the implementation, the bit strings are encoded into arrays of 64-bit unsigned integers of length $n_{64} = 2\lceil n/64 \rceil$. Therefore, we can represent a sum of $N_{\rm Pauli}$ Pauli operators as a $N_{\rm Pauli} \times n_{64}$ array $p$ of Pauli operators, a $N_{\rm Pauli}$ integer array of phases $q$, and an array $a$ of $N_{\rm Pauli}$ complex coefficients. The Pauli operator array is sorted so that finding a Pauli in $p$ takes $\mathcal{O}(n_{64}\log_{2}(N_{\rm Pauli}))$ time. Insertion and deletion of Pauli operators is implemented in a way that preserves this ordering. Overall, the algorithm to apply a single Pauli rotation gate $U_{\sigma}(\theta)$ can be described as follows (see Fig.~\ref{fig:SPDScheme}):
\begin{enumerate}
    \item Find Pauli operators representing the observable (i.e., rows of the array $p$) that anticommute with $\sigma$, compute new Pauli operators $\sigma \cdot P$ for all $P$ that anticommute with $\sigma$, and store them in a separate array $p_{\rm new}$.
    \item Check which rows of $p_{\rm new}$ are already in $p$.
    \item For new Pauli operators compute the coefficients as $a_{\sigma P}^{\prime} = i \sin(\theta) a_P$ and store these in a new array $a_{\rm new}$. For all Paulis that are already in the representation of the observable, update the coefficients according to Eq.~\ref{eq:update_coeffs}.
    \item Delete rows of $p$ and elements of the phase ($q$) and coefficient ($a$) arrays where $|a|<\delta$.
    \item Insert rows of $p_{\rm new}$ into $p$ (and analogously for $q$ and $a$ arrays) where $|a_{\rm new}| \ge \delta$.
\end{enumerate}
Finally, the expectation value is the sum of elements of $a$ that correspond to Pauli operators with $x_{i\in \{1, \dots, n\}} = 0$ (i.e., those consisting only of identity and $Z$ matrices) and the Frobenius norm of an operator in Pauli representation is the 2-norm of its vector of coefficients, $N_O = \lVert a \rVert$. Figure~\ref{fig:SPD_Scaling} shows how the computational cost of the method and the Frobenius norm of the evolved operator scale with the threshold.

\subsection*{Tensor Network simulations}

Our tensor network (TN) method involves three basic stages: i) Schr\"odinger evolution of a projected entangled pair state (PEPS) forwards in time, $|\psi(0)\rangle \rightarrow |\psi(\tau)\rangle$; ii) Heisenberg evolution of a projected entangled pair operator (PEPO) backwards in time, $\Phi(T+1) \rightarrow \Phi(\tau+1)$; and iii) contraction of the expectation (a tensor network ``sandwich'') between the two, $\langle \psi(\tau) | \Phi(\tau+1)| \psi(\tau) \rangle$. At the limit $\tau=T$ we call this the `PEPS method', when $\tau=0$ we call this the `PEPO method', and finally we also study the case $\tau=T / 2$ which we dub the `MIX method'.
This approach allows us to exploit low entanglement in both the Schr\"odinger and Heisenberg pictures~\cite{kennesExtendingRangeReal2016}, alongside the device geometry, yielding a very efficient description of the overall dynamics.

We employ two particular techniques that enable the steps above: i) lazy 2-norm belief propagation (L2BP) compression of both the PEPS and the PEPO as they evolve; and ii) lazy 1-norm belief propagation (L1BP) contraction of the final expectation. The first is equivalent to methods already demonstrated for wavefunctions \cite{jiangAccurateDeterminationTensor2008,ranOptimizedDecimationTensor2012,tindallGaugingTensorNetworks2023} and operators~\cite{kshetrimayumSimpleTensorNetwork2017} but with an efficient `lazy' implementation that allows us to globally compress large and complex tensor networks. The second is an extension of an idea that has only recently been introduced in the context of tensor networks~\cite{pancottiOnestepReplicaSymmetry2023}, and allows us to (approximately) compute non-local quantities such as high weight observables and the norm of the PEPS and PEPO very efficiently.

We first set up a tensor network representation of: 
$$
\langle O \rangle=
\langle 0| 
\prod_{t=1..T} U^\dagger_{t} 
O
\prod_{t=T..1} U_{t} 
| 0 \rangle~,
$$ where $O$ is the target observable, with $|0\rangle$ we denote $\otimes_{i=1..N}|0_i\rangle$ for sites $\{i\}$, and $U_t$ contains all gates in layer $t$ of a depth $T$ circuit. We decompose any entangling gates spatially so that every tensor is associated to a single site, $i$, and layer, $t$.
We take as our initial PEPS $|\psi(0)\rangle = |0\rangle$ and initial PEPO as $\Phi(T+1)=O$. For each PEPS evolution step we take the object $U_{t+1}|\psi(t)\rangle$ and use L2BP compression with maximum bond dimension $\chi$ to yield the next PEPS $|\psi(t + 1)\rangle$. Similarly for each PEPO evolution step we take the object $U^\dagger_{t - 1} \Phi(t) U_{t - 1}$ and use L2BP compression to yield $\Phi(t - 1)$.
This is sketched in Fig.~\ref{fig:tnlbp}A.
Finally we evaluate the resulting tensor network with L1BP contraction.

To define L2BP compression and L1BP contraction we first introduce `lazy' belief propagation (BP) for tensor networks, an iterative message passing algorithm.
Consider first a tensor network with a single effective tensor, $T^{[i]}$, at each site $i$, with bonds between pairs $\{(i, j)\}$. 
We define vector messages $m_{i\rightarrow j}$ and $m_{j \rightarrow i}$ for each bond (Fig.~\ref{fig:tnlbp}B), then iterate all messages according to the update contraction in Fig.~\ref{fig:tnlbp}C until convergence~\cite{hastingsQuantumBeliefPropagation2007,leiferQuantumGraphicalModels2008,mezardInformationPhysicsComputation2009,alkabetzTensorNetworksContraction2021,sahuEfficientTensorNetwork2022}.
Because the update is linear, we can allow $T^{[i]}$ to itself be a network of tensors per site, with multi-bond connections, as long as we can still lazily perform the contraction of the combined collection with input messages, thereby avoiding forming the dense version of $T^{[i]}$, (see Fig.~\ref{fig:tnlbp}B insets).
Exact contraction of these unstructured networks then becomes the dominant computational step, and we make use of recent advances to do so efficiently and with controlled memory~\cite{chenClassicalSimulationIntermediateSize2018,gray2021hyper}.
The lazy approach improves both the memory and cost scaling of BP, which we explicitly detail in the supplementary material.
Other recent works~\cite{wangTensorNetworkMessage2023,guoBlockBeliefPropagation2023a} have used a similar approach to BP where the sites are themselves networks, and Refs.~\cite{tindall2023efficient,tindallGaugingTensorNetworks2023}
can be thought of as using a lazy network of two tensors per site.

Now we describe L2BP \emph{compression}. To run BP and compress a tensor network, say $|\psi\rangle$ or $\Phi$, with dangling indices, we first have to form the 2-norm object $\langle \psi | \psi \rangle$ or $\langle\langle \Phi | \Phi \rangle\rangle$ by tracing each dangling index with its dual to form a scalar network,
shown in Figs.~\ref{fig:tnlbp}B(i)-(ii).
In our case we always have the PEPS or PEPO as well as the full next layer of gates. The messages in this case factorize as positive semidefinite matrices of size $d\times d$ with $d$ the total size of shared bonds between $i$ and $j$ in our original TN. Once converged we can take each pair $m_{i\rightarrow j}$ and $m_{j \rightarrow i}$ and compute $R_A$ and $R_B$ as the `reduced factors' of an approximate orthogonalization around that bond.
 This is done by eigen-decomposition, with $m_{i \rightarrow j} = W_A \lambda_A W^\dagger_A $ giving $R_A = \sqrt{\lambda_A} W_A^\dagger$ and similarly $m_{j \rightarrow i} = W_B \lambda_B W^\dagger_B $ giving $R_B = (\sqrt{\lambda_B} W_B^\dagger)^T$.
Taking the truncated SVD $R_A R_B \approx U \sigma V^\dagger$, we can define the optimal projectors $P_A=R_B V \sigma^{-1/2}$ and $P_B=\sigma^{-1/2} U^\dagger R_A$ and insert them into our original TN. Once this has been done for every bond, we can group and contract each site's tensors now with all adjacent projectors, yielding a single tensor per site with maximum single bond dimension~$\chi$~\cite{xieCoarsegrainingRenormalizationHigherorder2012,iinoBoundaryTensorRenormalization2019,tindallGaugingTensorNetworks2023}. This process is depicted in Fig.~\ref{fig:tnlbp}D.

Next we describe L1BP \emph{contraction}. Here we again consider that we have multiple tensors per site, but arranged in a tensor network that contracts to a scalar such as $\langle \psi(\tau) |\Phi(\tau + 1)|\psi(\tau) \rangle$, and that we can perform the message updates lazily. 
Note that this is a real but not strictly positive quantity. To estimate the contracted value (or `1-norm') we make use of an interpretation of the tensor network as the exponential of the Bethe free entropy $\exp(F)=Z$~\cite{mezardInformationPhysicsComputation2009,pancottiOnestepReplicaSymmetry2023}. Given converged messages, we can estimate:
$$
\langle O \rangle
\approx
Z
=
\left(
\prod_{i}
T^{[i]} \cdot \bigotimes_k m_{k \rightarrow i}
\right)
/
\left(
\prod_{\{(i, j)\}}
m_{i \rightarrow j } \cdot m_{j \rightarrow i}
\right)
$$
where the index $k$ runs over neighbors of $i$ and $\square \cdot \square$ denotes contraction.
This quantity is shown graphically in Fig.~\ref{fig:tnlbp}E. 
Note that each message appears exactly once in the numerator and once in the denominator, so its overall normalization does not matter.
Given the non-positive structure of our TN, it is not a priori clear that the messages should converge and $Z$ be accurate in this case, but as demonstrated in the main text, for this problem we find it to perform remarkably well.
Notably, it allows us to compute not just high weight observables, but also do so without relying on the renormalization of marginals/local environments. For both the PEPS and PEPO methods (with the exception of the PEPO benchmark in Fig.~\ref{fig:pepo-verify}), we start performing the L1BP contraction once the cost of doing so is similar to continuing compression. For PEPS this is at $\tau=T-2$ and for PEPO this is $\tau=\log_2(\chi) / 2$ (which likely results in the faster convergence compared to Ref.~\cite{liao2023simulation} seen in Fig.~\ref{fig:method-comparison}F). For the MIX method with $\chi=256$ we note the final overlap sandwich contraction, if flattened to a single 2D TN, would require a bond dimension of $256^3\sim 16,000,000$ (or if flattened to a 1D TN, a bond dimension of $256^{3\times4}\sim10^{29}$).

There are two sources of error with this TN method. The first is the bond truncation performed during the PEPS and PEPO evolution. This is a well understood source of error and we know that as $\chi\rightarrow \mathcal{O}(2^T)$ the tensor network becomes exact, with the L2BP aspect functioning simply as a gauge in which to perform the truncation when $\chi < \mathcal{O}(2^T)$. Heuristically, if a computation converges in $\chi$ before this point, it is usually assumed to be close to the exact value. We can also assess how much of the tensor network has been truncated away by computing $\langle \psi | \psi \rangle$ or $\langle\langle \Phi | \Phi \rangle \rangle / 2^n$, i.e. the norm, after all compressions. Since the action of the projectors $P_A, P_B$ can only decrease this quantity from the exact value of 1, it serves as a reasonable proxy of fidelity. In our case, using L1BP contraction we can compute this norm, and find that in all exactly accessible contexts, as it approaches 1 it also heralds high accuracy. The second source of error comes from the L1BP {contraction} of the final scalar tensor network. This method also has an exact limit when the underlying network becomes a tree. Away from this regime, it becomes only a heuristic, but can be remarkably accurate, for example in cases with sparse connectivity without strongly correlated loops. 

To verify the accuracy of L1BP, one approach in the main text for the PEPS method is to use Monte Carlo sampling of the expectation value: Fig.~\ref{fig:pepo-verify}A. 
We do so efficiently by sampling from the approximate distribution 
$|\langle x | \psi \rangle|^2 \approx \omega(x)$ where $\omega$ is the probability of sampling configuration $x$ via BP decimation sampling. This works as follows: we compute all approximate qubit reduced density matrices, $\{\rho_i\}$ using BP; we then sample a value for the qubit with $\max_i |\mathrm{Tr}(Z \rho_i)|$ according to $\mathrm{diag}(\rho_i)$. We then condition on that value and restart BP on the remaining qubits. When all qubits have been fixed we have a full string $x$, approximate probability $\omega(x)$, and from exact contraction can obtain the true probability $p(x) = |\langle x | \psi \rangle|^2$ (which is much easier than the exact observable due to only having a single layer).
This is sufficient to perform efficient importance sampling allowing us to compute unbiased Monte Carlo estimates of observables, to verify the accuracy of L1BP contraction.

Overall, we note that the methods presented here are easily implemented for arbitrary geometries and gates, and thus should be a useful tool in a wide variety of both real and imaginary time evolution settings.
There are also various obvious pathways to improving the fidelity, for example by matching the PEPS/PEPO norms in order to dynamically set the $\tau$ balance in the MIX method. 
One could also compress much larger blocks of gates or use MPS as the messages as in Ref.~\cite{guoBlockBeliefPropagation2023a}.
Given the equivalence of BP compression to simple update, itself equivalent to cluster update~\cite{lubasch2014unifying,lubasch2014algorithms} with `zero cluster size', an interesting question is whether the environment correlations can be systematically increased in a similar manner.
However, we find that we can reach sufficient fidelities across the relevant range of observables with the MIX method without invoking such improvements. The maximum values of $\chi$ utilized are listed in Table~\ref{tab:tn-chis}.

All tensor network simulations were performed using \texttt{quimb}~\cite{Quimb} and \texttt{cotengra}~\cite{gray2021hyper}. 
As well as controlling complexity through the bond dimension $\chi$, we use a truncation cutoff $\kappa=5\times10^{-6}$ throughout such that each local truncation error is $|AB - A P_A P_B B|<\kappa$. 
We also use light-cone cancellation (though this is also easily found in the PEPO picture by the truncation with a non-zero $\kappa$).
We performed the PEPS and MIX simulations in single precision, but did find that the PEPO simulation requires double precision.
To give an approximate sense of timings, all simulations up to 9 steps or 20 steps with $\chi \lesssim 64$ take on the order of minutes or less, including the cost of contraction path optimization.
For large $\chi$ we run calculations on a NVidia A100 GPU, with the most expensive points (MIX method with $\chi=192,~\theta_h > 8 \pi / 32$ for example) taking about a day. The `Mix 3min' method shown in Fig.~\ref{fig:SPD_Quick} was executed on a NVidia 4070 Ti.
A more detailed runtime analysis on a subset of the data can be found in the supplementary material.

\bibliography{bibliography}

\begin{thebibliography}{48}%
\makeatletter
\providecommand \@ifxundefined [1]{%
 \@ifx{#1\undefined}
}%
\providecommand \@ifnum [1]{%
 \ifnum #1\expandafter \@firstoftwo
 \else \expandafter \@secondoftwo
 \fi
}%
\providecommand \@ifx [1]{%
 \ifx #1\expandafter \@firstoftwo
 \else \expandafter \@secondoftwo
 \fi
}%
\providecommand \natexlab [1]{#1}%
\providecommand \enquote  [1]{``#1''}%
\providecommand \bibnamefont  [1]{#1}%
\providecommand \bibfnamefont [1]{#1}%
\providecommand \citenamefont [1]{#1}%
\providecommand \href@noop [0]{\@secondoftwo}%
\providecommand \href [0]{\begingroup \@sanitize@url \@href}%
\providecommand \@href[1]{\@@startlink{#1}\@@href}%
\providecommand \@@href[1]{\endgroup#1\@@endlink}%
\providecommand \@sanitize@url [0]{\catcode `\\12\catcode `\$12\catcode
  `\&12\catcode `\#12\catcode `\^12\catcode `\_12\catcode `\%12\relax}%
\providecommand \@@startlink[1]{}%
\providecommand \@@endlink[0]{}%
\providecommand \url  [0]{\begingroup\@sanitize@url \@url }%
\providecommand \@url [1]{\endgroup\@href {#1}{\urlprefix }}%
\providecommand \urlprefix  [0]{URL }%
\providecommand \Eprint [0]{\href }%
\providecommand \doibase [0]{https://doi.org/}%
\providecommand \selectlanguage [0]{\@gobble}%
\providecommand \bibinfo  [0]{\@secondoftwo}%
\providecommand \bibfield  [0]{\@secondoftwo}%
\providecommand \translation [1]{[#1]}%
\providecommand \BibitemOpen [0]{}%
\providecommand \bibitemStop [0]{}%
\providecommand \bibitemNoStop [0]{.\EOS\space}%
\providecommand \EOS [0]{\spacefactor3000\relax}%
\providecommand \BibitemShut  [1]{\csname bibitem#1\endcsname}%
\let\auto@bib@innerbib\@empty
\bibitem [{\citenamefont {Boixo}\ \emph {et~al.}(2017)\citenamefont {Boixo},
  \citenamefont {Isakov}, \citenamefont {Smelyanskiy},\ and\ \citenamefont
  {Neven}}]{boixoSimulationLowdepthQuantum2017}%
  \BibitemOpen
  \bibfield  {author} {\bibinfo {author} {\bibfnamefont {S.}~\bibnamefont
  {Boixo}}, \bibinfo {author} {\bibfnamefont {S.~V.}\ \bibnamefont {Isakov}},
  \bibinfo {author} {\bibfnamefont {V.~N.}\ \bibnamefont {Smelyanskiy}},\ and\
  \bibinfo {author} {\bibfnamefont {H.}~\bibnamefont {Neven}},\ }\href@noop {}
  {\bibinfo {title} {Simulation of low-depth quantum circuits as complex
  undirected graphical models}} (\bibinfo {year} {2017}),\ \Eprint
  {https://arxiv.org/abs/1712.05384} {arxiv:1712.05384 [quant-ph]} \BibitemShut
  {NoStop}%
\bibitem [{\citenamefont {Chen}\ \emph {et~al.}(2018)\citenamefont {Chen},
  \citenamefont {Zhang}, \citenamefont {Huang}, \citenamefont {Newman},\ and\
  \citenamefont {Shi}}]{chenClassicalSimulationIntermediateSize2018}%
  \BibitemOpen
  \bibfield  {author} {\bibinfo {author} {\bibfnamefont {J.}~\bibnamefont
  {Chen}}, \bibinfo {author} {\bibfnamefont {F.}~\bibnamefont {Zhang}},
  \bibinfo {author} {\bibfnamefont {C.}~\bibnamefont {Huang}}, \bibinfo
  {author} {\bibfnamefont {M.}~\bibnamefont {Newman}},\ and\ \bibinfo {author}
  {\bibfnamefont {Y.}~\bibnamefont {Shi}},\ }\href@noop {} {\bibinfo {title}
  {Classical {{Simulation}} of {{Intermediate-Size Quantum Circuits}}}}
  (\bibinfo {year} {2018}),\ \Eprint {https://arxiv.org/abs/1805.01450}
  {arxiv:1805.01450 [quant-ph]} \BibitemShut {NoStop}%
\bibitem [{\citenamefont {Huang}\ \emph {et~al.}(2020)\citenamefont {Huang},
  \citenamefont {Zhang}, \citenamefont {Newman}, \citenamefont {Cai},
  \citenamefont {Gao}, \citenamefont {Tian}, \citenamefont {Wu}, \citenamefont
  {Xu}, \citenamefont {Yu}, \citenamefont {Yuan}, \citenamefont {Szegedy},
  \citenamefont {Shi},\ and\ \citenamefont
  {Chen}}]{huangClassicalSimulationQuantum2020}%
  \BibitemOpen
  \bibfield  {author} {\bibinfo {author} {\bibfnamefont {C.}~\bibnamefont
  {Huang}}, \bibinfo {author} {\bibfnamefont {F.}~\bibnamefont {Zhang}},
  \bibinfo {author} {\bibfnamefont {M.}~\bibnamefont {Newman}}, \bibinfo
  {author} {\bibfnamefont {J.}~\bibnamefont {Cai}}, \bibinfo {author}
  {\bibfnamefont {X.}~\bibnamefont {Gao}}, \bibinfo {author} {\bibfnamefont
  {Z.}~\bibnamefont {Tian}}, \bibinfo {author} {\bibfnamefont {J.}~\bibnamefont
  {Wu}}, \bibinfo {author} {\bibfnamefont {H.}~\bibnamefont {Xu}}, \bibinfo
  {author} {\bibfnamefont {H.}~\bibnamefont {Yu}}, \bibinfo {author}
  {\bibfnamefont {B.}~\bibnamefont {Yuan}}, \bibinfo {author} {\bibfnamefont
  {M.}~\bibnamefont {Szegedy}}, \bibinfo {author} {\bibfnamefont
  {Y.}~\bibnamefont {Shi}},\ and\ \bibinfo {author} {\bibfnamefont
  {J.}~\bibnamefont {Chen}},\ }\bibfield  {title} {\bibinfo {title} {Classical
  {{Simulation}} of {{Quantum Supremacy Circuits}}},\ }\href@noop {} {\bibfield
   {journal} {\bibinfo  {journal} {arXiv:2005.06787 [quant-ph]}\ } (\bibinfo
  {year} {2020})},\ \Eprint {https://arxiv.org/abs/2005.06787}
  {arxiv:2005.06787 [quant-ph]} \BibitemShut {NoStop}%
\bibitem [{\citenamefont {Wu}\ \emph {et~al.}(2021)\citenamefont {Wu},
  \citenamefont {Bao}, \citenamefont {Cao}, \citenamefont {Chen}, \citenamefont
  {Chen}, \citenamefont {Chen}, \citenamefont {Chung}, \citenamefont {Deng},
  \citenamefont {Du}, \citenamefont {Fan}, \citenamefont {Gong}, \citenamefont
  {Guo}, \citenamefont {Guo}, \citenamefont {Guo}, \citenamefont {Han},
  \citenamefont {Hong}, \citenamefont {Huang}, \citenamefont {Huo},
  \citenamefont {Li}, \citenamefont {Li}, \citenamefont {Li}, \citenamefont
  {Li}, \citenamefont {Liang}, \citenamefont {Lin}, \citenamefont {Lin},
  \citenamefont {Qian}, \citenamefont {Qiao}, \citenamefont {Rong},
  \citenamefont {Su}, \citenamefont {Sun}, \citenamefont {Wang}, \citenamefont
  {Wang}, \citenamefont {Wu}, \citenamefont {Xu}, \citenamefont {Yan},
  \citenamefont {Yang}, \citenamefont {Yang}, \citenamefont {Ye}, \citenamefont
  {Yin}, \citenamefont {Ying}, \citenamefont {Yu}, \citenamefont {Zha},
  \citenamefont {Zhang}, \citenamefont {Zhang}, \citenamefont {Zhang},
  \citenamefont {Zhang}, \citenamefont {Zhao}, \citenamefont {Zhao},
  \citenamefont {Zhou}, \citenamefont {Zhu}, \citenamefont {Lu}, \citenamefont
  {Peng}, \citenamefont {Zhu},\ and\ \citenamefont
  {Pan}}]{wuStrongQuantumComputational2021}%
  \BibitemOpen
  \bibfield  {author} {\bibinfo {author} {\bibfnamefont {Y.}~\bibnamefont
  {Wu}}, \bibinfo {author} {\bibfnamefont {W.-S.}\ \bibnamefont {Bao}},
  \bibinfo {author} {\bibfnamefont {S.}~\bibnamefont {Cao}}, \bibinfo {author}
  {\bibfnamefont {F.}~\bibnamefont {Chen}}, \bibinfo {author} {\bibfnamefont
  {M.-C.}\ \bibnamefont {Chen}}, \bibinfo {author} {\bibfnamefont
  {X.}~\bibnamefont {Chen}}, \bibinfo {author} {\bibfnamefont {T.-H.}\
  \bibnamefont {Chung}}, \bibinfo {author} {\bibfnamefont {H.}~\bibnamefont
  {Deng}}, \bibinfo {author} {\bibfnamefont {Y.}~\bibnamefont {Du}}, \bibinfo
  {author} {\bibfnamefont {D.}~\bibnamefont {Fan}}, \bibinfo {author}
  {\bibfnamefont {M.}~\bibnamefont {Gong}}, \bibinfo {author} {\bibfnamefont
  {C.}~\bibnamefont {Guo}}, \bibinfo {author} {\bibfnamefont {C.}~\bibnamefont
  {Guo}}, \bibinfo {author} {\bibfnamefont {S.}~\bibnamefont {Guo}}, \bibinfo
  {author} {\bibfnamefont {L.}~\bibnamefont {Han}}, \bibinfo {author}
  {\bibfnamefont {L.}~\bibnamefont {Hong}}, \bibinfo {author} {\bibfnamefont
  {H.-L.}\ \bibnamefont {Huang}}, \bibinfo {author} {\bibfnamefont {Y.-H.}\
  \bibnamefont {Huo}}, \bibinfo {author} {\bibfnamefont {L.}~\bibnamefont
  {Li}}, \bibinfo {author} {\bibfnamefont {N.}~\bibnamefont {Li}}, \bibinfo
  {author} {\bibfnamefont {S.}~\bibnamefont {Li}}, \bibinfo {author}
  {\bibfnamefont {Y.}~\bibnamefont {Li}}, \bibinfo {author} {\bibfnamefont
  {F.}~\bibnamefont {Liang}}, \bibinfo {author} {\bibfnamefont
  {C.}~\bibnamefont {Lin}}, \bibinfo {author} {\bibfnamefont {J.}~\bibnamefont
  {Lin}}, \bibinfo {author} {\bibfnamefont {H.}~\bibnamefont {Qian}}, \bibinfo
  {author} {\bibfnamefont {D.}~\bibnamefont {Qiao}}, \bibinfo {author}
  {\bibfnamefont {H.}~\bibnamefont {Rong}}, \bibinfo {author} {\bibfnamefont
  {H.}~\bibnamefont {Su}}, \bibinfo {author} {\bibfnamefont {L.}~\bibnamefont
  {Sun}}, \bibinfo {author} {\bibfnamefont {L.}~\bibnamefont {Wang}}, \bibinfo
  {author} {\bibfnamefont {S.}~\bibnamefont {Wang}}, \bibinfo {author}
  {\bibfnamefont {D.}~\bibnamefont {Wu}}, \bibinfo {author} {\bibfnamefont
  {Y.}~\bibnamefont {Xu}}, \bibinfo {author} {\bibfnamefont {K.}~\bibnamefont
  {Yan}}, \bibinfo {author} {\bibfnamefont {W.}~\bibnamefont {Yang}}, \bibinfo
  {author} {\bibfnamefont {Y.}~\bibnamefont {Yang}}, \bibinfo {author}
  {\bibfnamefont {Y.}~\bibnamefont {Ye}}, \bibinfo {author} {\bibfnamefont
  {J.}~\bibnamefont {Yin}}, \bibinfo {author} {\bibfnamefont {C.}~\bibnamefont
  {Ying}}, \bibinfo {author} {\bibfnamefont {J.}~\bibnamefont {Yu}}, \bibinfo
  {author} {\bibfnamefont {C.}~\bibnamefont {Zha}}, \bibinfo {author}
  {\bibfnamefont {C.}~\bibnamefont {Zhang}}, \bibinfo {author} {\bibfnamefont
  {H.}~\bibnamefont {Zhang}}, \bibinfo {author} {\bibfnamefont
  {K.}~\bibnamefont {Zhang}}, \bibinfo {author} {\bibfnamefont
  {Y.}~\bibnamefont {Zhang}}, \bibinfo {author} {\bibfnamefont
  {H.}~\bibnamefont {Zhao}}, \bibinfo {author} {\bibfnamefont {Y.}~\bibnamefont
  {Zhao}}, \bibinfo {author} {\bibfnamefont {L.}~\bibnamefont {Zhou}}, \bibinfo
  {author} {\bibfnamefont {Q.}~\bibnamefont {Zhu}}, \bibinfo {author}
  {\bibfnamefont {C.-Y.}\ \bibnamefont {Lu}}, \bibinfo {author} {\bibfnamefont
  {C.-Z.}\ \bibnamefont {Peng}}, \bibinfo {author} {\bibfnamefont
  {X.}~\bibnamefont {Zhu}},\ and\ \bibinfo {author} {\bibfnamefont {J.-W.}\
  \bibnamefont {Pan}},\ }\bibfield  {title} {\bibinfo {title} {Strong {{Quantum
  Computational Advantage Using}} a {{Superconducting Quantum Processor}}},\
  }\href {https://doi.org/10.1103/PhysRevLett.127.180501} {\bibfield  {journal}
  {\bibinfo  {journal} {Physical Review Letters}\ }\textbf {\bibinfo {volume}
  {127}},\ \bibinfo {pages} {180501} (\bibinfo {year} {2021})}\BibitemShut
  {NoStop}%
\bibitem [{\citenamefont {Pan}\ \emph {et~al.}(2022)\citenamefont {Pan},
  \citenamefont {Chen},\ and\ \citenamefont
  {Zhang}}]{panSolvingSamplingProblem2022}%
  \BibitemOpen
  \bibfield  {author} {\bibinfo {author} {\bibfnamefont {F.}~\bibnamefont
  {Pan}}, \bibinfo {author} {\bibfnamefont {K.}~\bibnamefont {Chen}},\ and\
  \bibinfo {author} {\bibfnamefont {P.}~\bibnamefont {Zhang}},\ }\bibfield
  {title} {\bibinfo {title} {Solving the {{Sampling Problem}} of the {{Sycamore
  Quantum Circuits}}},\ }\href {https://doi.org/10.1103/PhysRevLett.129.090502}
  {\bibfield  {journal} {\bibinfo  {journal} {Physical Review Letters}\
  }\textbf {\bibinfo {volume} {129}},\ \bibinfo {pages} {090502} (\bibinfo
  {year} {2022})}\BibitemShut {NoStop}%
\bibitem [{\citenamefont {Morvan}\ \emph {et~al.}(2023)\citenamefont {Morvan},
  \citenamefont {Villalonga}, \citenamefont {Mi}, \citenamefont {Mandr{\`a}},
  \citenamefont {Bengtsson}, \citenamefont {Klimov}, \citenamefont {Chen},
  \citenamefont {Hong}, \citenamefont {Erickson}, \citenamefont {Drozdov},
  \citenamefont {Chau}, \citenamefont {Laun}, \citenamefont {Movassagh},
  \citenamefont {Asfaw}, \citenamefont {Brand{\~a}o}, \citenamefont {Peralta},
  \citenamefont {Abanin}, \citenamefont {Acharya}, \citenamefont {Allen},
  \citenamefont {Andersen}, \citenamefont {Anderson}, \citenamefont {Ansmann},
  \citenamefont {Arute}, \citenamefont {Arya}, \citenamefont {Atalaya},
  \citenamefont {Bardin}, \citenamefont {Bilmes}, \citenamefont {Bortoli},
  \citenamefont {Bourassa}, \citenamefont {Bovaird}, \citenamefont {Brill},
  \citenamefont {Broughton}, \citenamefont {Buckley}, \citenamefont {Buell},
  \citenamefont {Burger}, \citenamefont {Burkett}, \citenamefont {Bushnell},
  \citenamefont {Campero}, \citenamefont {Chang}, \citenamefont {Chiaro},
  \citenamefont {Chik}, \citenamefont {Chou}, \citenamefont {Cogan},
  \citenamefont {Collins}, \citenamefont {Conner}, \citenamefont {Courtney},
  \citenamefont {Crook}, \citenamefont {Curtin}, \citenamefont {Debroy},
  \citenamefont {Barba}, \citenamefont {Demura}, \citenamefont {Di~Paolo},
  \citenamefont {Dunsworth}, \citenamefont {Faoro}, \citenamefont {Farhi},
  \citenamefont {Fatemi}, \citenamefont {Ferreira}, \citenamefont {Burgos},
  \citenamefont {Forati}, \citenamefont {Fowler}, \citenamefont {Foxen},
  \citenamefont {Garcia}, \citenamefont {Genois}, \citenamefont {Giang},
  \citenamefont {Gidney}, \citenamefont {Gilboa}, \citenamefont {Giustina},
  \citenamefont {Gosula}, \citenamefont {Dau}, \citenamefont {Gross},
  \citenamefont {Habegger}, \citenamefont {Hamilton}, \citenamefont {Hansen},
  \citenamefont {Harrigan}, \citenamefont {Harrington}, \citenamefont {Heu},
  \citenamefont {Hoffmann}, \citenamefont {Huang}, \citenamefont {Huff},
  \citenamefont {Huggins}, \citenamefont {Ioffe}, \citenamefont {Isakov},
  \citenamefont {Iveland}, \citenamefont {Jeffrey}, \citenamefont {Jiang},
  \citenamefont {Jones}, \citenamefont {Juhas}, \citenamefont {Kafri},
  \citenamefont {Khattar}, \citenamefont {Khezri}, \citenamefont
  {Kieferov{\'a}}, \citenamefont {Kim}, \citenamefont {Kitaev}, \citenamefont
  {Klots}, \citenamefont {Korotkov}, \citenamefont {Kostritsa}, \citenamefont
  {Kreikebaum}, \citenamefont {Landhuis}, \citenamefont {Laptev}, \citenamefont
  {Lau}, \citenamefont {Laws}, \citenamefont {Lee}, \citenamefont {Lee},
  \citenamefont {Lensky}, \citenamefont {Lester}, \citenamefont {Lill},
  \citenamefont {Liu}, \citenamefont {Locharla}, \citenamefont {Malone},
  \citenamefont {Martin}, \citenamefont {Martin}, \citenamefont {McClean},
  \citenamefont {McEwen}, \citenamefont {Miao}, \citenamefont {Mieszala},
  \citenamefont {Montazeri}, \citenamefont {Mruczkiewicz}, \citenamefont
  {Naaman}, \citenamefont {Neeley}, \citenamefont {Neill}, \citenamefont
  {Nersisyan}, \citenamefont {Newman}, \citenamefont {Ng}, \citenamefont
  {Nguyen}, \citenamefont {Nguyen}, \citenamefont {Niu}, \citenamefont
  {O'Brien}, \citenamefont {Omonije}, \citenamefont {Opremcak}, \citenamefont
  {Petukhov}, \citenamefont {Potter}, \citenamefont {Pryadko}, \citenamefont
  {Quintana}, \citenamefont {Rhodes}, \citenamefont {Rocque}, \citenamefont
  {Roushan}, \citenamefont {Rubin}, \citenamefont {Saei}, \citenamefont {Sank},
  \citenamefont {Sankaragomathi}, \citenamefont {Satzinger}, \citenamefont
  {Schurkus}, \citenamefont {Schuster}, \citenamefont {Shearn}, \citenamefont
  {Shorter}, \citenamefont {Shutty}, \citenamefont {Shvarts}, \citenamefont
  {Sivak}, \citenamefont {Skruzny}, \citenamefont {Smith}, \citenamefont
  {Somma}, \citenamefont {Sterling}, \citenamefont {Strain}, \citenamefont
  {Szalay}, \citenamefont {Thor}, \citenamefont {Torres}, \citenamefont
  {Vidal}, \citenamefont {Heidweiller}, \citenamefont {White}, \citenamefont
  {Woo}, \citenamefont {Xing}, \citenamefont {Yao}, \citenamefont {Yeh},
  \citenamefont {Yoo}, \citenamefont {Young}, \citenamefont {Zalcman},
  \citenamefont {Zhang}, \citenamefont {Zhu}, \citenamefont {Zobrist},
  \citenamefont {Rieffel}, \citenamefont {Biswas}, \citenamefont {Babbush},
  \citenamefont {Bacon}, \citenamefont {Hilton}, \citenamefont {Lucero},
  \citenamefont {Neven}, \citenamefont {Megrant}, \citenamefont {Kelly},
  \citenamefont {Aleiner}, \citenamefont {Smelyanskiy}, \citenamefont
  {Kechedzhi}, \citenamefont {Chen},\ and\ \citenamefont
  {Boixo}}]{morvanPhaseTransitionRandom2023}%
  \BibitemOpen
  \bibfield  {author} {\bibinfo {author} {\bibfnamefont {A.}~\bibnamefont
  {Morvan}}, \bibinfo {author} {\bibfnamefont {B.}~\bibnamefont {Villalonga}},
  \bibinfo {author} {\bibfnamefont {X.}~\bibnamefont {Mi}}, \bibinfo {author}
  {\bibfnamefont {S.}~\bibnamefont {Mandr{\`a}}}, \bibinfo {author}
  {\bibfnamefont {A.}~\bibnamefont {Bengtsson}}, \bibinfo {author}
  {\bibfnamefont {P.~V.}\ \bibnamefont {Klimov}}, \bibinfo {author}
  {\bibfnamefont {Z.}~\bibnamefont {Chen}}, \bibinfo {author} {\bibfnamefont
  {S.}~\bibnamefont {Hong}}, \bibinfo {author} {\bibfnamefont {C.}~\bibnamefont
  {Erickson}}, \bibinfo {author} {\bibfnamefont {I.~K.}\ \bibnamefont
  {Drozdov}}, \bibinfo {author} {\bibfnamefont {J.}~\bibnamefont {Chau}},
  \bibinfo {author} {\bibfnamefont {G.}~\bibnamefont {Laun}}, \bibinfo {author}
  {\bibfnamefont {R.}~\bibnamefont {Movassagh}}, \bibinfo {author}
  {\bibfnamefont {A.}~\bibnamefont {Asfaw}}, \bibinfo {author} {\bibfnamefont
  {L.~T. A.~N.}\ \bibnamefont {Brand{\~a}o}}, \bibinfo {author} {\bibfnamefont
  {R.}~\bibnamefont {Peralta}}, \bibinfo {author} {\bibfnamefont
  {D.}~\bibnamefont {Abanin}}, \bibinfo {author} {\bibfnamefont
  {R.}~\bibnamefont {Acharya}}, \bibinfo {author} {\bibfnamefont
  {R.}~\bibnamefont {Allen}}, \bibinfo {author} {\bibfnamefont {T.~I.}\
  \bibnamefont {Andersen}}, \bibinfo {author} {\bibfnamefont {K.}~\bibnamefont
  {Anderson}}, \bibinfo {author} {\bibfnamefont {M.}~\bibnamefont {Ansmann}},
  \bibinfo {author} {\bibfnamefont {F.}~\bibnamefont {Arute}}, \bibinfo
  {author} {\bibfnamefont {K.}~\bibnamefont {Arya}}, \bibinfo {author}
  {\bibfnamefont {J.}~\bibnamefont {Atalaya}}, \bibinfo {author} {\bibfnamefont
  {J.~C.}\ \bibnamefont {Bardin}}, \bibinfo {author} {\bibfnamefont
  {A.}~\bibnamefont {Bilmes}}, \bibinfo {author} {\bibfnamefont
  {G.}~\bibnamefont {Bortoli}}, \bibinfo {author} {\bibfnamefont
  {A.}~\bibnamefont {Bourassa}}, \bibinfo {author} {\bibfnamefont
  {J.}~\bibnamefont {Bovaird}}, \bibinfo {author} {\bibfnamefont
  {L.}~\bibnamefont {Brill}}, \bibinfo {author} {\bibfnamefont
  {M.}~\bibnamefont {Broughton}}, \bibinfo {author} {\bibfnamefont {B.~B.}\
  \bibnamefont {Buckley}}, \bibinfo {author} {\bibfnamefont {D.~A.}\
  \bibnamefont {Buell}}, \bibinfo {author} {\bibfnamefont {T.}~\bibnamefont
  {Burger}}, \bibinfo {author} {\bibfnamefont {B.}~\bibnamefont {Burkett}},
  \bibinfo {author} {\bibfnamefont {N.}~\bibnamefont {Bushnell}}, \bibinfo
  {author} {\bibfnamefont {J.}~\bibnamefont {Campero}}, \bibinfo {author}
  {\bibfnamefont {H.~S.}\ \bibnamefont {Chang}}, \bibinfo {author}
  {\bibfnamefont {B.}~\bibnamefont {Chiaro}}, \bibinfo {author} {\bibfnamefont
  {D.}~\bibnamefont {Chik}}, \bibinfo {author} {\bibfnamefont {C.}~\bibnamefont
  {Chou}}, \bibinfo {author} {\bibfnamefont {J.}~\bibnamefont {Cogan}},
  \bibinfo {author} {\bibfnamefont {R.}~\bibnamefont {Collins}}, \bibinfo
  {author} {\bibfnamefont {P.}~\bibnamefont {Conner}}, \bibinfo {author}
  {\bibfnamefont {W.}~\bibnamefont {Courtney}}, \bibinfo {author}
  {\bibfnamefont {A.~L.}\ \bibnamefont {Crook}}, \bibinfo {author}
  {\bibfnamefont {B.}~\bibnamefont {Curtin}}, \bibinfo {author} {\bibfnamefont
  {D.~M.}\ \bibnamefont {Debroy}}, \bibinfo {author} {\bibfnamefont {A.~D.~T.}\
  \bibnamefont {Barba}}, \bibinfo {author} {\bibfnamefont {S.}~\bibnamefont
  {Demura}}, \bibinfo {author} {\bibfnamefont {A.}~\bibnamefont {Di~Paolo}},
  \bibinfo {author} {\bibfnamefont {A.}~\bibnamefont {Dunsworth}}, \bibinfo
  {author} {\bibfnamefont {L.}~\bibnamefont {Faoro}}, \bibinfo {author}
  {\bibfnamefont {E.}~\bibnamefont {Farhi}}, \bibinfo {author} {\bibfnamefont
  {R.}~\bibnamefont {Fatemi}}, \bibinfo {author} {\bibfnamefont {V.~S.}\
  \bibnamefont {Ferreira}}, \bibinfo {author} {\bibfnamefont {L.~F.}\
  \bibnamefont {Burgos}}, \bibinfo {author} {\bibfnamefont {E.}~\bibnamefont
  {Forati}}, \bibinfo {author} {\bibfnamefont {A.~G.}\ \bibnamefont {Fowler}},
  \bibinfo {author} {\bibfnamefont {B.}~\bibnamefont {Foxen}}, \bibinfo
  {author} {\bibfnamefont {G.}~\bibnamefont {Garcia}}, \bibinfo {author}
  {\bibfnamefont {E.}~\bibnamefont {Genois}}, \bibinfo {author} {\bibfnamefont
  {W.}~\bibnamefont {Giang}}, \bibinfo {author} {\bibfnamefont
  {C.}~\bibnamefont {Gidney}}, \bibinfo {author} {\bibfnamefont
  {D.}~\bibnamefont {Gilboa}}, \bibinfo {author} {\bibfnamefont
  {M.}~\bibnamefont {Giustina}}, \bibinfo {author} {\bibfnamefont
  {R.}~\bibnamefont {Gosula}}, \bibinfo {author} {\bibfnamefont {A.~G.}\
  \bibnamefont {Dau}}, \bibinfo {author} {\bibfnamefont {J.~A.}\ \bibnamefont
  {Gross}}, \bibinfo {author} {\bibfnamefont {S.}~\bibnamefont {Habegger}},
  \bibinfo {author} {\bibfnamefont {M.~C.}\ \bibnamefont {Hamilton}}, \bibinfo
  {author} {\bibfnamefont {M.}~\bibnamefont {Hansen}}, \bibinfo {author}
  {\bibfnamefont {M.~P.}\ \bibnamefont {Harrigan}}, \bibinfo {author}
  {\bibfnamefont {S.~D.}\ \bibnamefont {Harrington}}, \bibinfo {author}
  {\bibfnamefont {P.}~\bibnamefont {Heu}}, \bibinfo {author} {\bibfnamefont
  {M.~R.}\ \bibnamefont {Hoffmann}}, \bibinfo {author} {\bibfnamefont
  {T.}~\bibnamefont {Huang}}, \bibinfo {author} {\bibfnamefont
  {A.}~\bibnamefont {Huff}}, \bibinfo {author} {\bibfnamefont {W.~J.}\
  \bibnamefont {Huggins}}, \bibinfo {author} {\bibfnamefont {L.~B.}\
  \bibnamefont {Ioffe}}, \bibinfo {author} {\bibfnamefont {S.~V.}\ \bibnamefont
  {Isakov}}, \bibinfo {author} {\bibfnamefont {J.}~\bibnamefont {Iveland}},
  \bibinfo {author} {\bibfnamefont {E.}~\bibnamefont {Jeffrey}}, \bibinfo
  {author} {\bibfnamefont {Z.}~\bibnamefont {Jiang}}, \bibinfo {author}
  {\bibfnamefont {C.}~\bibnamefont {Jones}}, \bibinfo {author} {\bibfnamefont
  {P.}~\bibnamefont {Juhas}}, \bibinfo {author} {\bibfnamefont
  {D.}~\bibnamefont {Kafri}}, \bibinfo {author} {\bibfnamefont
  {T.}~\bibnamefont {Khattar}}, \bibinfo {author} {\bibfnamefont
  {M.}~\bibnamefont {Khezri}}, \bibinfo {author} {\bibfnamefont
  {M.}~\bibnamefont {Kieferov{\'a}}}, \bibinfo {author} {\bibfnamefont
  {S.}~\bibnamefont {Kim}}, \bibinfo {author} {\bibfnamefont {A.}~\bibnamefont
  {Kitaev}}, \bibinfo {author} {\bibfnamefont {A.~R.}\ \bibnamefont {Klots}},
  \bibinfo {author} {\bibfnamefont {A.~N.}\ \bibnamefont {Korotkov}}, \bibinfo
  {author} {\bibfnamefont {F.}~\bibnamefont {Kostritsa}}, \bibinfo {author}
  {\bibfnamefont {J.~M.}\ \bibnamefont {Kreikebaum}}, \bibinfo {author}
  {\bibfnamefont {D.}~\bibnamefont {Landhuis}}, \bibinfo {author}
  {\bibfnamefont {P.}~\bibnamefont {Laptev}}, \bibinfo {author} {\bibfnamefont
  {K.-M.}\ \bibnamefont {Lau}}, \bibinfo {author} {\bibfnamefont
  {L.}~\bibnamefont {Laws}}, \bibinfo {author} {\bibfnamefont {J.}~\bibnamefont
  {Lee}}, \bibinfo {author} {\bibfnamefont {K.~W.}\ \bibnamefont {Lee}},
  \bibinfo {author} {\bibfnamefont {Y.~D.}\ \bibnamefont {Lensky}}, \bibinfo
  {author} {\bibfnamefont {B.~J.}\ \bibnamefont {Lester}}, \bibinfo {author}
  {\bibfnamefont {A.~T.}\ \bibnamefont {Lill}}, \bibinfo {author}
  {\bibfnamefont {W.}~\bibnamefont {Liu}}, \bibinfo {author} {\bibfnamefont
  {A.}~\bibnamefont {Locharla}}, \bibinfo {author} {\bibfnamefont {F.~D.}\
  \bibnamefont {Malone}}, \bibinfo {author} {\bibfnamefont {O.}~\bibnamefont
  {Martin}}, \bibinfo {author} {\bibfnamefont {S.}~\bibnamefont {Martin}},
  \bibinfo {author} {\bibfnamefont {J.~R.}\ \bibnamefont {McClean}}, \bibinfo
  {author} {\bibfnamefont {M.}~\bibnamefont {McEwen}}, \bibinfo {author}
  {\bibfnamefont {K.~C.}\ \bibnamefont {Miao}}, \bibinfo {author}
  {\bibfnamefont {A.}~\bibnamefont {Mieszala}}, \bibinfo {author}
  {\bibfnamefont {S.}~\bibnamefont {Montazeri}}, \bibinfo {author}
  {\bibfnamefont {W.}~\bibnamefont {Mruczkiewicz}}, \bibinfo {author}
  {\bibfnamefont {O.}~\bibnamefont {Naaman}}, \bibinfo {author} {\bibfnamefont
  {M.}~\bibnamefont {Neeley}}, \bibinfo {author} {\bibfnamefont
  {C.}~\bibnamefont {Neill}}, \bibinfo {author} {\bibfnamefont
  {A.}~\bibnamefont {Nersisyan}}, \bibinfo {author} {\bibfnamefont
  {M.}~\bibnamefont {Newman}}, \bibinfo {author} {\bibfnamefont {J.~H.}\
  \bibnamefont {Ng}}, \bibinfo {author} {\bibfnamefont {A.}~\bibnamefont
  {Nguyen}}, \bibinfo {author} {\bibfnamefont {M.}~\bibnamefont {Nguyen}},
  \bibinfo {author} {\bibfnamefont {M.~Y.}\ \bibnamefont {Niu}}, \bibinfo
  {author} {\bibfnamefont {T.~E.}\ \bibnamefont {O'Brien}}, \bibinfo {author}
  {\bibfnamefont {S.}~\bibnamefont {Omonije}}, \bibinfo {author} {\bibfnamefont
  {A.}~\bibnamefont {Opremcak}}, \bibinfo {author} {\bibfnamefont
  {A.}~\bibnamefont {Petukhov}}, \bibinfo {author} {\bibfnamefont
  {R.}~\bibnamefont {Potter}}, \bibinfo {author} {\bibfnamefont {L.~P.}\
  \bibnamefont {Pryadko}}, \bibinfo {author} {\bibfnamefont {C.}~\bibnamefont
  {Quintana}}, \bibinfo {author} {\bibfnamefont {D.~M.}\ \bibnamefont
  {Rhodes}}, \bibinfo {author} {\bibfnamefont {C.}~\bibnamefont {Rocque}},
  \bibinfo {author} {\bibfnamefont {P.}~\bibnamefont {Roushan}}, \bibinfo
  {author} {\bibfnamefont {N.~C.}\ \bibnamefont {Rubin}}, \bibinfo {author}
  {\bibfnamefont {N.}~\bibnamefont {Saei}}, \bibinfo {author} {\bibfnamefont
  {D.}~\bibnamefont {Sank}}, \bibinfo {author} {\bibfnamefont {K.}~\bibnamefont
  {Sankaragomathi}}, \bibinfo {author} {\bibfnamefont {K.~J.}\ \bibnamefont
  {Satzinger}}, \bibinfo {author} {\bibfnamefont {H.~F.}\ \bibnamefont
  {Schurkus}}, \bibinfo {author} {\bibfnamefont {C.}~\bibnamefont {Schuster}},
  \bibinfo {author} {\bibfnamefont {M.~J.}\ \bibnamefont {Shearn}}, \bibinfo
  {author} {\bibfnamefont {A.}~\bibnamefont {Shorter}}, \bibinfo {author}
  {\bibfnamefont {N.}~\bibnamefont {Shutty}}, \bibinfo {author} {\bibfnamefont
  {V.}~\bibnamefont {Shvarts}}, \bibinfo {author} {\bibfnamefont
  {V.}~\bibnamefont {Sivak}}, \bibinfo {author} {\bibfnamefont
  {J.}~\bibnamefont {Skruzny}}, \bibinfo {author} {\bibfnamefont {W.~C.}\
  \bibnamefont {Smith}}, \bibinfo {author} {\bibfnamefont {R.~D.}\ \bibnamefont
  {Somma}}, \bibinfo {author} {\bibfnamefont {G.}~\bibnamefont {Sterling}},
  \bibinfo {author} {\bibfnamefont {D.}~\bibnamefont {Strain}}, \bibinfo
  {author} {\bibfnamefont {M.}~\bibnamefont {Szalay}}, \bibinfo {author}
  {\bibfnamefont {D.}~\bibnamefont {Thor}}, \bibinfo {author} {\bibfnamefont
  {A.}~\bibnamefont {Torres}}, \bibinfo {author} {\bibfnamefont
  {G.}~\bibnamefont {Vidal}}, \bibinfo {author} {\bibfnamefont {C.~V.}\
  \bibnamefont {Heidweiller}}, \bibinfo {author} {\bibfnamefont
  {T.}~\bibnamefont {White}}, \bibinfo {author} {\bibfnamefont {B.~W.~K.}\
  \bibnamefont {Woo}}, \bibinfo {author} {\bibfnamefont {C.}~\bibnamefont
  {Xing}}, \bibinfo {author} {\bibfnamefont {Z.~J.}\ \bibnamefont {Yao}},
  \bibinfo {author} {\bibfnamefont {P.}~\bibnamefont {Yeh}}, \bibinfo {author}
  {\bibfnamefont {J.}~\bibnamefont {Yoo}}, \bibinfo {author} {\bibfnamefont
  {G.}~\bibnamefont {Young}}, \bibinfo {author} {\bibfnamefont
  {A.}~\bibnamefont {Zalcman}}, \bibinfo {author} {\bibfnamefont
  {Y.}~\bibnamefont {Zhang}}, \bibinfo {author} {\bibfnamefont
  {N.}~\bibnamefont {Zhu}}, \bibinfo {author} {\bibfnamefont {N.}~\bibnamefont
  {Zobrist}}, \bibinfo {author} {\bibfnamefont {E.~G.}\ \bibnamefont
  {Rieffel}}, \bibinfo {author} {\bibfnamefont {R.}~\bibnamefont {Biswas}},
  \bibinfo {author} {\bibfnamefont {R.}~\bibnamefont {Babbush}}, \bibinfo
  {author} {\bibfnamefont {D.}~\bibnamefont {Bacon}}, \bibinfo {author}
  {\bibfnamefont {J.}~\bibnamefont {Hilton}}, \bibinfo {author} {\bibfnamefont
  {E.}~\bibnamefont {Lucero}}, \bibinfo {author} {\bibfnamefont
  {H.}~\bibnamefont {Neven}}, \bibinfo {author} {\bibfnamefont
  {A.}~\bibnamefont {Megrant}}, \bibinfo {author} {\bibfnamefont
  {J.}~\bibnamefont {Kelly}}, \bibinfo {author} {\bibfnamefont
  {I.}~\bibnamefont {Aleiner}}, \bibinfo {author} {\bibfnamefont
  {V.}~\bibnamefont {Smelyanskiy}}, \bibinfo {author} {\bibfnamefont
  {K.}~\bibnamefont {Kechedzhi}}, \bibinfo {author} {\bibfnamefont
  {Y.}~\bibnamefont {Chen}},\ and\ \bibinfo {author} {\bibfnamefont
  {S.}~\bibnamefont {Boixo}},\ }\href
  {https://doi.org/10.48550/arXiv.2304.11119} {\bibinfo {title} {Phase
  transition in {{Random Circuit Sampling}}}} (\bibinfo {year} {2023}),\
  \Eprint {https://arxiv.org/abs/2304.11119} {arxiv:2304.11119 [quant-ph]}
  \BibitemShut {NoStop}%
\bibitem [{\citenamefont {Markov}\ and\ \citenamefont
  {Shi}(2008)}]{Markov_Shi:2008}%
  \BibitemOpen
  \bibfield  {author} {\bibinfo {author} {\bibfnamefont {I.~L.}\ \bibnamefont
  {Markov}}\ and\ \bibinfo {author} {\bibfnamefont {Y.}~\bibnamefont {Shi}},\
  }\bibfield  {title} {\bibinfo {title} {{Simulating Quantum Computation by
  Contracting Tensor Networks}},\ }\href {https://doi.org/10.1137/050644756}
  {\bibfield  {journal} {\bibinfo  {journal} {SIAM J. Comput.}\ }\textbf
  {\bibinfo {volume} {38}},\ \bibinfo {pages} {963} (\bibinfo {year}
  {2008})}\BibitemShut {NoStop}%
\bibitem [{\citenamefont {Kim}\ \emph {et~al.}(2023)\citenamefont {Kim},
  \citenamefont {Eddins}, \citenamefont {Anand}, \citenamefont {Wei},
  \citenamefont {Van Den~Berg}, \citenamefont {Rosenblatt}, \citenamefont
  {Nayfeh}, \citenamefont {Wu}, \citenamefont {Zaletel}, \citenamefont
  {Temme},\ and\ \citenamefont {Kandala}}]{kim2023evidence}%
  \BibitemOpen
  \bibfield  {author} {\bibinfo {author} {\bibfnamefont {Y.}~\bibnamefont
  {Kim}}, \bibinfo {author} {\bibfnamefont {A.}~\bibnamefont {Eddins}},
  \bibinfo {author} {\bibfnamefont {S.}~\bibnamefont {Anand}}, \bibinfo
  {author} {\bibfnamefont {K.~X.}\ \bibnamefont {Wei}}, \bibinfo {author}
  {\bibfnamefont {E.}~\bibnamefont {Van Den~Berg}}, \bibinfo {author}
  {\bibfnamefont {S.}~\bibnamefont {Rosenblatt}}, \bibinfo {author}
  {\bibfnamefont {H.}~\bibnamefont {Nayfeh}}, \bibinfo {author} {\bibfnamefont
  {Y.}~\bibnamefont {Wu}}, \bibinfo {author} {\bibfnamefont {M.}~\bibnamefont
  {Zaletel}}, \bibinfo {author} {\bibfnamefont {K.}~\bibnamefont {Temme}},\
  and\ \bibinfo {author} {\bibfnamefont {A.}~\bibnamefont {Kandala}},\
  }\bibfield  {title} {\bibinfo {title} {Evidence for the utility of quantum
  computing before fault tolerance},\ }\href
  {https://doi.org/10.1038/s41586-023-06096-3} {\bibfield  {journal} {\bibinfo
  {journal} {Nature}\ }\textbf {\bibinfo {volume} {618}},\ \bibinfo {pages}
  {500} (\bibinfo {year} {2023})}\BibitemShut {NoStop}%
\bibitem [{\citenamefont {Temme}\ \emph {et~al.}(2017)\citenamefont {Temme},
  \citenamefont {Bravyi},\ and\ \citenamefont {Gambetta}}]{temme2017error}%
  \BibitemOpen
  \bibfield  {author} {\bibinfo {author} {\bibfnamefont {K.}~\bibnamefont
  {Temme}}, \bibinfo {author} {\bibfnamefont {S.}~\bibnamefont {Bravyi}},\ and\
  \bibinfo {author} {\bibfnamefont {J.~M.}\ \bibnamefont {Gambetta}},\
  }\bibfield  {title} {\bibinfo {title} {Error mitigation for short-depth
  quantum circuits},\ }\href {https://doi.org/10.1103/PhysRevLett.119.180509}
  {\bibfield  {journal} {\bibinfo  {journal} {Phys. Rev. Lett.}\ }\textbf
  {\bibinfo {volume} {119}},\ \bibinfo {pages} {180509} (\bibinfo {year}
  {2017})}\BibitemShut {NoStop}%
\bibitem [{\citenamefont {Zaletel}\ and\ \citenamefont
  {Pollmann}(2020)}]{zaletel2020isometric}%
  \BibitemOpen
  \bibfield  {author} {\bibinfo {author} {\bibfnamefont {M.~P.}\ \bibnamefont
  {Zaletel}}\ and\ \bibinfo {author} {\bibfnamefont {F.}~\bibnamefont
  {Pollmann}},\ }\bibfield  {title} {\bibinfo {title} {Isometric tensor network
  states in two dimensions},\ }\href
  {https://doi.org/10.1103/PhysRevLett.124.037201} {\bibfield  {journal}
  {\bibinfo  {journal} {Phys. Rev. Lett.}\ }\textbf {\bibinfo {volume} {124}},\
  \bibinfo {pages} {037201} (\bibinfo {year} {2020})}\BibitemShut {NoStop}%
\bibitem [{\citenamefont {Begu{\v{s}}i{\'{c}}}\ and\ \citenamefont
  {Chan}(2023)}]{Begusic2023}%
  \BibitemOpen
  \bibfield  {author} {\bibinfo {author} {\bibfnamefont {T.}~\bibnamefont
  {Begu{\v{s}}i{\'{c}}}}\ and\ \bibinfo {author} {\bibfnamefont {G.~K.-L.}\
  \bibnamefont {Chan}},\ }\href {http://arxiv.org/abs/2306.16372} {\bibinfo
  {title} {{Fast classical simulation of evidence for the utility of quantum
  computing before fault tolerance}}} (\bibinfo {year} {2023}),\ \Eprint
  {https://arxiv.org/abs/2306.16372} {arXiv:2306.16372} \BibitemShut {NoStop}%
\bibitem [{\citenamefont {Tindall}\ \emph {et~al.}(2023)\citenamefont
  {Tindall}, \citenamefont {Fishman}, \citenamefont {Stoudenmire},\ and\
  \citenamefont {Sels}}]{tindall2023efficient}%
  \BibitemOpen
  \bibfield  {author} {\bibinfo {author} {\bibfnamefont {J.}~\bibnamefont
  {Tindall}}, \bibinfo {author} {\bibfnamefont {M.}~\bibnamefont {Fishman}},
  \bibinfo {author} {\bibfnamefont {M.}~\bibnamefont {Stoudenmire}},\ and\
  \bibinfo {author} {\bibfnamefont {D.}~\bibnamefont {Sels}},\ }\href@noop {}
  {\bibinfo {title} {{Efficient tensor network simulation of IBM's kicked Ising
  experiment}}} (\bibinfo {year} {2023}),\ \Eprint
  {https://arxiv.org/abs/2306.14887} {arXiv:2306.14887 [quant-ph]} \BibitemShut
  {NoStop}%
\bibitem [{\citenamefont {Kechedzhi}\ \emph {et~al.}(2023)\citenamefont
  {Kechedzhi}, \citenamefont {Isakov}, \citenamefont {Mandr{\`{a}}},
  \citenamefont {Villalonga}, \citenamefont {Mi}, \citenamefont {Boixo},\ and\
  \citenamefont {Smelyanskiy}}]{Kechedzhi2023}%
  \BibitemOpen
  \bibfield  {author} {\bibinfo {author} {\bibfnamefont {K.}~\bibnamefont
  {Kechedzhi}}, \bibinfo {author} {\bibfnamefont {S.~V.}\ \bibnamefont
  {Isakov}}, \bibinfo {author} {\bibfnamefont {S.}~\bibnamefont
  {Mandr{\`{a}}}}, \bibinfo {author} {\bibfnamefont {B.}~\bibnamefont
  {Villalonga}}, \bibinfo {author} {\bibfnamefont {X.}~\bibnamefont {Mi}},
  \bibinfo {author} {\bibfnamefont {S.}~\bibnamefont {Boixo}},\ and\ \bibinfo
  {author} {\bibfnamefont {V.}~\bibnamefont {Smelyanskiy}},\ }\href
  {http://arxiv.org/abs/2306.15970} {\bibinfo {title} {{Effective quantum
  volume, fidelity and computational cost of noisy quantum processing
  experiments}}} (\bibinfo {year} {2023}),\ \Eprint
  {https://arxiv.org/abs/2306.15970} {arXiv:2306.15970} \BibitemShut {NoStop}%
\bibitem [{\citenamefont {Shao}\ \emph {et~al.}(2023)\citenamefont {Shao},
  \citenamefont {Wei}, \citenamefont {Cheng},\ and\ \citenamefont
  {Liu}}]{Shao2023}%
  \BibitemOpen
  \bibfield  {author} {\bibinfo {author} {\bibfnamefont {Y.}~\bibnamefont
  {Shao}}, \bibinfo {author} {\bibfnamefont {F.}~\bibnamefont {Wei}}, \bibinfo
  {author} {\bibfnamefont {S.}~\bibnamefont {Cheng}},\ and\ \bibinfo {author}
  {\bibfnamefont {Z.}~\bibnamefont {Liu}},\ }\href
  {http://arxiv.org/abs/2306.05804} {\bibinfo {title} {{Simulating Quantum Mean
  Values in Noisy Variational Quantum Algorithms: A Polynomial-Scale
  Approach}}} (\bibinfo {year} {2023}),\ \Eprint
  {https://arxiv.org/abs/2306.05804} {arXiv:2306.05804} \BibitemShut {NoStop}%
\bibitem [{\citenamefont {Liao}\ \emph {et~al.}(2023)\citenamefont {Liao},
  \citenamefont {Wang}, \citenamefont {Zhou}, \citenamefont {Zhang},\ and\
  \citenamefont {Xiang}}]{liao2023simulation}%
  \BibitemOpen
  \bibfield  {author} {\bibinfo {author} {\bibfnamefont {H.-J.}\ \bibnamefont
  {Liao}}, \bibinfo {author} {\bibfnamefont {K.}~\bibnamefont {Wang}}, \bibinfo
  {author} {\bibfnamefont {Z.-S.}\ \bibnamefont {Zhou}}, \bibinfo {author}
  {\bibfnamefont {P.}~\bibnamefont {Zhang}},\ and\ \bibinfo {author}
  {\bibfnamefont {T.}~\bibnamefont {Xiang}},\ }\href@noop {} {\bibinfo {title}
  {Simulation of ibm's kicked ising experiment with projected entangled pair
  operator}} (\bibinfo {year} {2023}),\ \Eprint
  {https://arxiv.org/abs/2308.03082} {arXiv:2308.03082 [quant-ph]} \BibitemShut
  {NoStop}%
\bibitem [{\citenamefont {Rudolph}\ \emph {et~al.}(2023)\citenamefont
  {Rudolph}, \citenamefont {Fontana}, \citenamefont {Holmes},\ and\
  \citenamefont {Cincio}}]{rudolph2023classical}%
  \BibitemOpen
  \bibfield  {author} {\bibinfo {author} {\bibfnamefont {M.~S.}\ \bibnamefont
  {Rudolph}}, \bibinfo {author} {\bibfnamefont {E.}~\bibnamefont {Fontana}},
  \bibinfo {author} {\bibfnamefont {Z.}~\bibnamefont {Holmes}},\ and\ \bibinfo
  {author} {\bibfnamefont {L.}~\bibnamefont {Cincio}},\ }\href@noop {}
  {\bibinfo {title} {Classical surrogate simulation of quantum systems with
  lowesa}} (\bibinfo {year} {2023}),\ \Eprint
  {https://arxiv.org/abs/2308.09109} {arXiv:2308.09109 [quant-ph]} \BibitemShut
  {NoStop}%
\bibitem [{\citenamefont {Anand}\ \emph {et~al.}(2023)\citenamefont {Anand},
  \citenamefont {Temme}, \citenamefont {Kandala},\ and\ \citenamefont
  {Zaletel}}]{Anand2023}%
  \BibitemOpen
  \bibfield  {author} {\bibinfo {author} {\bibfnamefont {S.}~\bibnamefont
  {Anand}}, \bibinfo {author} {\bibfnamefont {K.}~\bibnamefont {Temme}},
  \bibinfo {author} {\bibfnamefont {A.}~\bibnamefont {Kandala}},\ and\ \bibinfo
  {author} {\bibfnamefont {M.}~\bibnamefont {Zaletel}},\ }\href
  {http://arxiv.org/abs/2306.17839} {\bibinfo {title} {{Classical benchmarking
  of zero noise extrapolation beyond the exactly-verifiable regime}}} (\bibinfo
  {year} {2023}),\ \Eprint {https://arxiv.org/abs/2306.17839}
  {arXiv:2306.17839} \BibitemShut {NoStop}%
\bibitem [{\citenamefont {Begu\v{s}i\'{c}}\ \emph {et~al.}(2023)\citenamefont
  {Begu\v{s}i\'{c}}, \citenamefont {Hejazi},\ and\ \citenamefont
  {Chan}}]{begusic2023simulating}%
  \BibitemOpen
  \bibfield  {author} {\bibinfo {author} {\bibfnamefont {T.}~\bibnamefont
  {Begu\v{s}i\'{c}}}, \bibinfo {author} {\bibfnamefont {K.}~\bibnamefont
  {Hejazi}},\ and\ \bibinfo {author} {\bibfnamefont {G.~K.-L.}\ \bibnamefont
  {Chan}},\ }\href@noop {} {\bibinfo {title} {{Simulating quantum circuit
  expectation values by Clifford perturbation theory}}} (\bibinfo {year}
  {2023}),\ \Eprint {https://arxiv.org/abs/2306.04797} {arXiv:2306.04797
  [quant-ph]} \BibitemShut {NoStop}%
\bibitem [{\citenamefont {Pancotti}\ and\ \citenamefont
  {Gray}(2023)}]{pancottiOnestepReplicaSymmetry2023}%
  \BibitemOpen
  \bibfield  {author} {\bibinfo {author} {\bibfnamefont {N.}~\bibnamefont
  {Pancotti}}\ and\ \bibinfo {author} {\bibfnamefont {J.}~\bibnamefont
  {Gray}},\ }\href {https://doi.org/10.48550/arXiv.2306.15004} {\bibinfo
  {title} {One-step replica symmetry breaking in the language of tensor
  networks}} (\bibinfo {year} {2023}),\ \Eprint
  {https://arxiv.org/abs/2306.15004} {arxiv:2306.15004 [quant-ph]} \BibitemShut
  {NoStop}%
\bibitem [{\citenamefont {Aaronson}\ and\ \citenamefont
  {Gottesman}(2004)}]{Aaronson_Gottesman:2004}%
  \BibitemOpen
  \bibfield  {author} {\bibinfo {author} {\bibfnamefont {S.}~\bibnamefont
  {Aaronson}}\ and\ \bibinfo {author} {\bibfnamefont {D.}~\bibnamefont
  {Gottesman}},\ }\bibfield  {title} {\bibinfo {title} {{Improved simulation of
  stabilizer circuits}},\ }\href {https://doi.org/10.1103/PhysRevA.70.052328}
  {\bibfield  {journal} {\bibinfo  {journal} {Phys. Rev. A}\ }\textbf {\bibinfo
  {volume} {70}},\ \bibinfo {pages} {052328} (\bibinfo {year}
  {2004})}\BibitemShut {NoStop}%
\bibitem [{\citenamefont {Bravyi}\ and\ \citenamefont
  {Gosset}(2016)}]{Bravyi_Gosset:2016}%
  \BibitemOpen
  \bibfield  {author} {\bibinfo {author} {\bibfnamefont {S.}~\bibnamefont
  {Bravyi}}\ and\ \bibinfo {author} {\bibfnamefont {D.}~\bibnamefont
  {Gosset}},\ }\bibfield  {title} {\bibinfo {title} {{Improved Classical
  Simulation of Quantum Circuits Dominated by Clifford Gates}},\ }\href
  {https://doi.org/10.1103/PhysRevLett.116.250501} {\bibfield  {journal}
  {\bibinfo  {journal} {Phys. Rev. Lett.}\ }\textbf {\bibinfo {volume} {116}},\
  \bibinfo {pages} {250501} (\bibinfo {year} {2016})}\BibitemShut {NoStop}%
\bibitem [{\citenamefont {Bennink}\ \emph {et~al.}(2017)\citenamefont
  {Bennink}, \citenamefont {Ferragut}, \citenamefont {Humble}, \citenamefont
  {Laska}, \citenamefont {Nutaro}, \citenamefont {Pleszkoch},\ and\
  \citenamefont {Pooser}}]{Bennink_Pooser:2017}%
  \BibitemOpen
  \bibfield  {author} {\bibinfo {author} {\bibfnamefont {R.~S.}\ \bibnamefont
  {Bennink}}, \bibinfo {author} {\bibfnamefont {E.~M.}\ \bibnamefont
  {Ferragut}}, \bibinfo {author} {\bibfnamefont {T.~S.}\ \bibnamefont
  {Humble}}, \bibinfo {author} {\bibfnamefont {J.~A.}\ \bibnamefont {Laska}},
  \bibinfo {author} {\bibfnamefont {J.~J.}\ \bibnamefont {Nutaro}}, \bibinfo
  {author} {\bibfnamefont {M.~G.}\ \bibnamefont {Pleszkoch}},\ and\ \bibinfo
  {author} {\bibfnamefont {R.~C.}\ \bibnamefont {Pooser}},\ }\bibfield  {title}
  {\bibinfo {title} {{Unbiased simulation of near-Clifford quantum circuits}},\
  }\href {https://doi.org/10.1103/PhysRevA.95.062337} {\bibfield  {journal}
  {\bibinfo  {journal} {Phys. Rev. A}\ }\textbf {\bibinfo {volume} {95}},\
  \bibinfo {pages} {062337} (\bibinfo {year} {2017})}\BibitemShut {NoStop}%
\bibitem [{\citenamefont {Bravyi}\ \emph {et~al.}(2019)\citenamefont {Bravyi},
  \citenamefont {Browne}, \citenamefont {Calpin}, \citenamefont {Campbell},
  \citenamefont {Gosset},\ and\ \citenamefont {Howard}}]{Bravyi_Howard:2019}%
  \BibitemOpen
  \bibfield  {author} {\bibinfo {author} {\bibfnamefont {S.}~\bibnamefont
  {Bravyi}}, \bibinfo {author} {\bibfnamefont {D.}~\bibnamefont {Browne}},
  \bibinfo {author} {\bibfnamefont {P.}~\bibnamefont {Calpin}}, \bibinfo
  {author} {\bibfnamefont {E.}~\bibnamefont {Campbell}}, \bibinfo {author}
  {\bibfnamefont {D.}~\bibnamefont {Gosset}},\ and\ \bibinfo {author}
  {\bibfnamefont {M.}~\bibnamefont {Howard}},\ }\bibfield  {title} {\bibinfo
  {title} {{Simulation of quantum circuits by low-rank stabilizer
  decompositions}},\ }\href {https://doi.org/10.22331/q-2019-09-02-181}
  {\bibfield  {journal} {\bibinfo  {journal} {Quantum}\ }\textbf {\bibinfo
  {volume} {3}},\ \bibinfo {pages} {181} (\bibinfo {year} {2019})}\BibitemShut
  {NoStop}%
\bibitem [{\citenamefont {Rall}\ \emph {et~al.}(2019)\citenamefont {Rall},
  \citenamefont {Liang}, \citenamefont {Cook},\ and\ \citenamefont
  {Kretschmer}}]{rall2019simulation}%
  \BibitemOpen
  \bibfield  {author} {\bibinfo {author} {\bibfnamefont {P.}~\bibnamefont
  {Rall}}, \bibinfo {author} {\bibfnamefont {D.}~\bibnamefont {Liang}},
  \bibinfo {author} {\bibfnamefont {J.}~\bibnamefont {Cook}},\ and\ \bibinfo
  {author} {\bibfnamefont {W.}~\bibnamefont {Kretschmer}},\ }\bibfield  {title}
  {\bibinfo {title} {Simulation of qubit quantum circuits via pauli
  propagation},\ }\href {https://doi.org/10.1103/PhysRevA.99.062337} {\bibfield
   {journal} {\bibinfo  {journal} {Phys. Rev. A}\ }\textbf {\bibinfo {volume}
  {99}},\ \bibinfo {pages} {062337} (\bibinfo {year} {2019})}\BibitemShut
  {NoStop}%
\bibitem [{\citenamefont {Nemkov}\ \emph {et~al.}(2023)\citenamefont {Nemkov},
  \citenamefont {Kiktenko},\ and\ \citenamefont
  {Fedorov}}]{Nemkov_Fedorov:2023}%
  \BibitemOpen
  \bibfield  {author} {\bibinfo {author} {\bibfnamefont {N.~A.}\ \bibnamefont
  {Nemkov}}, \bibinfo {author} {\bibfnamefont {E.~O.}\ \bibnamefont
  {Kiktenko}},\ and\ \bibinfo {author} {\bibfnamefont {A.~K.}\ \bibnamefont
  {Fedorov}},\ }\bibfield  {title} {\bibinfo {title} {Fourier expansion in
  variational quantum algorithms},\ }\href
  {https://doi.org/10.1103/PhysRevA.108.032406} {\bibfield  {journal} {\bibinfo
   {journal} {Phys. Rev. A}\ }\textbf {\bibinfo {volume} {108}},\ \bibinfo
  {pages} {032406} (\bibinfo {year} {2023})}\BibitemShut {NoStop}%
\bibitem [{\citenamefont {Fontana}\ \emph {et~al.}(2023)\citenamefont
  {Fontana}, \citenamefont {Rudolph}, \citenamefont {Duncan}, \citenamefont
  {Rungger},\ and\ \citenamefont {Cîrstoiu}}]{fontana2023classical}%
  \BibitemOpen
  \bibfield  {author} {\bibinfo {author} {\bibfnamefont {E.}~\bibnamefont
  {Fontana}}, \bibinfo {author} {\bibfnamefont {M.~S.}\ \bibnamefont
  {Rudolph}}, \bibinfo {author} {\bibfnamefont {R.}~\bibnamefont {Duncan}},
  \bibinfo {author} {\bibfnamefont {I.}~\bibnamefont {Rungger}},\ and\ \bibinfo
  {author} {\bibfnamefont {C.}~\bibnamefont {Cîrstoiu}},\ }\href@noop {}
  {\bibinfo {title} {Classical simulations of noisy variational quantum
  circuits}} (\bibinfo {year} {2023}),\ \Eprint
  {https://arxiv.org/abs/2306.05400} {arXiv:2306.05400 [quant-ph]} \BibitemShut
  {NoStop}%
\bibitem [{\citenamefont {Jiang}\ \emph {et~al.}(2008)\citenamefont {Jiang},
  \citenamefont {Weng},\ and\ \citenamefont
  {Xiang}}]{jiangAccurateDeterminationTensor2008}%
  \BibitemOpen
  \bibfield  {author} {\bibinfo {author} {\bibfnamefont {H.~C.}\ \bibnamefont
  {Jiang}}, \bibinfo {author} {\bibfnamefont {Z.~Y.}\ \bibnamefont {Weng}},\
  and\ \bibinfo {author} {\bibfnamefont {T.}~\bibnamefont {Xiang}},\ }\bibfield
   {title} {\bibinfo {title} {Accurate {{Determination}} of {{Tensor Network
  State}} of {{Quantum Lattice Models}} in {{Two Dimensions}}},\ }\href
  {https://doi.org/10.1103/PhysRevLett.101.090603} {\bibfield  {journal}
  {\bibinfo  {journal} {Phys. Rev. Lett.}\ }\textbf {\bibinfo {volume} {101}},\
  \bibinfo {pages} {090603} (\bibinfo {year} {2008})}\BibitemShut {NoStop}%
\bibitem [{\citenamefont {Lubasch}\ \emph
  {et~al.}(2014{\natexlab{a}})\citenamefont {Lubasch}, \citenamefont {Cirac},\
  and\ \citenamefont {Banuls}}]{lubasch2014algorithms}%
  \BibitemOpen
  \bibfield  {author} {\bibinfo {author} {\bibfnamefont {M.}~\bibnamefont
  {Lubasch}}, \bibinfo {author} {\bibfnamefont {J.~I.}\ \bibnamefont {Cirac}},\
  and\ \bibinfo {author} {\bibfnamefont {M.-C.}\ \bibnamefont {Banuls}},\
  }\bibfield  {title} {\bibinfo {title} {Algorithms for finite projected
  entangled pair states},\ }\href@noop {} {\bibfield  {journal} {\bibinfo
  {journal} {Physical Review B}\ }\textbf {\bibinfo {volume} {90}},\ \bibinfo
  {pages} {064425} (\bibinfo {year} {2014}{\natexlab{a}})}\BibitemShut
  {NoStop}%
\bibitem [{\citenamefont {Lubasch}\ \emph
  {et~al.}(2014{\natexlab{b}})\citenamefont {Lubasch}, \citenamefont {Cirac},\
  and\ \citenamefont {Banuls}}]{lubasch2014unifying}%
  \BibitemOpen
  \bibfield  {author} {\bibinfo {author} {\bibfnamefont {M.}~\bibnamefont
  {Lubasch}}, \bibinfo {author} {\bibfnamefont {J.~I.}\ \bibnamefont {Cirac}},\
  and\ \bibinfo {author} {\bibfnamefont {M.-C.}\ \bibnamefont {Banuls}},\
  }\bibfield  {title} {\bibinfo {title} {Unifying projected entangled pair
  state contractions},\ }\href@noop {} {\bibfield  {journal} {\bibinfo
  {journal} {New Journal of Physics}\ }\textbf {\bibinfo {volume} {16}},\
  \bibinfo {pages} {033014} (\bibinfo {year} {2014}{\natexlab{b}})}\BibitemShut
  {NoStop}%
\bibitem [{\citenamefont {Alkabetz}\ and\ \citenamefont
  {Arad}(2021)}]{alkabetzTensorNetworksContraction2021}%
  \BibitemOpen
  \bibfield  {author} {\bibinfo {author} {\bibfnamefont {R.}~\bibnamefont
  {Alkabetz}}\ and\ \bibinfo {author} {\bibfnamefont {I.}~\bibnamefont
  {Arad}},\ }\bibfield  {title} {\bibinfo {title} {Tensor {{Networks}}
  contraction and the {{Belief Propagation}} algorithm},\ }\href
  {https://doi.org/10.1103/PhysRevResearch.3.023073} {\bibfield  {journal}
  {\bibinfo  {journal} {Phys. Rev. Research}\ }\textbf {\bibinfo {volume}
  {3}},\ \bibinfo {pages} {023073} (\bibinfo {year} {2021})}\BibitemShut
  {NoStop}%
\bibitem [{\citenamefont {Tindall}\ and\ \citenamefont
  {Fishman}(2023)}]{tindallGaugingTensorNetworks2023}%
  \BibitemOpen
  \bibfield  {author} {\bibinfo {author} {\bibfnamefont {J.}~\bibnamefont
  {Tindall}}\ and\ \bibinfo {author} {\bibfnamefont {M.}~\bibnamefont
  {Fishman}},\ }\href {https://doi.org/10.48550/arXiv.2306.17837} {\bibinfo
  {title} {Gauging tensor networks with belief propagation}} (\bibinfo {year}
  {2023}),\ \Eprint {https://arxiv.org/abs/2306.17837} {arxiv:2306.17837
  [quant-ph]} \BibitemShut {NoStop}%
\bibitem [{\citenamefont {Ran}\ \emph {et~al.}(2012)\citenamefont {Ran},
  \citenamefont {Li}, \citenamefont {Xi}, \citenamefont {Zhang},\ and\
  \citenamefont {Su}}]{ranOptimizedDecimationTensor2012}%
  \BibitemOpen
  \bibfield  {author} {\bibinfo {author} {\bibfnamefont {S.-J.}\ \bibnamefont
  {Ran}}, \bibinfo {author} {\bibfnamefont {W.}~\bibnamefont {Li}}, \bibinfo
  {author} {\bibfnamefont {B.}~\bibnamefont {Xi}}, \bibinfo {author}
  {\bibfnamefont {Z.}~\bibnamefont {Zhang}},\ and\ \bibinfo {author}
  {\bibfnamefont {G.}~\bibnamefont {Su}},\ }\bibfield  {title} {\bibinfo
  {title} {Optimized decimation of tensor networks with super-orthogonalization
  for two-dimensional quantum lattice models},\ }\href
  {https://doi.org/10.1103/PhysRevB.86.134429} {\bibfield  {journal} {\bibinfo
  {journal} {Phys. Rev. B}\ }\textbf {\bibinfo {volume} {86}},\ \bibinfo
  {pages} {134429} (\bibinfo {year} {2012})}\BibitemShut {NoStop}%
\bibitem [{\citenamefont {Wang}\ \emph {et~al.}(2023)\citenamefont {Wang},
  \citenamefont {Zhang}, \citenamefont {Pan},\ and\ \citenamefont
  {Zhang}}]{wangTensorNetworkMessage2023}%
  \BibitemOpen
  \bibfield  {author} {\bibinfo {author} {\bibfnamefont {Y.}~\bibnamefont
  {Wang}}, \bibinfo {author} {\bibfnamefont {Y.~E.}\ \bibnamefont {Zhang}},
  \bibinfo {author} {\bibfnamefont {F.}~\bibnamefont {Pan}},\ and\ \bibinfo
  {author} {\bibfnamefont {P.}~\bibnamefont {Zhang}},\ }\href
  {https://doi.org/10.48550/arXiv.2305.01874} {\bibinfo {title} {Tensor
  {{Network Message Passing}}}} (\bibinfo {year} {2023}),\ \Eprint
  {https://arxiv.org/abs/2305.01874} {arxiv:2305.01874 [cond-mat.stat-mech]}
  \BibitemShut {NoStop}%
\bibitem [{\citenamefont {Guo}\ \emph {et~al.}(2023)\citenamefont {Guo},
  \citenamefont {Poletti},\ and\ \citenamefont
  {Arad}}]{guoBlockBeliefPropagation2023a}%
  \BibitemOpen
  \bibfield  {author} {\bibinfo {author} {\bibfnamefont {C.}~\bibnamefont
  {Guo}}, \bibinfo {author} {\bibfnamefont {D.}~\bibnamefont {Poletti}},\ and\
  \bibinfo {author} {\bibfnamefont {I.}~\bibnamefont {Arad}},\ }\href
  {https://doi.org/10.48550/arXiv.2301.05844} {\bibinfo {title} {Block {{Belief
  Propagation Algorithm}} for {{2D Tensor Networks}}}} (\bibinfo {year}
  {2023}),\ \Eprint {https://arxiv.org/abs/2301.05844} {arxiv:2301.05844
  [quant-ph]} \BibitemShut {NoStop}%
\bibitem [{\citenamefont {M{\'e}zard}\ and\ \citenamefont
  {Montanari}(2009)}]{mezardInformationPhysicsComputation2009}%
  \BibitemOpen
  \bibfield  {author} {\bibinfo {author} {\bibfnamefont {M.}~\bibnamefont
  {M{\'e}zard}}\ and\ \bibinfo {author} {\bibfnamefont {A.}~\bibnamefont
  {Montanari}},\ }\href@noop {} {\emph {\bibinfo {title} {Information,
  {{Physics}}, and {{Computation}}}}}\ (\bibinfo  {publisher} {{Oxford
  University Press}},\ \bibinfo {year} {2009})\BibitemShut {NoStop}%
\bibitem [{\citenamefont {Gray}\ and\ \citenamefont
  {Chan}(2022)}]{gray2022hyper}%
  \BibitemOpen
  \bibfield  {author} {\bibinfo {author} {\bibfnamefont {J.}~\bibnamefont
  {Gray}}\ and\ \bibinfo {author} {\bibfnamefont {G.~K.-L.}\ \bibnamefont
  {Chan}},\ }\href@noop {} {\bibinfo {title} {Hyper-optimized compressed
  contraction of tensor networks with arbitrary geometry}} (\bibinfo {year}
  {2022}),\ \Eprint {https://arxiv.org/abs/2206.07044} {arXiv:2206.07044
  [quant-ph]} \BibitemShut {NoStop}%
\bibitem [{\citenamefont {Qassim}\ \emph {et~al.}(2019)\citenamefont {Qassim},
  \citenamefont {Wallman},\ and\ \citenamefont
  {Emerson}}]{Qassim_Emerson:2019}%
  \BibitemOpen
  \bibfield  {author} {\bibinfo {author} {\bibfnamefont {H.}~\bibnamefont
  {Qassim}}, \bibinfo {author} {\bibfnamefont {J.~J.}\ \bibnamefont
  {Wallman}},\ and\ \bibinfo {author} {\bibfnamefont {J.}~\bibnamefont
  {Emerson}},\ }\bibfield  {title} {\bibinfo {title} {{Clifford recompilation
  for faster classical simulation of quantum circuits}},\ }\href
  {https://doi.org/10.22331/q-2019-08-05-170} {\bibfield  {journal} {\bibinfo
  {journal} {Quantum}\ }\textbf {\bibinfo {volume} {3}},\ \bibinfo {pages}
  {170} (\bibinfo {year} {2019})}\BibitemShut {NoStop}%
\bibitem [{\citenamefont {Kuprov}\ \emph {et~al.}(2007)\citenamefont {Kuprov},
  \citenamefont {Wagner-Rundell},\ and\ \citenamefont
  {Hore}}]{kuprov2007polynomially}%
  \BibitemOpen
  \bibfield  {author} {\bibinfo {author} {\bibfnamefont {I.}~\bibnamefont
  {Kuprov}}, \bibinfo {author} {\bibfnamefont {N.}~\bibnamefont
  {Wagner-Rundell}},\ and\ \bibinfo {author} {\bibfnamefont {P.}~\bibnamefont
  {Hore}},\ }\bibfield  {title} {\bibinfo {title} {Polynomially scaling spin
  dynamics simulation algorithm based on adaptive state-space restriction},\
  }\href {https://doi.org/10.1016/j.jmr.2007.09.014} {\bibfield  {journal}
  {\bibinfo  {journal} {J. Magn. Reson.}\ }\textbf {\bibinfo {volume} {189}},\
  \bibinfo {pages} {241} (\bibinfo {year} {2007})}\BibitemShut {NoStop}%
\bibitem [{\citenamefont {Rakovszky}\ \emph {et~al.}(2022)\citenamefont
  {Rakovszky}, \citenamefont {von Keyserlingk},\ and\ \citenamefont
  {Pollmann}}]{rakovszky2022dissipation}%
  \BibitemOpen
  \bibfield  {author} {\bibinfo {author} {\bibfnamefont {T.}~\bibnamefont
  {Rakovszky}}, \bibinfo {author} {\bibfnamefont {C.}~\bibnamefont {von
  Keyserlingk}},\ and\ \bibinfo {author} {\bibfnamefont {F.}~\bibnamefont
  {Pollmann}},\ }\bibfield  {title} {\bibinfo {title} {Dissipation-assisted
  operator evolution method for capturing hydrodynamic transport},\ }\href
  {https://doi.org/10.1103/PhysRevB.105.075131} {\bibfield  {journal} {\bibinfo
   {journal} {Phys. Rev. B}\ }\textbf {\bibinfo {volume} {105}},\ \bibinfo
  {pages} {075131} (\bibinfo {year} {2022})}\BibitemShut {NoStop}%
\bibitem [{\citenamefont {Kennes}\ and\ \citenamefont
  {Karrasch}(2016)}]{kennesExtendingRangeReal2016}%
  \BibitemOpen
  \bibfield  {author} {\bibinfo {author} {\bibfnamefont {D.~M.}\ \bibnamefont
  {Kennes}}\ and\ \bibinfo {author} {\bibfnamefont {C.}~\bibnamefont
  {Karrasch}},\ }\bibfield  {title} {\bibinfo {title} {Extending the range of
  real time density matrix renormalization group simulations},\ }\href
  {https://doi.org/10.1016/j.cpc.2015.10.019} {\bibfield  {journal} {\bibinfo
  {journal} {Comput. Phys. Commun.}\ }\textbf {\bibinfo {volume} {200}},\
  \bibinfo {pages} {37} (\bibinfo {year} {2016})}\BibitemShut {NoStop}%
\bibitem [{\citenamefont {Kshetrimayum}\ \emph {et~al.}(2017)\citenamefont
  {Kshetrimayum}, \citenamefont {Weimer},\ and\ \citenamefont
  {Or{\'u}s}}]{kshetrimayumSimpleTensorNetwork2017}%
  \BibitemOpen
  \bibfield  {author} {\bibinfo {author} {\bibfnamefont {A.}~\bibnamefont
  {Kshetrimayum}}, \bibinfo {author} {\bibfnamefont {H.}~\bibnamefont
  {Weimer}},\ and\ \bibinfo {author} {\bibfnamefont {R.}~\bibnamefont
  {Or{\'u}s}},\ }\bibfield  {title} {\bibinfo {title} {A simple tensor network
  algorithm for two-dimensional steady states},\ }\href
  {https://doi.org/10.1038/s41467-017-01511-6} {\bibfield  {journal} {\bibinfo
  {journal} {Nat. Commun.}\ }\textbf {\bibinfo {volume} {8}},\ \bibinfo {pages}
  {1291} (\bibinfo {year} {2017})}\BibitemShut {NoStop}%
\bibitem [{\citenamefont
  {Hastings}(2007)}]{hastingsQuantumBeliefPropagation2007}%
  \BibitemOpen
  \bibfield  {author} {\bibinfo {author} {\bibfnamefont {M.~B.}\ \bibnamefont
  {Hastings}},\ }\bibfield  {title} {\bibinfo {title} {Quantum belief
  propagation: {{An}} algorithm for thermal quantum systems},\ }\href
  {https://doi.org/10.1103/PhysRevB.76.201102} {\bibfield  {journal} {\bibinfo
  {journal} {Phys. Rev. B}\ }\textbf {\bibinfo {volume} {76}},\ \bibinfo
  {pages} {201102} (\bibinfo {year} {2007})}\BibitemShut {NoStop}%
\bibitem [{\citenamefont {Leifer}\ and\ \citenamefont
  {Poulin}(2008)}]{leiferQuantumGraphicalModels2008}%
  \BibitemOpen
  \bibfield  {author} {\bibinfo {author} {\bibfnamefont {M.~S.}\ \bibnamefont
  {Leifer}}\ and\ \bibinfo {author} {\bibfnamefont {D.}~\bibnamefont
  {Poulin}},\ }\bibfield  {title} {\bibinfo {title} {Quantum {{Graphical
  Models}} and {{Belief Propagation}}},\ }\href
  {https://doi.org/10.1016/j.aop.2007.10.001} {\bibfield  {journal} {\bibinfo
  {journal} {Ann. Phys. (N. Y.)}\ }\textbf {\bibinfo {volume} {323}},\ \bibinfo
  {pages} {1899} (\bibinfo {year} {2008})}\BibitemShut {NoStop}%
\bibitem [{\citenamefont {Sahu}\ and\ \citenamefont
  {Swingle}(2022)}]{sahuEfficientTensorNetwork2022}%
  \BibitemOpen
  \bibfield  {author} {\bibinfo {author} {\bibfnamefont {S.}~\bibnamefont
  {Sahu}}\ and\ \bibinfo {author} {\bibfnamefont {B.}~\bibnamefont {Swingle}},\
  }\href {https://doi.org/10.48550/arXiv.2206.04701} {\bibinfo {title}
  {Efficient tensor network simulation of quantum many-body physics on sparse
  graphs}} (\bibinfo {year} {2022}),\ \Eprint
  {https://arxiv.org/abs/2206.04701} {arxiv:2206.04701 [quant-ph]} \BibitemShut
  {NoStop}%
\bibitem [{\citenamefont {Gray}\ and\ \citenamefont
  {Kourtis}(2021)}]{gray2021hyper}%
  \BibitemOpen
  \bibfield  {author} {\bibinfo {author} {\bibfnamefont {J.}~\bibnamefont
  {Gray}}\ and\ \bibinfo {author} {\bibfnamefont {S.}~\bibnamefont {Kourtis}},\
  }\bibfield  {title} {\bibinfo {title} {Hyper-optimized tensor network
  contraction},\ }\href@noop {} {\bibfield  {journal} {\bibinfo  {journal}
  {Quantum}\ }\textbf {\bibinfo {volume} {5}},\ \bibinfo {pages} {410}
  (\bibinfo {year} {2021})}\BibitemShut {NoStop}%
\bibitem [{\citenamefont {Xie}\ \emph {et~al.}(2012)\citenamefont {Xie},
  \citenamefont {Chen}, \citenamefont {Qin}, \citenamefont {Zhu}, \citenamefont
  {Yang},\ and\ \citenamefont
  {Xiang}}]{xieCoarsegrainingRenormalizationHigherorder2012}%
  \BibitemOpen
  \bibfield  {author} {\bibinfo {author} {\bibfnamefont {Z.~Y.}\ \bibnamefont
  {Xie}}, \bibinfo {author} {\bibfnamefont {J.}~\bibnamefont {Chen}}, \bibinfo
  {author} {\bibfnamefont {M.~P.}\ \bibnamefont {Qin}}, \bibinfo {author}
  {\bibfnamefont {J.~W.}\ \bibnamefont {Zhu}}, \bibinfo {author} {\bibfnamefont
  {L.~P.}\ \bibnamefont {Yang}},\ and\ \bibinfo {author} {\bibfnamefont
  {T.}~\bibnamefont {Xiang}},\ }\bibfield  {title} {\bibinfo {title}
  {Coarse-graining renormalization by higher-order singular value
  decomposition},\ }\href {https://doi.org/10.1103/PhysRevB.86.045139}
  {\bibfield  {journal} {\bibinfo  {journal} {Phys. Rev. B}\ }\textbf {\bibinfo
  {volume} {86}},\ \bibinfo {pages} {045139} (\bibinfo {year}
  {2012})}\BibitemShut {NoStop}%
\bibitem [{\citenamefont {Iino}\ \emph {et~al.}(2019)\citenamefont {Iino},
  \citenamefont {Morita},\ and\ \citenamefont
  {Kawashima}}]{iinoBoundaryTensorRenormalization2019}%
  \BibitemOpen
  \bibfield  {author} {\bibinfo {author} {\bibfnamefont {S.}~\bibnamefont
  {Iino}}, \bibinfo {author} {\bibfnamefont {S.}~\bibnamefont {Morita}},\ and\
  \bibinfo {author} {\bibfnamefont {N.}~\bibnamefont {Kawashima}},\ }\bibfield
  {title} {\bibinfo {title} {Boundary {{Tensor Renormalization Group}}},\
  }\href {https://doi.org/10.1103/PhysRevB.100.035449} {\bibfield  {journal}
  {\bibinfo  {journal} {Phys. Rev. B}\ }\textbf {\bibinfo {volume} {100}},\
  \bibinfo {pages} {035449} (\bibinfo {year} {2019})}\BibitemShut {NoStop}%
\bibitem [{\citenamefont {Gray}(2018)}]{Quimb}%
  \BibitemOpen
  \bibfield  {author} {\bibinfo {author} {\bibfnamefont {J.}~\bibnamefont
  {Gray}},\ }\bibfield  {title} {\bibinfo {title} {quimb: A python package for
  quantum information and many-body calculations},\ }\href
  {https://doi.org/10.21105/joss.00819} {\bibfield  {journal} {\bibinfo
  {journal} {J. Open Source Softw.}\ }\textbf {\bibinfo {volume} {3}},\
  \bibinfo {pages} {819} (\bibinfo {year} {2018})}\BibitemShut {NoStop}%
\end{thebibliography}%

\section*{Acknowledgments}
JG thanks Nicola Pancotti for many fruitful discussions of BP and TNs. We thank the authors of Refs.~\cite{Anand2023,tindall2023efficient,Kechedzhi2023,Shao2023,liao2023simulation} for sharing their data presented in Fig.~\ref{fig:method-comparison}.

\textbf{Funding:} The authors were supported by the US Department of Energy, Office of Science, Office of Advanced Scientific Computing Research and Office of Basic Energy Sciences, Scientific Discovery through Advanced Computing (SciDAC) program under Award Number DE-SC0022088. TB acknowledges financial support from the Swiss National Science Foundation through the Postdoc Mobility Fellowship (grant number P500PN-214214). GKC is a Simons Investigator in Physics. Computations presented here were partly conducted in the Resnick High Performance Computing Center, a facility supported by Resnick Sustainability Institute at the California Institute of Technology. This research used resources of the National Energy Research Scientific Computing Center (NERSC), a U.S. Department of Energy Office of Science User Facility located at Lawrence Berkeley National Laboratory, operated under Contract No. DE-AC02-05CH11231 using NERSC award BES-ERCAP0024087.

\textbf{Author contributions:} GKC conceptualized and supervised the work. JG and TB participated in conceptualization, implemented the computational methods, carried out simulations, and prepared the data and figures. All authors contributed to the analysis of the results and writing of the manuscript.

\textbf{Competing interests:} GKC is part owner of QSimulate Inc and has served as a scientific consultant to the Flatiron Institute. Between 2019-2021 he was on the advisory board of the Qingdao Institute of Theoretical and Computational Chemistry. 
He is the Chief Scientific Advisor to QSimulate and an Associate Researcher of the HK Quantum AI lab. All other authors declare that they have no competing interests.

\textbf{Data availability:} All data needed to evaluate the conclusions in the paper are present in the paper and the Supplementary Materials. Data presented in Figs.~\ref{fig:SPD_Quick}--\ref{fig:Z_62} are available at \url{www.github.com/tbegusic/arxiv-2308.05077-data} (Zenodo access code \url{www.doi.org/10.5281/zenodo.10223349}). Code for running the SPD and TN simulations is available on GitHub (\url{www.github.com/tbegusic/spd} and \url{www.github.com/jcmgray/mixpic-bp-quantum-dynamics} respectively).

\clearpage

\setcounter{page}{1}
\setcounter{section}{0}
\setcounter{equation}{0}
\setcounter{figure}{0}
\renewcommand{\thefigure}{S\arabic{figure}}
\setcounter{table}{0}
\renewcommand{\thetable}{S\arabic{table}}

\title{Supplementary materials for: Fast and converged classical simulations of evidence for the utility of quantum computing before fault tolerance}

\maketitle

\clearpage

\subsection*{Supplementary Text}

\subsubsection*{Detailed descriptions of the lazy TN contractions}

We here give a detailed description  of the tensor networks formed and contracted in Fig.~\ref{fig:tnlbp}. A generic requirement for us to run lazy BP on a tensor network is that every tensor belongs uniquely to a single site only. The PEPS and PEPO parts of the TN clearly satisfy this by construction, as do the single qubit gates that appear in $U$. For the two-qubit gates we perform a SVD, grouping indices spatially, to yield a single tensor per site, connected by a bond. In the case of the $ZZ$ rotation here, the gate is exactly low-rank with a bond dimension of 2. If we further contract all tensors with the same site, $j$, and layer (time-step), $t$, (effectively fusing single and two-qubit gates), then we are left with a TN that looks like Fig.~\ref{fig:tnlbp}A, with each unitary layer of gates, $U_t$, appearing as a PEPO. This step, which yields a TN with exactly one tensor per site \emph{and} layer, $T_{j,t}$, is not strictly necessary but helps in clarity of presentation.
We thus initially have
\begin{align}
    \langle 0^{\otimes N} | U^\dagger O U  | 0^{\otimes N} \rangle = 
    \sum_{\{x\}}
    \prod_{j \in 1 \ldots N}
    {T}^{\psi \dagger}_{j, 0} T^{\psi}_{j,0}
    \prod_{j \in 1 \ldots N, t \in 1 \ldots T}
    T^{U \dagger}_{j,t} T^{U}_{j,t}
    \prod_{j \in 1 \ldots N}
     T^{\Phi}_{j,T + 1}~,
\end{align}
where $T^{\psi}_{j,0}$ is the tensor of site $j$ in the PEPS $\psi$ representing the initial state at $t=0$,
$T^{U}_{j,t}$ is the tensor of site $j$ and layer $t$ of the unitary $U$ presented as a PEPO, and 
$T^{\Phi}_{j,T + 1}$ is the tensor at site $j$ of the PEPO representation of initial operator $\Phi=O$, and the sum over $\{x\}$ refers to the sum over all indices appearing on each tensor which for brevity we do not show here. 
As we evolve the PEPS and PEPO we will generate intermediate tensors $T^{\psi}_{j,t}$ and $T^{\Phi}_{j,t}$.
Each tensor carries a subset of all indices $\bar x_T$, either physical (connecting between layers of the same site) or virtual (connecting between sites of the same layer -- green bonds in Fig.~\ref{fig:tnlbp}A).
The physical indices and the virtual indices of $U$ here are always of dimension 2 whereas the virtual indices of the $\psi$ and $\Phi$ tensors generically (after some evolution) will be between $1$ and $\chi$.

There are three different tensor networks we run need to run `lazy' BP on. The first is when we evolve the PEPS by one step and want to combine and compress the next layer of gates:
\begin{align}
    U(t + 1)|\psi(t)\rangle \rightarrow |\psi(t+1)\rangle~.
\end{align}
We first form a sub network of the tensors in layers $\{t, t+1\}$. Since there are dangling indices (which would connect to layer $t+2$) and we want to compress, we need to run `2-norm' BP.
Specifically we form the 2-norm TN:
\begin{align}
    \langle \psi(t) | U^\dagger(t+1)
    U(t + 1) | \psi(t) \rangle
    =
    \sum_{\{x : \delta[t+1,t+2] \} }
    \prod_{j \in 1\ldots N}
    {T}^{\psi \dagger}_{j, t} 
    T^{U \dagger}_{j,t + 1}
    T^{U }_{j,t + 1}
    T^{\psi}_{j,t}~,
\end{align}
where with the sum and delta we imply that only indices connecting layers $t+1$ to $t+2$ are traced over and the remaining indices are summed normally (as shown in Fig.~\ref{fig:tnlbp}B(i)).
If we consider site $j$ and neighbor $k$ then each of the four tensors at each site is connected by a virtual bond to its matching partner: two of size $\leq \chi$ and two of size 2. The two BP messages for this bond, $m_{j\rightarrow k}, m_{k \rightarrow j}$, should thus should be of size $\leq 4\chi^2$, and we will be able to factorize them as matrices, by fusing all the `bra' ($\dagger$) and all the `ket' indices separately.
The expression to compute an updated message (which we initialize as the uniform distribution), is given by:
\begin{align}
    m'_{j \rightarrow k} = 
    \sum_{\{x\} \setminus {\{j \rightarrow k\}}}
    {T}^{\psi \dagger}_{j, t} 
    T^{U \dagger}_{j,t + 1}
    T^{U }_{j,t + 1}
    T^{\psi}_{j,t}
    \prod_{l \in \textrm{neighbors}(j) : l \neq k}
    m_{l \rightarrow j}
    ~,
\end{align}
where the sum denotes that the indices shared between site $j$ and $k$ are not summed over.
Performing this message update contraction lazily, i.e. making using of associativity to find an optimized sequence of intermediates, is much cheaper than explicitly forming a single tensor site, which (for our degree 3 lattice) would have size $64\chi^6$, whereas this lazy method scales like 
$\mathcal{O}(\chi^4)$.
This is the same scaling as Refs.~\cite{tindall2023efficient,tindallGaugingTensorNetworks2023}.

The second case we run lazy BP on is the PEPO evolution step. Here we want to compress the operator PEPO along with the next layer of gates:
\begin{align}
    U^\dagger(t - 1) \Phi(t) U(t- 1) \rightarrow \Phi(t-1)
    ~.
\end{align}
Again we have dangling indices (two per site that would connect to layer $t-2$) and thus must form the 2-norm of this operator in order to run BP. Specifically, we form:
\begin{align}
    \left \langle \left \langle 
    U^\dagger(t - 1) \Phi(t) U(t- 1)
    |
    U^\dagger(t - 1) \Phi(t) U(t- 1)
    \right \rangle \right \rangle
    =
    \sum_{\{x : \delta[t-1,t-2] \} }
    \prod_{j \in 1\ldots N}
    T^{U\dagger}_{j,t-1}
    T^{\Phi \dagger}_{j,t}
    T^{U}_{j,t-1}
    T^{U \dagger}_{j,t-1}
    T^{\Phi}_{j,t}
    T^{U}_{j,t-1}
\end{align}
where with the sum and delta we imply that only indices connecting layers $t-1$ to $t-2$ are traced over and the remaining indices are summed normally (as shown in Fig.~\ref{fig:tnlbp}B(ii)).
In this case, considering two neighboring sites $j$ and $k$, we have connecting bonds: two of dimension up to $\chi$ coming from the central operator PEPOs $\Phi$ and $\Phi^\dagger$ and four of dimension 2 coming from the unitary PEPOs. The two BP messages in this case will thus be of size $16 \chi^2$ and will again be able to be factorized as matrices due to the 2-norm structure. The messages are updated according to the contraction:
\begin{align}
    m'_{j \rightarrow k} = 
    \sum_{\{x\} \setminus {\{j \rightarrow k\}}}
    T^{U\dagger}_{j,t-1}
    T^{\Phi \dagger}_{j,t}
    T^{U}_{j,t-1}
    T^{U \dagger}_{j,t-1}
    T^{\Phi}_{j,t}
    T^{U}_{j,t-1}
    \prod_{l \in \textrm{neighbors}(j) : l \neq k}
    m_{l \rightarrow j}
    ~,
\end{align}
which again we perform as an optimized sequence of pair-wise contractions, to find a scaling of $\mathcal{O}(\chi^4)$.

The final TN we run lazy BP on is the operator expectation once the PEPS $\psi$ and PEPO $\Phi$ have reached each other at time $\tau \sim T / 2$. In this case the network $\langle \psi(\tau) | \Phi(\tau + 1) | \psi(\tau) \rangle$ has no dangling indices and so we do not need to form the 2-norm, instead approximating the contraction directly.
This TN takes the form:
\begin{align}
    \langle \psi(\tau) | \Phi(\tau + 1) | \psi(\tau) \rangle 
    = 
    \sum_{\{x\}}
    \prod_{j \in 1 \ldots N}
    T^{\psi \dagger}_{j, \tau} 
    T^{\Phi}_{j, \tau + 1}
    T^{\psi}_{j, \tau}
     ~,
\end{align}
(as shown in Fig.~\ref{fig:tnlbp}B(iii)).
In this case there are three bonds between any two neighbors $j$ and $k$, each of dimension up to $\chi$, and thus the two BP messages will have total size $\chi^3$ and will \emph{not} be factorizable as matrices.
The message update is given by:
\begin{align}
    m'_{j \rightarrow k} = 
    \sum_{\{x\} \setminus {\{j \rightarrow k\}}}
    T^{\psi \dagger}_{j, \tau} 
    T^{\Phi}_{j, \tau + 1}
    T^{\psi}_{j, \tau}
    \prod_{l \in \textrm{neighbors}(j) : l \neq k}
    m_{l \rightarrow j}
    ~.
\end{align}
Here contracting the three tensors at a single site would yield a tensor with size $\chi^9$, whereas the the optimized sequence of contractions scales like $\mathcal{O}(\chi^6)$. Note this is only for the MIX method where both $\psi$ and $\Phi$ have bond dimensions $\chi$.

In the current implementation we perform a single iteration of BP by updating the messages in parallel, which is to say, each new message is computed by contraction with messages from the previous round only (see Fig.~\ref{fig:tnlbp}C).
We measure the change in a message as:
\begin{align}
    \Delta_{j, k} = |m'_{j,k} - m_{j,k}|_1
\end{align}
and stop iterating when the maximum change, $\Delta_{j, k}$, over all messages drops below a specified threshold, here typically taken as $5\times10^{-6}$ for single precision.

\subsubsection*{Error analysis}

\begin{figure}[H]
    \centering
    \includegraphics[width=\textwidth]{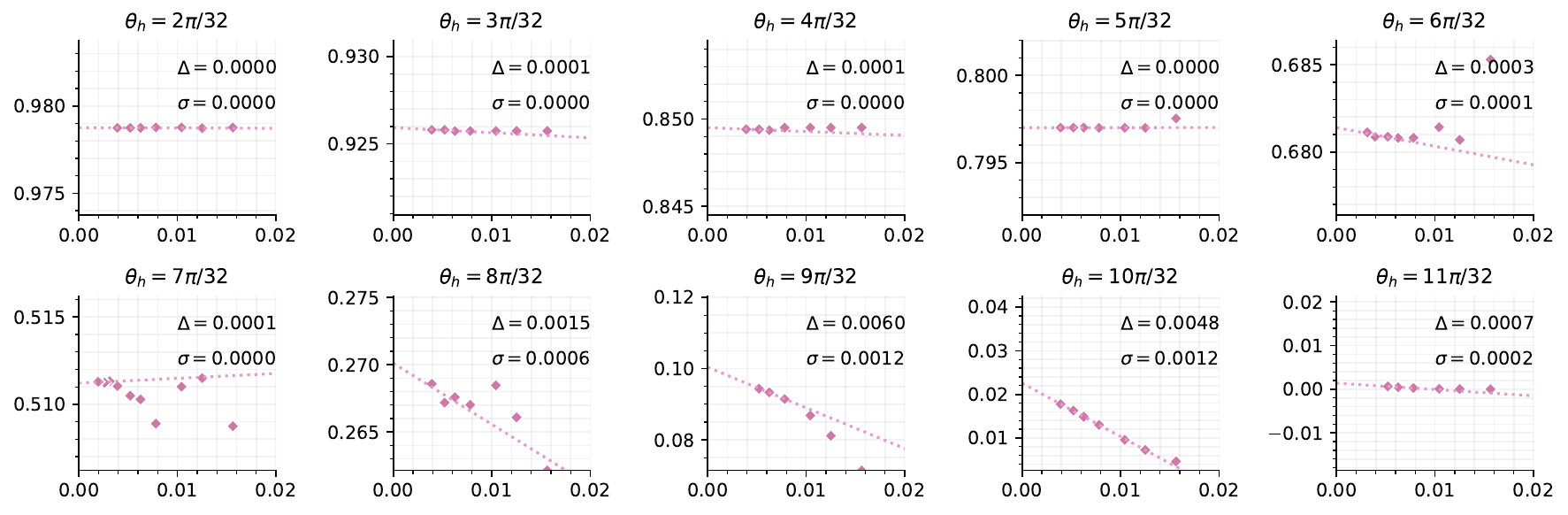}
    \caption{\textbf{Convergence of the MIX expectation value $\mathbf{\langle Z_{62} \rangle}$ after 20 steps and linear extrapolation (dashed line) using the last three points, i.e., three points with the highest available $\mathbf{\chi}$.} $\Delta = |\langle Z_{62} \rangle - \langle Z_{62} \rangle_{\chi \rightarrow  \infty}|$ is the difference between the expectation value at the highest $\chi$ and the extrapolated value, while $\sigma$ corresponds to the standard deviation of the last three points.}
    \label{fig:convergence-extrapolation-error-mix}
\end{figure}

To estimate the convergence error of the MIX method, we propose three independent metrics. First, for every $\theta_h$ we consider the standard deviation $\sigma$ of three results with the highest available $\chi$. These are presented in Fig.~\ref{fig:convergence-extrapolation-error-mix}. The highest $\sigma$ is found at $\theta_h=10\pi/32$ with the value of $1.2 \times 10^{-3}$.

Second, a linear extrapolation using the last three points is compared to the most converged (highest $\chi$) result for each $\theta_h$ and the error is computed as
\begin{equation*}
    \Delta = |\langle O \rangle - \langle O \rangle_{\chi \rightarrow \infty}|.
\end{equation*}
Figure~\ref{fig:convergence-extrapolation-error-mix} shows the data at different $\theta_h$ and different $\chi$, as well as the extrapolation fits and the associated error estimates $\Delta$. Although the extrapolation approach estimates larger errors than the standard deviation metric, all error estimates remain well below $0.01$. 

\begin{figure}[H]
    \centering
    \includegraphics[width=\textwidth]{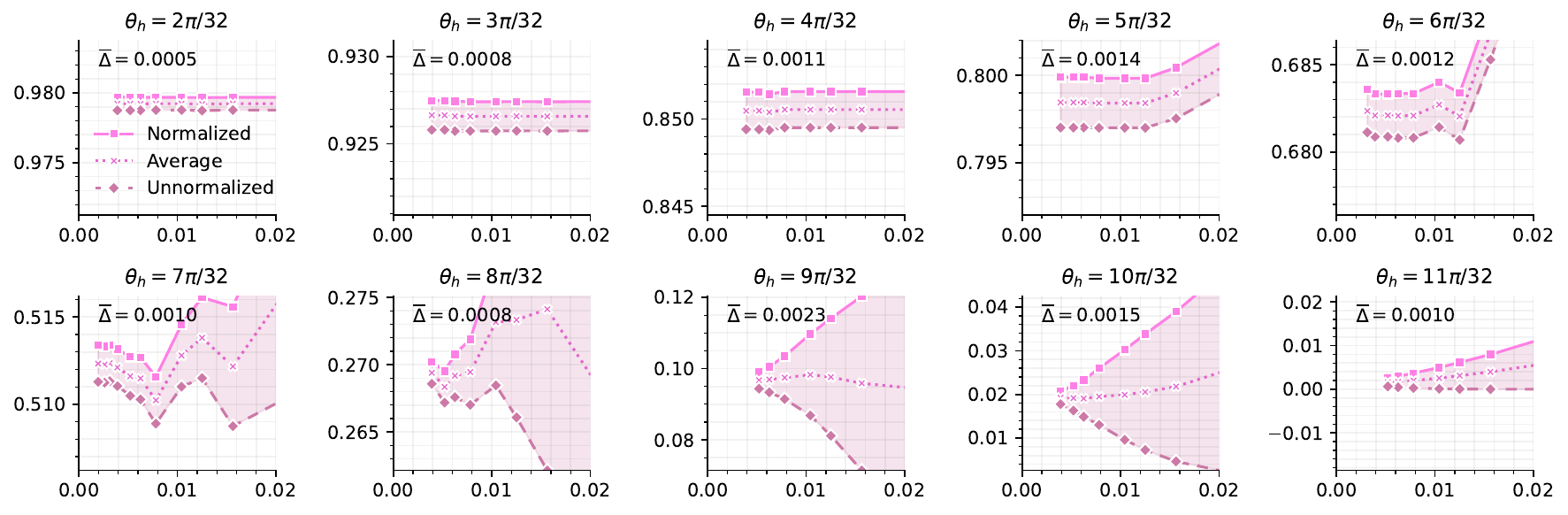}
    \caption{\textbf{Convergence of the unnormalized MIX expectation value $\mathbf{\langle Z_{62} \rangle}$ after 20 steps compared with the normalized $\mathbf{\langle Z_{62} \rangle / N_{\rm MIX}}$ and average $\mathbf{\langle Z_{62} \rangle_{\rm \textbf{average}} = (\langle Z_{62} \rangle + \langle Z_{62} \rangle / N_{\rm MIX})/2}$ expectation values.} $\bar{\Delta} = |\langle Z_{62} \rangle - \langle Z_{62} \rangle_{\rm average}|$ for $\langle Z_{62} \rangle$ evaluated at the highest available $\chi$ for each $\theta_h$.
    Note for $\theta_h \leq6 \pi / 32$ $\bar\Delta$ stops decreasing as the  
    truncation cutoff $\kappa=5\times10^{-6}$ starts to dominate over the maximum bond dimension $\chi$.
    }
    \label{fig:convergence-extrapolation-error-mix-av}
\end{figure}

Third, we recall that the expectation values presented in our results are computed without any renormalization of the wavefunction or operator. Here, we compare these expectation values to their normalized counterparts $\langle O \rangle / N$, where $N$ is the appropriate norm (e.g., $N_{\rm MIX} = N_{O} N_{\psi}$ for the mixed Schr\"odinger-Heisenberg approach). Figure~\ref{fig:convergence-extrapolation-error-mix-av} presents these two types of expectation values, along with their average $\langle O \rangle_{\rm av} = (\langle O \rangle + \langle O \rangle / N)/2$ for the MIX result of the 20-step expectation value $\langle Z_{62} \rangle$. Generally, the normalized and unnormalized methods approach the converged result from above and below, respectively, whereas their average converges faster by cancellation of errors. We can therefore use $\langle O \rangle_{\rm av}$ as an estimate of the exact value and $\bar{\Delta} = |\langle O \rangle - \langle O \rangle_{\rm av}|$ to estimate the error of the unnormalized result. Again, we find that the errors are conservatively below $0.01$, with the highest $\bar{\Delta} = 0.0023$ for $\theta_h=9\pi/32$. 

\subsubsection{TN timing analysis}

\begin{figure}
    \centering
    \includegraphics[width=0.8\linewidth]{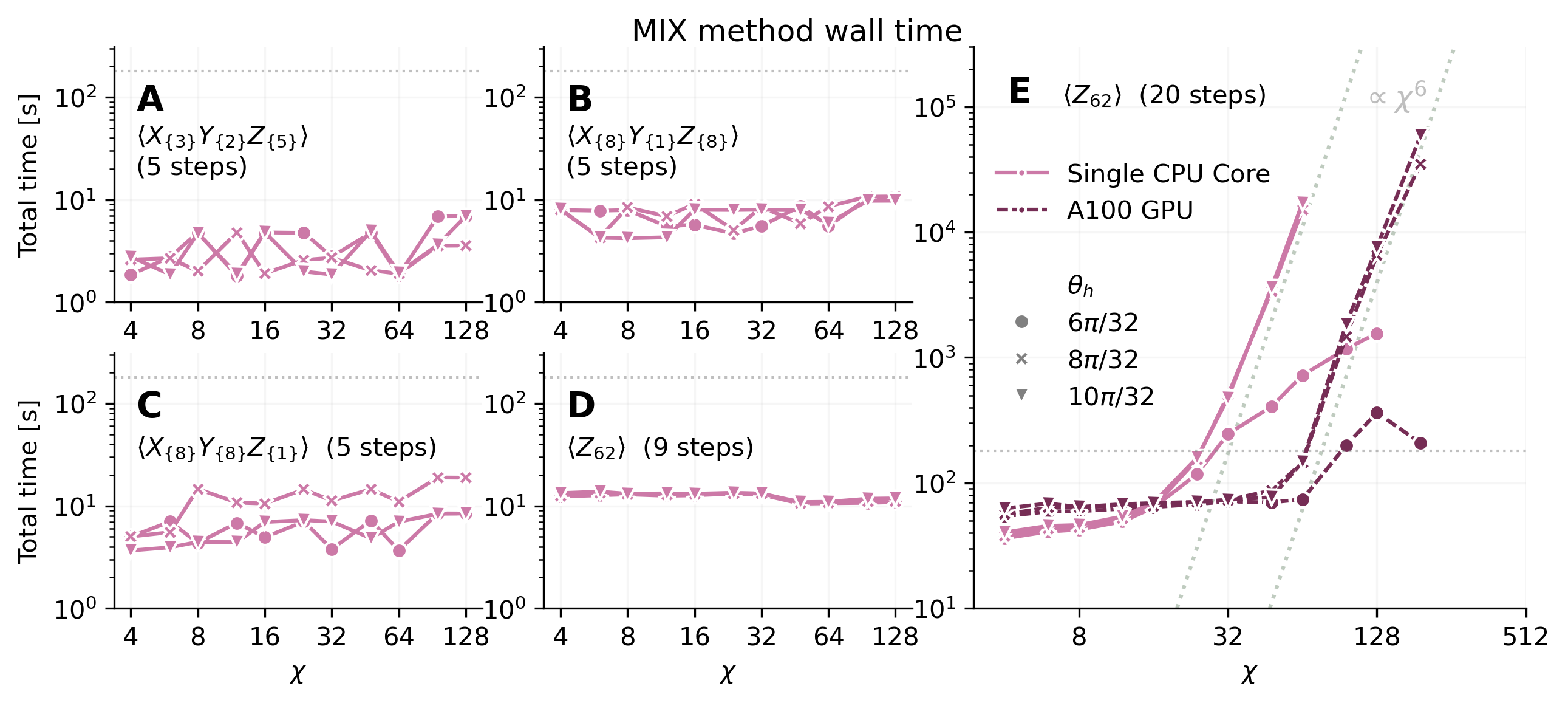}
    \caption{
    \textbf{Timings for the MIX TN algorithm.} \textbf{A}-\textbf{C}: the multi qubit observables at depth 5; \textbf{D}, \textbf{E}: the $\langle Z_{62}\rangle$ observable at depth 9 and 20. The time includes all steps including: circuit construction, L2BP compression, L1BP contraction of both the norms and observable, automatic contraction tree optimization (which can be cached from run to run). The horizontal line denotes 3 minutes, and the diagonal lines in panel \textbf{E} show a $\chi^6$ scaling as a guide.
    Three different angles, $\theta_h$, are shown, and two different contraction backends are shown - either a single core of a AMD EPYC 7742 or a single NVidia A100.
    }
    \label{fig:timings-mix}
\end{figure}

\begin{figure}
    \centering
    \includegraphics[width=0.8\linewidth]{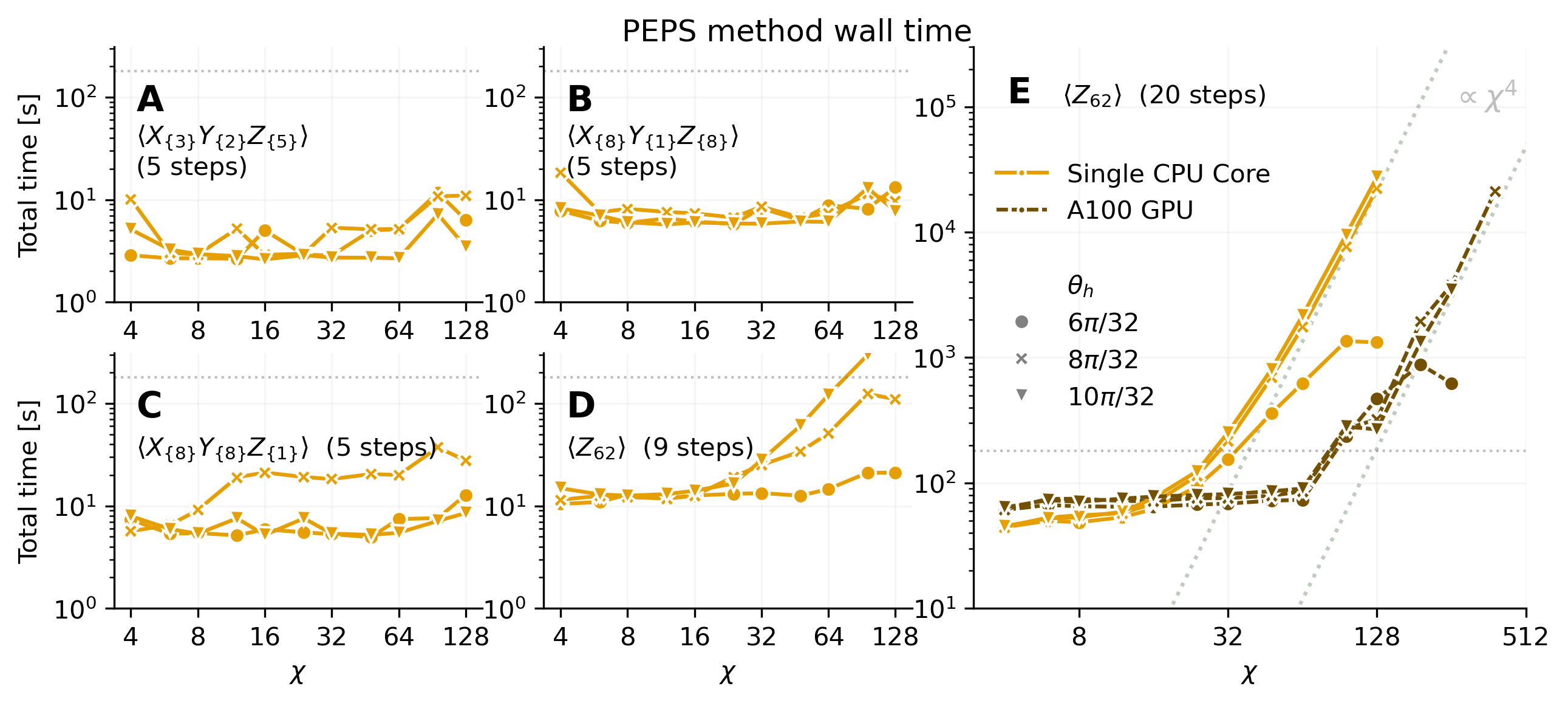}
    \caption{
    \textbf{Timings for the PEPS TN algorithm.} \textbf{A}-\textbf{C}: the multi qubit observables at depth 5; \textbf{D}, \textbf{E}: the $\langle Z_{62}\rangle$ observable at depth 9 and 20. The time includes all steps including: circuit construction, L2BP compression, L1BP contraction of both the norm and observable, automatic contraction tree optimization (which can be cached from run to run). The horizontal line denotes 3 minutes, and the diagonal lines in panel \textbf{E} show a $\chi^4$ scaling as a guide.
    Three different angles, $\theta_h$, are shown, and two different contraction backends are shown - either a single core of a AMD EPYC 7742 or a single NVidia A100.
    }
    \label{fig:timings-peps}
\end{figure}

\begin{figure}
    \centering
    \includegraphics[width=0.8\linewidth]{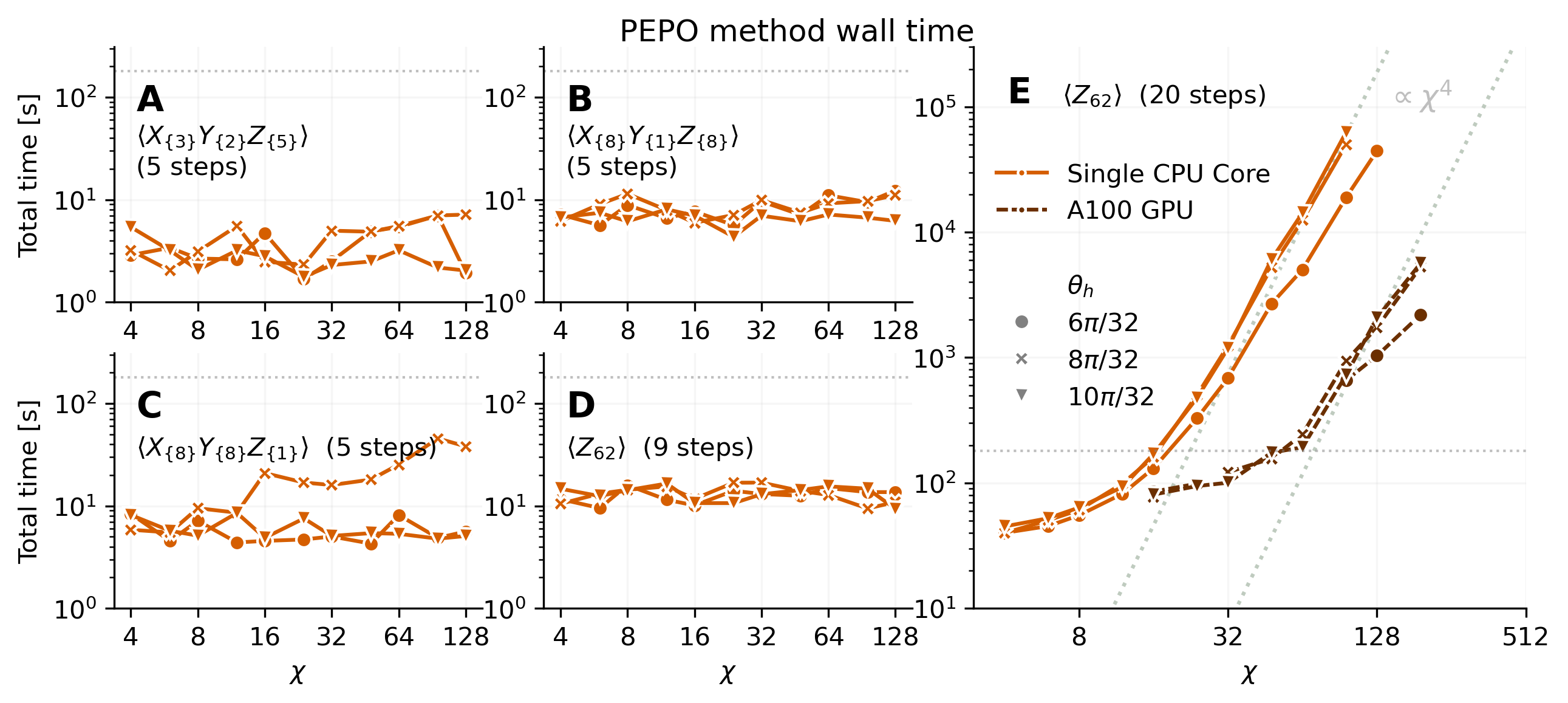}
    \caption{
    \textbf{Timings for the PEPO TN algorithm.} \textbf{A}-\textbf{C}: the multi qubit observables at depth 5; \textbf{D}, \textbf{E}: the $\langle Z_{62}\rangle$ observable at depth 9 and 20. The time includes all steps including: circuit construction, L2BP compression, L1BP contraction of both the norm and observable, automatic contraction tree optimization (which can be cached from run to run). The horizontal line denotes 3 minutes, and the diagonal lines in panel \textbf{E} show a $\chi^4$ scaling as a guide.
    Three different angles, $\theta_h$, are shown, and two different contraction backends are shown - either a single core of a AMD EPYC 7742 or a single NVidia A100.
    }
    \label{fig:timings-pepo}
\end{figure}

In Fig.~\ref{fig:timings-mix}, Fig.~\ref{fig:timings-peps}, and Fig.~\ref{fig:timings-pepo}, we show the total time to run the MIX, PEPS and PEPO TN methods respectively. Note that not all data in appearing in the main text was timed or used the same CPU/GPU and thus not all the data appears here.
We include in the total time all aspects of the algorithm. 
After the actual contractions involved in the L2BP compression and L1BP contraction, the second most time consuming part comes from automatic contraction tree optimization.
We note that for all the methods and 5 step observables the time is about 10 seconds for a single CPU core, and this is dominated by overhead.
It is only the 20 step $\langle Z_{62} \rangle$ simulations where one can see the scaling with $\chi$, and here we show both total time using a single CPU core (of a AMD EPYC 7742) time vs using a single GPU (NVIDIA A100).
We show three different angles, $\theta_h=6\pi/32, 8\pi/32, 10\pi/32$. 
Notably at $\theta_h=6\pi/32$ the entanglement growth is slowest and in fact the dynamic singular value cutoff ($\kappa=5\times10^{-6}$) limits the bond dimension of the PEPS or PEPO, whereas at the two higher values the bond dimension becomes limited by the maximum allowed value $\chi$ and we see the asymptotic complexity regime.
Due to the fact that the final TN that we L1BP contract in the MIX method has regions connected by 3 bonds of size up to $\chi$ (see Fig.~\ref{fig:tnlbp}B(iii), this method has a higher scaling of $\chi^6$, which is evident in Fig.~\ref{fig:timings-mix}E, as compared to the $\chi^4$ scaling evident in Fig.~\ref{fig:timings-peps}E and Fig.~\ref{fig:timings-pepo}E for the PEPS and PEPO methods.
It offers better performance despite this since it effectively doubles the accessible evolution time for a given $\chi$.

\newpage
\subsection*{Supplementary Figures}

\begin{figure}[H]
    \centering
    \includegraphics[width=\textwidth]{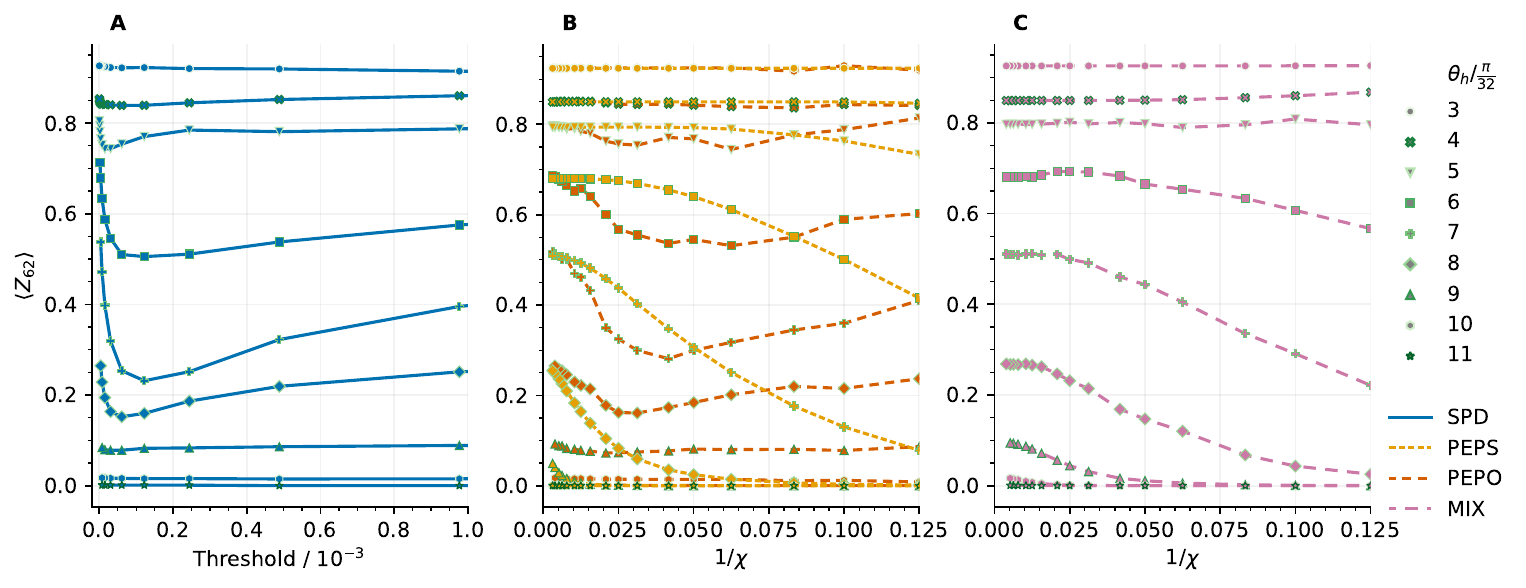}
    \caption{\textbf{Same as Fig.~\ref{fig:convergence} but with an extended $x$-axis.} 
    }
    \label{fig:convergence-wide_xrange}
\end{figure}

\begin{figure}[H]
    \centering
    \includegraphics[width=\textwidth]{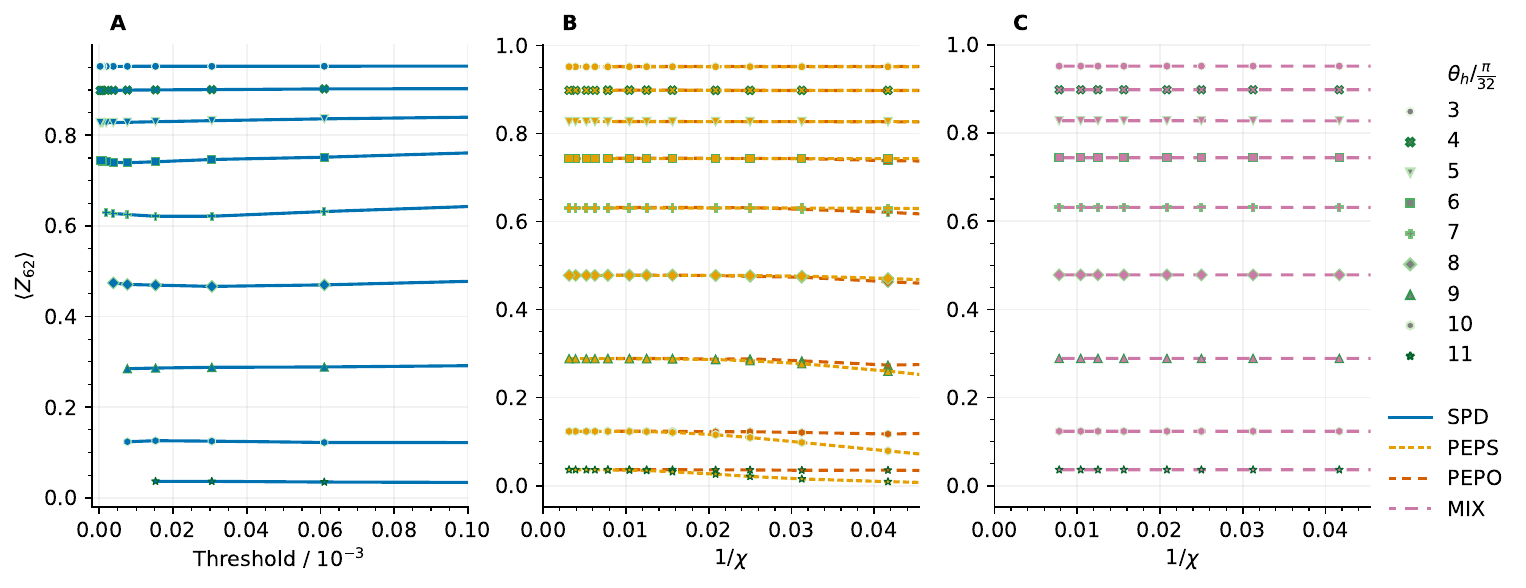}
    \caption{\textbf{Same as Fig.~\ref{fig:convergence-wide_xrange} but for the 9-step circuit.}}
    \label{fig:convergence-9steps}
\end{figure}

\begin{figure}[H]
    \centering
    \includegraphics[width=\textwidth]{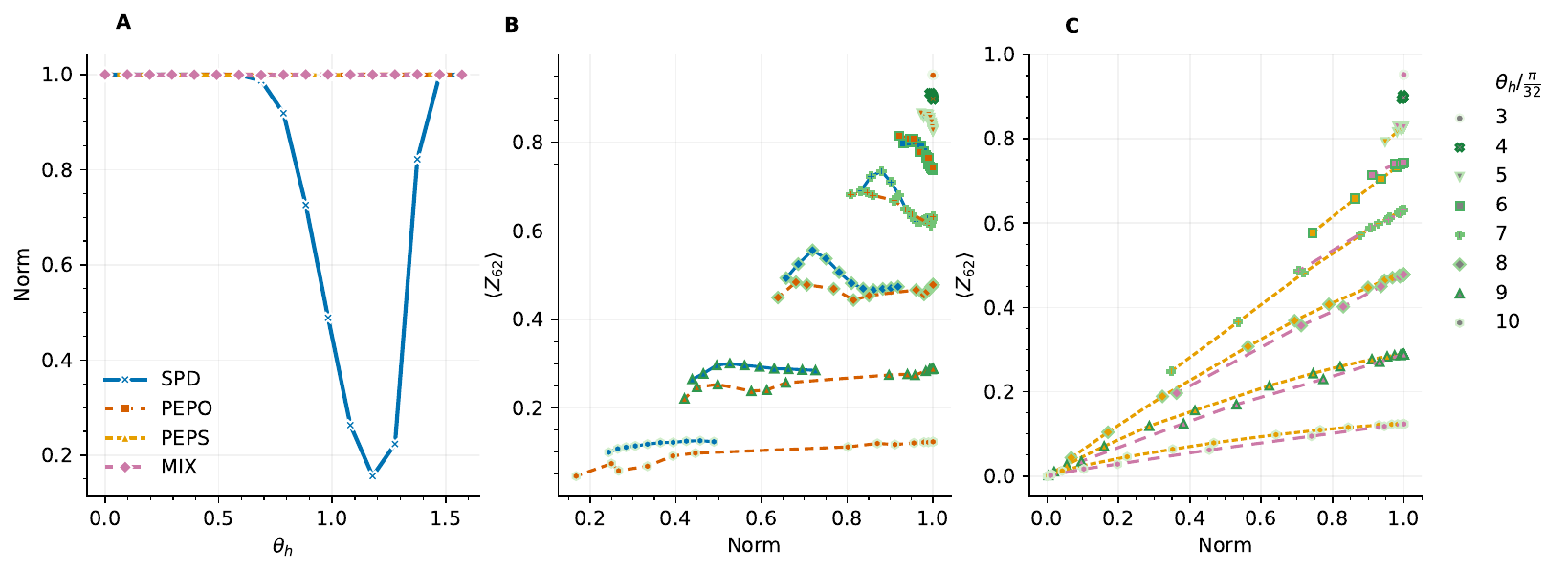}
    \caption{\textbf{Same as Fig.~\ref{fig:convergence-norm} of the main text but for the 9-step circuit.}}
    \label{fig:Convergence_9steps_norm}
\end{figure}

\begin{figure}[H]
    \centering
    \includegraphics[width=\textwidth]{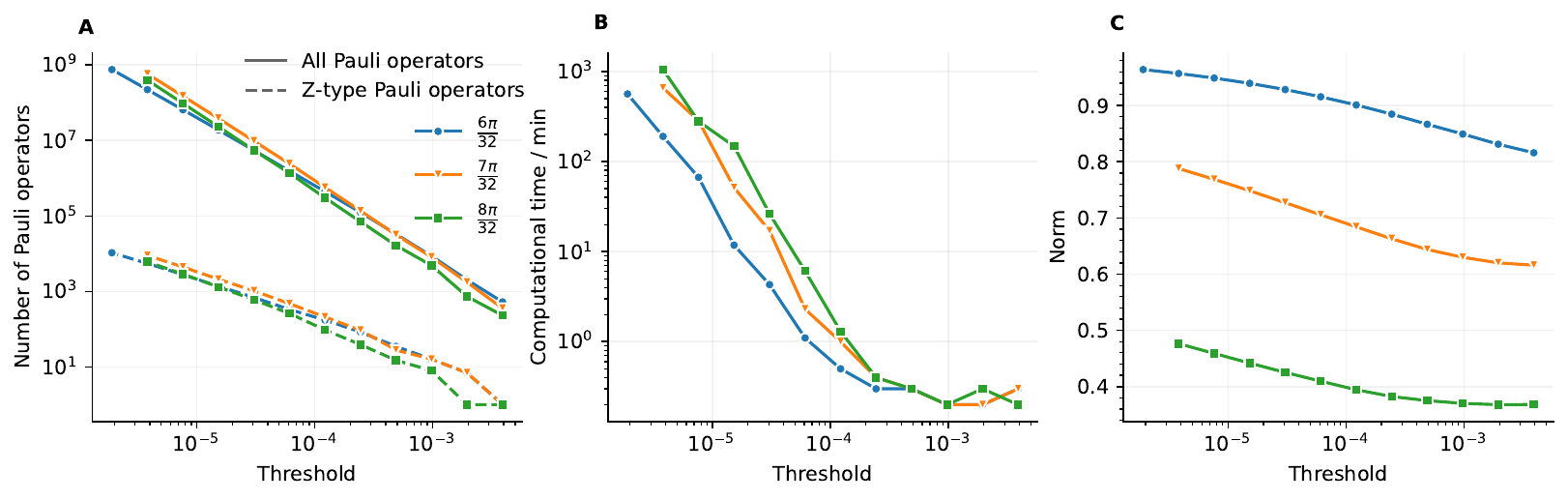}
    \caption{\textbf{Computational cost scaling of SPD.} Total number of Pauli operators (solid, A), the number of Z-type Pauli operators (dashed, A), computational time (B), and the Frobenius norm of the observable (C) at the end of the SPD simulation of $\langle Z_{62} \rangle$ (20-step quantum circuit) for $\theta_h = 6\pi/32, 7\pi/32, 8\pi/32$. The total number of Paulis and computational time scale roughly quadratically with the inverse of the threshold, the number of $Z$-type Paulis is inversely proportional to the threshold, whereas the norm is approximately logarithmic in this range of thresholds and for $\theta_h$ values presented.}
    \label{fig:SPD_Scaling}
\end{figure}

\begin{figure}[H]
    \centering
    \includegraphics[width=0.4\textwidth]{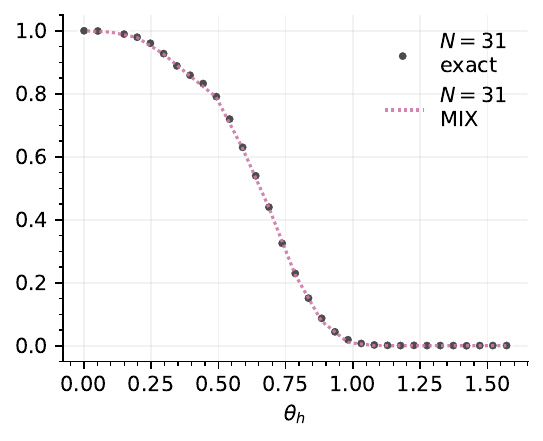}
    \caption{\textbf{Exact and MIX simulations of the 31-qubit model \cite{Kechedzhi2023}.}}
    \label{fig:31qubit-mix}
\end{figure}

\newpage
\subsection*{Supplementary Tables}

\begin{table}[H]
    \caption{\textbf{Thresholds for SPD simulations presented in Figs.~\ref{fig:SPD_Quick} and \ref{fig:Errors}.}}
    \centering
    \begin{ruledtabular}
    \begin{tabular}{lcc}
        & SPD  10 s & SPD \\
    \midrule
        Magnetization $M_Z$ & $10^{-3}$ & 0 \\
        $X_{13,29,31}Y_{9,30}Z_{8,12,17,28,32}$ & $1.5 \times 10^{-4}$  & $5 \times 10^{-6}$ \\ $X_{37,41,52,56,57,58,62,79}Y_{75}Z_{38,40,42,63,72,80,90,91}$ & $3.5 \times 10^{-4}$ & $10^{-5}$ \\
        $X_{37,41,52,56,57,58,62,79}Y_{38,40,42,63,72,80,90,91}Z_{75}$ & $3.5 \times 10^{-4}$ & $10^{-5}$ \\
        $Z_{62}$ & $8 \times 10^{-4}$
    \end{tabular}
    \end{ruledtabular}
    \label{tab:Thresholds1}
\end{table}

\begin{table}[H]
    \caption{\textbf{Timings (wall time in minutes using 4 CPU cores) for SPD simulations of high-weight Pauli observables with the thresholds reported in the third column of Table~\ref{tab:Thresholds1}.}}
    \centering
    \begin{ruledtabular}
    \begin{tabular}{cccc}
        $\theta_h / (\pi/32)$  & $X_{\{3\}}Y_{\{2\}}Z_{\{5\}}$ & $X_{\{8\}}Y_{\{1\}}Z_{\{8\}}$ & $X_{\{8\}}Y_{\{8\}}Z_{\{1\}}$  \\
    \midrule
      1   &  0.4   &  0.3   &  0.3   \\
      2   &  0.4   &  0.6   &  1.4   \\
      3   &  0.4   &  3.0   &  12.6  \\
      4   &  0.8   &  10.5  &  40.9  \\
      5   &  1.7   &  36.8  &  54.4  \\
      6   &  5.0   &  78.0  &  154.3 \\
      7   &  12.4  &  141.0 &  212.0 \\
      8   &  20.3  &  263.2 &  362.6 \\
      9   &  44.8  &  217.1 &  356.1 \\
      10  &  31.7  &  294.7 &  258.1 \\
      11  &  21.2  &  215.1 &  277.4 \\
      12  &  7.7   &  108.1 &  198.1 \\
      13  &  1.5   &  20.4  &  114.6 \\
      14  &  0.4   &  3.0   &  14.8  \\
      15  &  0.3   &  0.3   &  0.4
    \end{tabular}
    \end{ruledtabular}
    \label{tab:timings-spd}
\end{table}

\begin{table}[H]
    \caption{\textbf{Thresholds, numbers of Pauli operators at the end of simulation, and wall time (on 6 CPU cores) for the 9-step and 20-step SPD simulations of Fig.~\ref{fig:Z_62}.} Note that the number of Pauli operators operated with during the simulation can be much greater, especially in the examples where few Paulis are left at the end of the simulation.}
    \centering
    \begin{ruledtabular}
    \begin{tabular}{ccrrcrr}
      \multirow{2}{*}{$\theta_h / (\pi/32)$} &
        \multicolumn{3}{c}{9 steps} &
        \multicolumn{3}{c}{20 steps} \\
        & Threshold & $N_{\rm Pauli}$ & Timing / min & Threshold & $N_{\rm Pauli}$ & Timing / min \\
        \midrule
      1   & $2^{-22}$ & 1,723       & 0.5    & $2^{-22}$ & 10,527      & 0.4 \\
      2   & $2^{-22}$ & 50,391      & 2.0    & $2^{-22}$ & 705,896     & 0.4 \\
      3   & $2^{-22}$ & 1,037,984   & 0.4    & $2^{-22}$ & 16,447,691  & 7.4 \\
      4   & $2^{-22}$ & 15,630,592  & 1.8    & $2^{-22}$ & 215,196,358 & 190.9 \\
      5   & $2^{-22}$ & 226,944,006 & 44.2   & $2^{-21}$ & 885,442,824 & 803.5 \\
      6   & $2^{-21}$ & 999,386,176 & 229.8  & $2^{-19}$ & 760,142,525 & 562.7 \\
      7   & $2^{-19}$ & 833,860,251 & 197.5  & $2^{-18}$ & 582,723,097 & 656.2 \\
      8   & $2^{-18}$ & 865,920,748 & 489.2  & $2^{-18}$ & 398,851,143 & 1050.2 \\
      9   & $2^{-17}$ & 357,673,648 & 216.9  & $2^{-17}$ & 22,894,192  & 861.9 \\
      10 & $2^{-17}$ & 285,852,981  & 288.7  & $2^{-17}$  & 2,372,374  & 820.2 \\
      11 & $2^{-16}$ & 33,062,799   & 117.0  & $2^{-17}$  & 88,810     & 799.5 \\
      12 & $2^{-16}$ & 16,659,256   & 120.2  & $2^{-16}$  & 172        & 181.4 \\
      13 & $2^{-16}$ & 54,559,484   & 121.0  & $2^{-16}$  & 0          & 144.3 \\
      14 & $2^{-17}$ & 580,390,756  & 261.4  & $2^{-16}$  & 0          & 294.7 \\
      15 & $2^{-20}$ & 375,828,476  & 64.4   & $2^{-16}$  & 73,030,357 & 635.5
    \end{tabular} 
    \end{ruledtabular}
    \label{tab:Thresholds2}
\end{table}

\begin{table}[H]
    \caption{\textbf{Maximum bond dimension $\chi$ used for the TN simulations presented in the main text.}
    Note for 9 steps, the norm reaches $\sim 1$ for all $\theta_h$ (see Fig.~\ref{fig:Convergence_9steps_norm}A), implying sufficient $\chi$, at 256, 64, and 16 for PEPS, PEPO and MIX methods respectively. The dynamic singular value truncation threshold in all cases was set to $\kappa=5\times10^{-6}$. This, combined with the different entanglement growth at different $\theta_h$, is what leads to slightly different accessible $\chi$ values for each method at 20 steps.
    The `Mix 3min' method shown in Fig.~\ref{fig:SPD_Quick} was executed on a NVidia 4070 Ti.
    }
    \centering
    \begin{ruledtabular}
    \begin{tabular}{lcccc}
        Method & 5 steps & 9 steps & 20 steps ($\theta_h \leq 8 \pi / 32 $) & 20 steps ($\theta_h \geq 9 \pi / 32 $) \\
    \midrule
        PEPS & 64 & 320 & 320 & 320 \\
        PEPO & 64 & 128 & 320 & 256 \\
        MIX  & 64 & 128 & 256 & 192 \\
        MIX 3min & 64 & n/a & 64 & 64
    \end{tabular}
    \end{ruledtabular}
    \label{tab:tn-chis}
\end{table}

\end{document}